\documentclass[%
 reprint,
jmp,
superscriptaddress,
nofootinbib,
 amsmath,amssymb,
table
]{revtex4-2}
\usepackage[english]{babel}
\usepackage{graphicx}
\usepackage{dcolumn}
\usepackage{bm}
\usepackage[version=4]{mhchem}
\usepackage{multirow}
\usepackage{comment}
\usepackage{tabularx}
\usepackage{array}
\usepackage{makecell}
\usepackage{amssymb}
\usepackage{float}
\usepackage{afterpage}
\usepackage[utf8]{inputenc}
\usepackage{braket}
\usepackage{amsmath}
\usepackage{amsfonts}
\usepackage{booktabs} 
\usepackage{titlesec}
\usepackage{dsfont}
\usepackage{pifont}
\usepackage{adjustbox}
\usepackage[normalem]{ulem}
\usepackage{tabularray}
\usepackage[left=0.8in,right=0.8in,top=0.8in,bottom=0.8in]{geometry}
\usepackage{upgreek}
\usepackage{textgreek}
\usepackage[super]{nth}
\usepackage{csquotes}
\usepackage{xcolor}
\usepackage{hhline}

\newcolumntype{Y}{>{\centering\arraybackslash}X}

\definecolor{myblue}{RGB}{71,120,207}
\definecolor{mygreen}{RGB}{106,204,100}
\definecolor{myred}{RGB}{213,95,95}

\date{\today}

\makeatletter
\def\@email#1#2{%
 \endgroup
 \patchcmd{\titleblock@produce}
  {\frontmatter@RRAPformat}
  {\frontmatter@RRAPformat{\produce@RRAP{*#1\href{mailto:#2}{#2}}}\frontmatter@RRAPformat}
  {}{}
}%
\makeatother
\begin{document}
\preprint{AIP/123-QED}

\title{MACE-OFF: Short Range Transferable Machine Learning \\ Force Fields for Organic Molecules}

\author{D\'{a}vid P\'{e}ter Kov\'{a}cs\textsuperscript{\textdagger}}
\affiliation{Engineering Laboratory, University of Cambridge, Cambridge, CB2 1PZ, UK}

\author{J. Harry Moore\textsuperscript{\textdagger}\textsuperscript{*}}
\affiliation{Engineering Laboratory, University of Cambridge, Cambridge, CB2 1PZ, UK}
\affiliation{Ångström AI, 2325 3rd Street, San Francisco, CA 94107, USA}

\author{Nicholas J. Browning}
\affiliation{Swiss National Supercomputing Centre (CSCS), 6900, Lugano, Switzerland}
 
\author{Ilyes Batatia}
\affiliation{Engineering Laboratory, University of Cambridge, Cambridge, CB2 1PZ, UK}

\author{Joshua T. Horton}
\affiliation{School of Natural and Environmental Sciences, Newcastle University, Newcastle upon Tyne NE1 7RU, UK}

\author{Yixuan Pu}
\affiliation{Department of Physics and Astronomy, University College, London WC1E 6BT, UK}

\author{Venkat Kapil}
\affiliation{Yusuf Hamied Department of Chemistry,  University of Cambridge,  Lensfield Road,  Cambridge,  CB2 1EW,UK}
\affiliation{Department of Physics and Astronomy, University College, London WC1E 6BT, UK}
\affiliation{Thomas Young Centre and London Centre for Nanotechnology, London WC1E 6BT, UK}

\author{William C. Witt}
\affiliation{Department of Materials Science and Metallurgy, University of Cambridge, 27 Charles Babbage Road, CB3 0FS, Cambridge, United Kingdom}

\author{Ioan-Bogdan Magdău}
\affiliation{School of Natural and Environmental Sciences, Newcastle University, Newcastle upon Tyne NE1 7RU, UK}

\author{Daniel J. Cole}
\affiliation{School of Natural and Environmental Sciences, Newcastle University, Newcastle upon Tyne NE1 7RU, UK}

\author{G\'abor Cs\'anyi}
\affiliation{Engineering Laboratory, University of Cambridge, Cambridge, CB2 1PZ, UK}
\affiliation{Ångström AI, 2325 3rd Street, San Francisco, CA 94107, USA}

\begin{abstract}
\quad
\section*{Abstract}
    Classical empirical force fields have dominated biomolecular simulation for over 50 years. 
    Although widely used in drug discovery, crystal structure prediction, and biomolecular dynamics, they generally lack the accuracy and transferability required for first-principles predictive modeling.
    In this paper, we introduce MACE-OFF, a series of short range transferable force fields for organic molecules created using state-of-the-art machine learning technology and first-principles reference data computed with a high level of quantum mechanical theory. 
    MACE-OFF demonstrates the remarkable capabilities of short range models by accurately predicting a wide variety of gas and condensed phase properties of molecular systems. 
    It produces accurate, easy-to-converge dihedral torsion scans of unseen molecules, as well as reliable descriptions  of molecular crystals and liquids, including quantum nuclear effects. We further demonstrate the capabilities of MACE-OFF by determining free energy surfaces in explicit solvent, as well as the folding dynamics of peptides and nanosecond simulation of a fully solvated protein. 
    These developments enable first-principles simulations of molecular systems for the broader chemistry community at high accuracy and relatively low computational cost. 
\end{abstract}

\maketitle

\renewcommand*{\thefootnote}{\fnsymbol{footnote}}
\footnotetext[2]{These authors contributed equally.}
\footnotetext[1]{Corresponding author: \href{mailto:jhm72@cam.ac.uk}{jhm72@cam.ac.uk}}
\renewcommand*{\thefootnote}{\arabic{footnote}}

\section{Introduction}
Machine learning (ML) force fields have recently undergone major improvements in accuracy, robustness, and computational speed~\cite{lysogorskiy2021performant, batatia2022mace, kovacs2023mace_eval, batzner2023nequip, zaverkin2023transfer, haghighatlari2022newtonnet, shapeev2016moment, anstine2023aimnet2, kabylda2023efficient, behler2021four, musil2021physics, deringer2021gaussian, huang2021ab}. They are now routinely used in materials chemistry contexts where density functional theory was previously the method of choice. In these applications, available empirical force fields, such as the embedded-atom method~\cite{daw1984embedded}, do not provide sufficient accuracy and transferability to describe many scientifically interesting and challenging phenomena. 
Successful applications of ML potentials include simulation of quenching of amorphous silicon~\cite{deringer2021origins}, determination of the phase diagrams of inorganic perovskites~\cite{baldwin2023dynamic} and alloys~\cite{rosenbrock2021machine}, and device-scale simulation of phase-change memory materials~\cite{zhou2023device}. 

In contrast, simulating bio-organic systems entails a different set of trade-offs, with greater emphasis on simulating large systems over long timescales.  This means that empirical force fields, which sacrifice accuracy for computational speed, continue to be used routinely to study molecular liquids, crystals, biological systems, and drug-like molecules~\cite{hagler-1, hagler-2, openff-sage, Gapsys2020LargeAlchemy}. 


Two alternatives to empirical force fields are available. The first is semi-empirical quantum mechanics, such as the series of extended tight-binding models~\cite{bannwarth2019gfn2}, which represents a low-cost solution for small molecules. The method is limited by its moderate accuracy compared to quantum chemistry methods, its restriction to modelling non-periodic systems (if not in principle, but in widely used implementations), and its cubic scaling with system size. 

More recently, a number of transferable machine learning force fields have also been developed for organic chemistry. By ``transferable'', here we mean that these potentials are capable of generalising in system size and both chemical and conformation space sufficiently well to perform stable molecular dynamics and accurate property prediction for a wide range of molecular systems beyond those on which the model was trained. The most notable are the series of ANI~\cite{smith2017ani, smith2018ani_1x, smith2019approaching_ani1ccx, devereux2020ani_2x} and AIMNet potentials~\cite{Zubatyuk2019aimnet, zubatyuk2021aimnet_nse, anstine2023aimnet2}. ANI potentials pioneered the use of local symmetry function-based feedforward neural networks~\cite{behler2007generalized} trained on a large dataset of organic molecular geometries~\cite{smith2017ani_data, smith2020ani_cc_data} to create transferable ML force fields. The ANI-2x model became the most widely adopted ML force field and therefore serves as one of the primary points of comparison in this paper. The ANI-2x model was recently combined with a polarizable electrostatic model~\cite{inizan2023scalable} in a hybrid ML/MM simulation setting, and also with a neural network based dispersion correction~\cite{tu2023neural}. The AIMNet models apply a message passing architecture~\cite{gilmer2017neural}, where the initial embeddings are the ANI symmetry functions. They have also extended the applicability of the models to a larger set of chemical elements, as well as to charged species. These models relax the locality assumptions by incorporating electrostatic and dispersion interactions. The PhysNet model uses a message passing architecture~\cite{gilmer2017neural} and in addition to the semi-local terms also includes long-range electrostatic and dispersion interactions~\cite{unke2019physnet}. 

Other recent bio-organic force fields include the FENNIX model, which combines a local equivariant machine learning model with a physical long-range functional form for electrostatics and dispersion~\cite{Ple2023fennix}. The model was trained to reproduce the CCSD(T)/CBS energies of small molecules and molecular dimers. It was shown that FENNIX can be used to run stable dynamics of liquid water, solvated alanine dipeptides, and an entire protein in the gas phase. However, further benchmarking of this model is required to assess the accuracy of the intramolecular potential outside the training set and of condensed phase molecular dynamics simulations. 

Similarly, the ANA2B potential employs a short-ranged ML potential, with long-ranged, classical multipolar electrostatics, polarization and dispersion interactions~\cite{ana2b}. Although this long-ranged model shows promising accuracy for condensed phase properties and crystal structure ranking, its accuracy and computational performance has not yet been demonstrated for larger biomolecules. 

The GEMS model~\cite{unke-gems-sciadv}, built on the SpookyNet architecture~\cite{unke2021spookynet}, is another recent ML force field for biomolecular simulations. Whilst a transferable version of this potential is capable of producing stable dynamics for challenging biomolecular systems, certain properties, for example the folding dynamics of small peptides, require the addition of system-specific training configurations in order to accurately reproduce experimental properties. 
More recently, the transferable SO3LR potential was introduced by the same group of researchers~\cite{kabyldaMolecularSimulationsPretrained2024}. This is based on the equivariant SO3krates architecture and also includes analytic dispersion and electrostatic long range contributions. Whilst this potential is capable of fast and stable simulations of biomolecular systems, we are not yet aware of more extensive quantitative benchmarking on off-equilibrium energetics or thermodynamic properties.

In this paper, we introduce a family of purely local (i.e.\ short range), transferable bio-organic machine learning force fields which we refer to collectively as MACE-OFF. The force fields are parameterized for the ten most important chemical elements for organic chemistry: H, C, N, O, F, P, S, Cl, Br, and I. They are capable of accurately describing intra- and intermolecular interactions of neutral closed-shell systems. This enables the simulation of a wide range of chemical systems, from molecular liquids and crystals to drug-like molecules and biopolymers. 

The models were validated on a number of tasks, including the prediction of small molecule torsion barriers, geometry optimizations, calculations of lattice parameters and enthalpies of formation of molecular crystals, and the calculation of the Raman spectra of molecular crystals, including nuclear quantum effects. We have also validated the models for predicting the densities and heats of vaporization of a range of molecular liquids. In particular, we study how well MACE-OFF reproduces the fundamental properties of water, including density and radial distribution functions. To further showcase the capabilities of the model, we computed the free energy surface and J-coupling parameters of alanine tripeptide (Ala\textsubscript{3}) in vacuum and explicit water, simulated the folding of Ala\textsubscript{15} at different temperatures, and 
carried out an all-atom simulation of the protein crambin in explicit water (18\,000 atoms).
Finally, we tested the computational speed of the current implementation in the LAMMPS and OpenMM simulation packages, demonstrating both strong and weak scaling.

While truly general purpose force fields for biomolecular modeling require explicit treatment of long-range Coulomb interactions, the purpose of this paper is to push the capabilities of short-range potentials as far as they can go, in part as preparation for inclusion of explicit polarizable electrostatic interactions in the future. 

\section{Methods}
\subsection{The MACE architecture}

The MACE model~\cite{batatia2022mace} is a force field that maps the positions and chemical elements of atoms to the system's potential energy. Linear scaling with system size is achieved by decomposing the total energy into atomic site energies. First, a graph is defined by connecting two nodes (atoms) by an edge if they are in each other's local environment. The local environment $\mathcal{N}(i)$ is the set of all atoms $j$ around the central atom $i$ for which $ \|\bm{r}_{ij}\| \leq r_{\text{cut}}$, where $\bm{r}_{ij}$ is the vector from atom $i$ to atom $j$ and $r_{\text{cut}}$ is the cutoff hyperparameter. The array of features of node $i$ is denoted by ${\bm h}_i^{(t)}$ and is expressed in the spherical harmonic basis, with its elements indexed by $l$ and $m$. The superscript in $t$ indicates the iteration steps (corresponding to ``layers'' of message passing in the language of graph neural networks). All MACE-OFF models are made up of two layers. The ${\bm h}^{(t)}_{i}$ features depend on the chemical environment of the atoms for $t > 0$. 

The node features ${\bm h}^{(0)}_{i}$ are initialized as a (learnable) embedding of the chemical elements with atomic numbers $z_i$ into $k$ channels:
\begin{equation}
\label{eq:element_embedding}
    h_{i,k00}^{(0)} = \sum_z W_{kz} \delta_{zz_{i}}
\end{equation}
The superscript $(0)$ in this case indicates the initial (0-th) layer. This type of mapping has been widely applied to graph neural networks~\cite{schutt2017schnet, schutt2021PAINN, gasteiger2021gemnet} and other models~\cite{willatt2018feature, darby2023tensor}.

Next, for each atom, the features of its neighbours are combined with the interatomic displacement vectors, $\bm r_{ij}$, to form the one-particle basis $\phi_{ij,k \eta_{1} l_{3}m_{3}}^{(t)}$. The radial distance is used as an input into a learnable radial function $R(r_{ij})$ with several outputs that correspond to the different ways in which the displacement vector and the node features can be combined while preserving equivariance~\cite{wigner2012group}:
\begin{align}
  \label{eq:phi-basis-t}
  \phi_{ij,k \eta_{1} l_{3}m_{3}}^{(t)} &= 
    \sum_{l_1l_2m_1m_2} C_{\eta_1,l_1m_1l_2m_2}^{l_3m_3}
      R_{k \eta_{1} l_{1}l_{2} l_{3}}^{(t)}(r_{ij}) \,\,\times \notag\\
      & \qquad\qquad \times Y^{m_{1}}_{l_{1}} (\boldsymbol{\hat{r}}_{ij}) h^{(t)}_{j,kl_2m_2} 
\end{align}
where $Y_l^m$ are the spherical harmonics, and $C_{\eta_1,l_1m_1l_2m_2}^{l_3m_3}$ denotes the Clebsch-Gordan coefficients.
There are multiple ways of constructing an equivariant combination with a given symmetry $(l_3,m_3)$, and these multiplicities are enumerated by the index $\eta_1$~\cite{dusson2022atomic}.  

The one-particle basis $\phi$ is summed over the neighborhood and the $k$ channels are mixed together with learnable weights to form the permutation invariant atomic basis $A_i$:
\begin{equation}
\label{eq:atomic-basis-t}
    A_{i,kl_{3}m_{3}}^{(t)} = \sum_{\tilde{k}, \eta_{1}} W_{k \tilde{k} \eta_{1}l_{3}}^{(t)}
    \sum_{j \in \mathcal{N}(i)}  \phi_{ij,\tilde{k} \eta_{1} l_{3}m_{3}}^{(t)} 
\end{equation}

Higher-order (many-body) symmetric features are created on each atom $i$  by taking tensor products of the atomic basis, $A_i$, with itself $\nu$ times, resulting in the ``product basis''. The product basis is then contracted with the generalized Clebsch-Gordan coefficients, $\mathcal{C}^{LM}_{\eta_{\nu} \bm l \bm m}$, to obtain the equivariant many-body basis, $\bm B_i$~\cite{dusson2022atomic},
\begin{equation}
\label{eq:symmbasis_L1}
  {\bm B}^{(t),\nu}_{i,\eta_{\nu} k LM}
  = \sum_{{\bm l}{\bm m}} \mathcal{C}^{LM}_{\eta_{\nu} \bm l \bm m} \prod_{\xi = 1}^{\nu} A_{i,k l_\xi  m_\xi}^{(t)} 
\end{equation}
where bold $\bm l \bm m$ denotes the $\nu$-tuple of $l$ and $m$ values and similarly to Eq.~\eqref{eq:phi-basis-t}, $\eta_{\nu}$ enumerates the number of possible couplings to create the features with equivariance $LM$. The maximum body-order is controlled by the parameter $\nu$ and is fixed at 3 (corresponding to 4-body terms, which includes the central atom) for all MACE-OFF models in this work.

Finally, a ``message'' $m_{i}$ is created on each atom as a learnable linear combination of the equivariant many-body features:
\begin{equation}
 \label{eq:message}
      m_{i,k LM}^{(t)} =  \sum_{\nu}\sum_{\eta_{\nu}} W_{z_{i} \eta_{\nu} k L}^{(t),\nu} {\bm B}^{(t),\nu}_{i,\eta_{\nu} k LM}
\end{equation}
The recursive update of the node features ($t:0\rightarrow 2$) is obtained by adding the message to the atoms' features from the previous iteration, with weights that depend on the chemical element ($z_i$):
\begin{equation}
 \label{eq:update}
  h^{(t+1)}_{i,k LM}
    = \sum_{\tilde{k}} W_{k L,\tilde{k}}^{(t)} m_{i,\tilde{k}LM}^{(t)}
  + \sum_{\tilde{k}} W_{kz_{i} L,\tilde{k}}^{(t)}
  h^{(t)}_{i,\tilde{k}LM} 
\end{equation}
Since the initial node features ${\bm h}^{(0)}$ are only dependent on the chemical element of the atom, the second term of Eq.~\eqref{eq:update} is not included in the first layer. This makes it possible to set the energy of the isolated atoms (that is, those with no neighbours) exactly~\cite{batatiaDesignSpaceE3equivariant2025}, which is often desirable~\cite{kovacs2021linear, Ple2023fennix}.
The MACE models in this work have an effective receptive field of $2 \times r_{\text{cut}}$ due to the two layers of message passing. 

The site energy is a sum of read-out functions applied to node features from the first and second layers. The read-out function is defined as a linear combination of rotationally invariant node features for the first layer, and as multi-layer perceptron (MLP) for the second layer.
\begin{equation}
  \label{eq:site-energy}
    E_i = \sum_{t=1}^2 E_i^{(t)} = \sum_{t=1}^2
    \mathcal{R}^{(t)} \left({\bm h}_{i}^{(t)}\right),
\end{equation}
\begin{equation}\label{eq:readout}
    \mathcal{R}^{(t)} \left( \boldsymbol{h}_i^{(t)} \right) = 
    \begin{cases}
      \sum_{k}W^{(t)}_{k}h^{(t)}_{i,k00}     & \text{for} \;\; t = 1 \\[13pt]
      {\rm MLP} \left( \left\{ h^{(t)}_{i,k00} \right\}_k \right)  &\text{for} \;\; t = 2
    \end{cases}
\end{equation}
The forces and stresses on the atoms are calculated by taking analytical derivatives of the total potential energy with respect to the positions of the atoms and the components of the deformation tensor in the periodic setting, as is usual.

\subsection{Training data}
The core of the training set is version 1 of the SPICE dataset~\cite{eastman2023spice} with 95\% of the data used for training and validation, and 5\% set aside for testing. Test/train splitting was performed at the molecule level, ensuring conformers of the same molecule do not appear in both train/validation and test sets. Table~\ref{tab:training_set} summarizes the training and test sets. The MACE-OFF models are trained to reproduce the energies and forces computed at the \textomega B97M-D3(BJ)/def2-TZVPPD level of quantum mechanics~\cite{najibi2018nonlocal, weigend2005balanced, rappoport2010property, grimme2011bj_damp, grimme2010d3}, as implemented in the PSI4 software package~\cite{smith2020psi4}. We have used a subset of SPICE that contains the ten chemical elements H, C, N, O, F, P, S, Cl, Br, and I, and has a neutral formal charge (see Table 2 in Ref.~\cite{eastman2023spice} for element coverage in the dataset). We have also removed the ion pairs subset. Overall, we used about 85\% of the full SPICE dataset. The geometries in the SPICE dataset have been generated by running molecular dynamics simulations using classical force fields~\cite{maier2015ff14sb}, performed at both 300K and 500K, and sampling maximally different conformations from the resulting trajectories~\cite{eastman2023spice}.

\begin{table*}[ht]
    \centering
    \caption{\textbf{Summary of training and test sets.} The columns `PubChem' to `Solvated Amino Acids' correspond to the original SPICE dataset~\cite{eastman2023spice}. }
    \label{tab:training_set}
    \begin{tabularx}{0.98\textwidth}{lYYYYYYYY}
        \toprule
        & PubChem & DES370K Monomers & DES370K Dimers \cite{donchev2021quantum} & Dipeptides & Solvated Amino Acids & Water & QMugs \cite{isert2022qmugs} & Tripeptides \\
        \midrule
        Chemical elements & H, C, N, O, F, P, S, Cl, Br, I & H, C, N, O, F, P, S, Cl, Br, I & H, C, N, O, F, P, S, Cl, Br, I & H, C, N, O, S & H, C, N, O, S & H, O & H, C, N, O, F, P, S, Cl, Br, I & H, C, N, O \\
        System size & 3-50 & 3-22 & 4-34 & 26-60 & 79-96 & 3-150 & 51-90 & 30-69 \\
        \# Train & 646821 & 16861 & 263065 & 19773 & 948 & 1597 & 2748 & 0 \\
        \# Test & 33884 & 889 & 13896 & 1025 & 52 & 84 & 144 & 898 \\
        \bottomrule
    \end{tabularx}
\end{table*}

The SPICE dataset only contains small molecules of up to 50 atoms. To facilitate the learning of intramolecular non-bonded interactions, we augmented SPICE with larger 50--90 atom molecules randomly selected from the QMugs dataset~\cite{isert2022qmugs}. The geometries were generated by running molecular dynamics simulations using GFN2-xTB~\cite{bannwarth2019gfn2} similarly to the protocol described in Ref.~\cite{darby2023TrAce}. The energies and forces were re-evaluated at the level of QM theory used in SPICE. Finally, to obtain a better description of water, the dataset was further augmented with a number of water clusters carved out of molecular dynamics simulations of liquid water~\cite{schran2021machine}, with sizes of up to 50 water molecules. 
In addition to the 5\% of each subset, part of the COMP6 tripeptide dataset~\cite{smith2018ani_1x} was also recomputed at the SPICE level of theory and used as part of the test evaluation, but not for training.


In order to understand the role additional data plays in the performance of the model, we also investigated the difference in performance due to inclusion of additional configurations released as part of the second version of SPICE~\cite{eastman_2024_10975225}. 
The updated dataset, which we incorporate into the MACE-OFF24(M) model (see Section \ref{sec:model_details}), contains among others, an additional 208,000 configurations comprising solvated PubChem molecules and amino acid-ligand pairs. The former were sampled from molecular dynamics using the same protocol as in the original dataset, whilst the amino acid-ligand pairs were extracted from structures in the PDB. This release also included 1000 water clusters, containing up to 90 atoms. These were also added and used to complement the set of water clusters already included in the extended dataset.


\subsection{Model details}
\label{sec:model_details}
The MACE model has parameters that enable systematic control of model expressivity (and subsequent accuracy) against computational cost. In this paper, we present three variants of the MACE-OFF23 model, a small, a medium, and a large one, denoted in the text as MACE-OFF23(S), MACE-OFF23(M) and MACE-OFF23(L) respectively. The models get more accurate with size, but also have an increasing computational cost. The small model is well suited for large-scale simulations, the medium model offers a good balance of speed and accuracy, and the large model is best used for small systems or when the highest possible accuracy is desirable. MACE-OFF24(M) is an update to the medium model that uses an extended 6~\AA{} cutoff and includes additional configurations from version 2 of the SPICE dataset~\cite{eastman_2024_10975225}. The hyperparameters of the models are displayed in Table~\ref{tab:mace_models}. All models were trained using the PyTorch~\cite{Paszke2019PyTorchAI} implementation of MACE, available at \texttt{https://github.com/ACEsuit/mace}. More information about the training is provided in {\bf Section~S1}.

\begin{table}[ht]
    \centering
    \caption{Hyperparameters of the MACE-OFF models}
    \label{tab:mace_models}
    \begin{tabular}{lccccc}
        \toprule
        & 23(S) & 23(M) & 23(L) & 24(M)  \\
        \toprule
        Cutoff radius (\AA) & 4.5 & 5.0 & 5.0 & 6.0\\ 
        Chemical channels & \multicolumn{1}{c}{\multirow{2}{*}{96}} & \multicolumn{1}{c}{\multirow{2}{*}{128}} & \multicolumn{1}{c}{\multirow{2}{*}{192}} & \multicolumn{1}{c}{\multirow{2}{*}{128}} \\
        $k$ (see Eq.~\eqref{eq:element_embedding}) & & \\ 
        max L (see Eq.~\eqref{eq:message}) & 0 & 1 & 2 & 1 \\ 
        SPICE version & 1 &1&1&2\\
        \bottomrule
    \end{tabular}
\end{table}

\section{Results}

\subsection{Extended SPICE Test Set}

\begin{figure}[ht!]
    \centering\includegraphics[width=1.0\columnwidth]{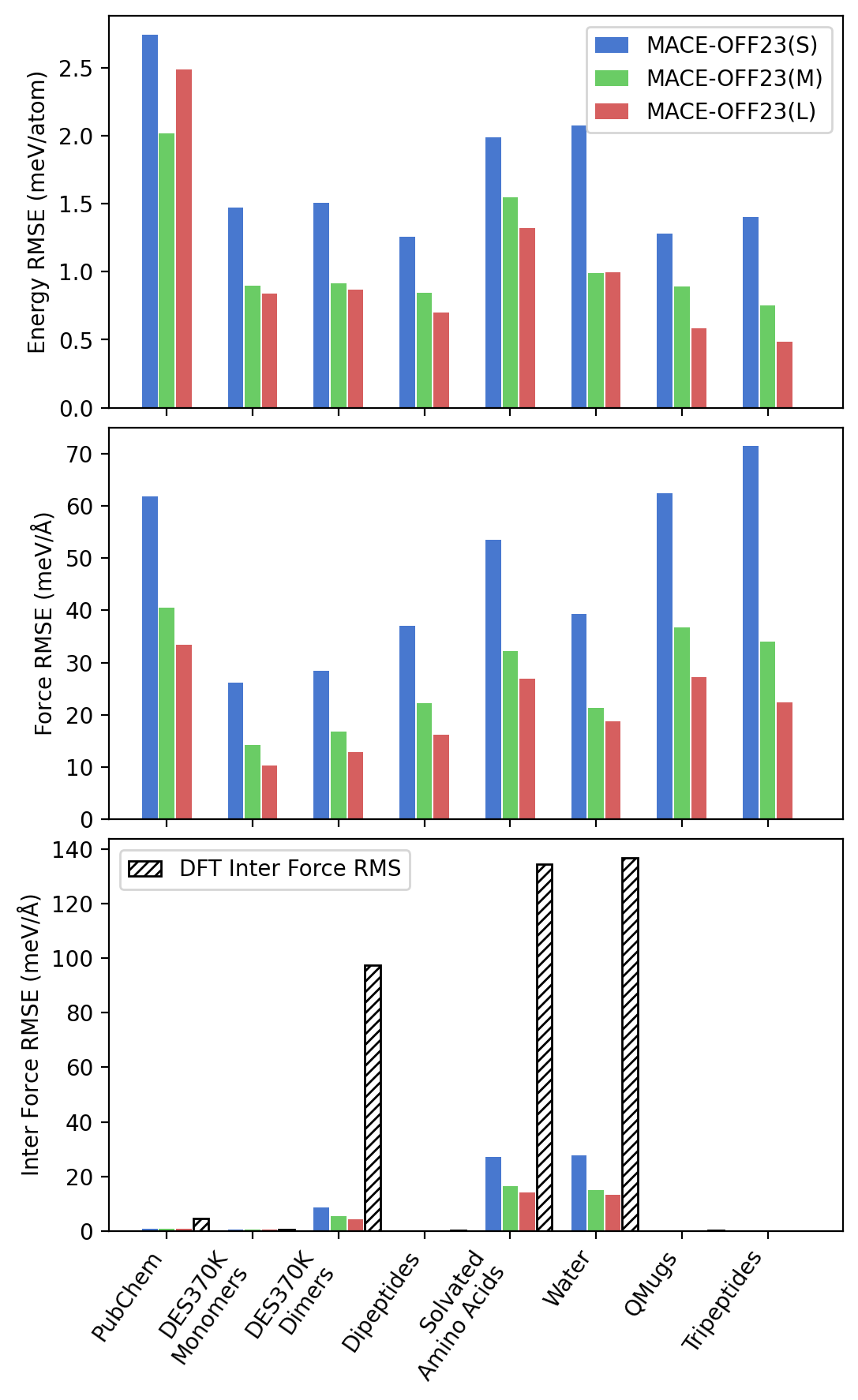}
    \caption{\textbf{Test set root mean square errors (RMSE).} Errors in the MACE-OFF23 models compared to the underlying DFT reference data, highlighting the relative accuracy of the three models. Bottom panels show specifically inter-molecular force errors compared to overall DFT inter-molecular force magnitudes (RMS). Note that for subsets comprising only single molecule configurations (DES370K Monomers, Dipeptides, QMugs, Tripeptides) inter-molecular contributions are expected to be zero. The slight deviation from zero arises because DFT forces do not obey translational and rotational symmetries with sufficient accuracy, while MACE models do.}
    \label{fig:test_mae}
\end{figure}

First, we look at the pointwise errors of the energy and force predictions on a held-out test set for each of the three baseline MACE-OFF23 models. Figure~\ref{fig:test_mae} shows the per-atom energy, force and intermolecular force root-mean-square errors (full statistics are provided in {\bf Section~S2}). As the size of the model increases, the models gradually become more accurate, with the large model generally achieving errors of around 0.5--1.0~meV/atom and 15--20~meV/\AA{}, well below the 1~kcal/mol (43~meV) chemical accuracy limit for the drug-like organic molecules studied here.

The last column represents an extrapolation task, the training set contains only dipeptides, and this test set looks at tripeptides, indicating that the models are able to extrapolate to larger fragments with more complex interactions, with no loss of accuracy. 

The bottom panel of Figure~\ref{fig:test_mae} shows specifically the intermolecular force errors, which were obtained by separating the force contributions to molecular translations and rotations (see {\bf Section~S2} and Ref.~\cite{magduau2023machine} for details). These interactions are crucial as they underpin the thermodynamics and transport properties in the organic condensed phase. MACE-OFF23(M) and MACE-OFF23(L) both yield similar errors of around 5--15 meV/\AA, which are about 1.5--3 times smaller than total force errors, and about 5--10\% of the typical DFT intermolecular force magnitude (RMS). For context, relative intramolecular force errors, which are roughly equivalent to total force errors, are around 1--2\%. The latter are routinely used across the literature to benchmark ML models, however, we advocate that intermolecular RMSE errors are a more appropriate  metric to assess model accuracy for organic condensed phase applications. The results here demonstrate the remarkable accuracy of MACE models in predicting these subtle but crucial interactions. We will investigate how these energy and force errors translate into condensed phase physical property predictions in later Sections.

%
%
%
%

\subsection{Dihedral scans}

Next, we evaluate the performance of the MACE-OFF23 models on dihedral scans of drug-like molecules. This task is routinely carried out using quantum mechanical methods to create reference data for parameterizing classical empirical force field dihedral terms~\cite{horton2022open}. The task is particularly challenging, as constrained geometry optimizations can be difficult to converge if the potential energy surface is not smooth, as has been observed for example for the ANI-1x potential~\cite{hao2022ani_geom_opt}.

\begin{figure*}
    \centering
    \includegraphics[width=1.0\linewidth]{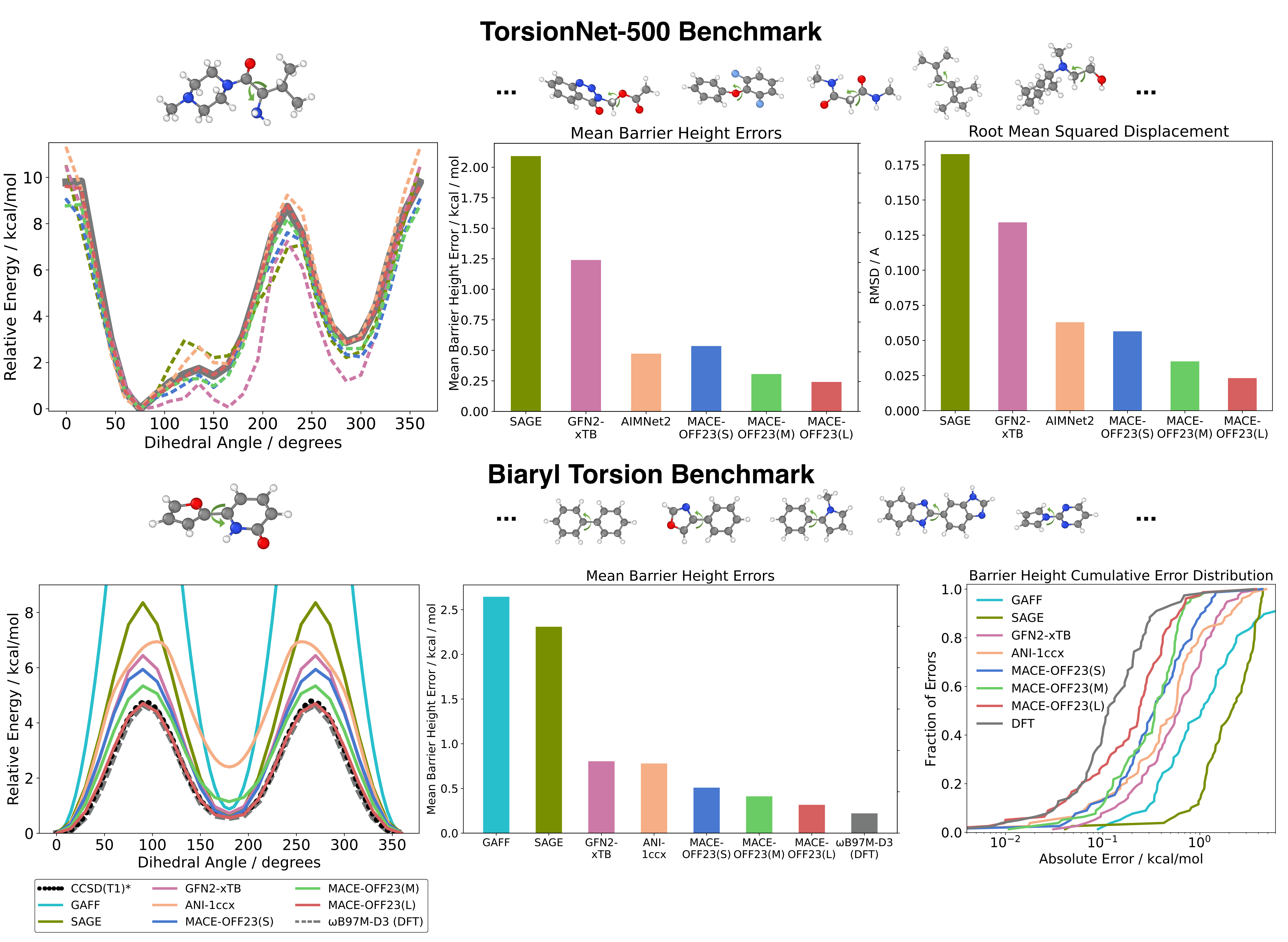}
    \caption{{\bf Dihedral benchmark scans.} The top panel shows torsion drive data for the TorsionNet-500 dataset, which has a wide chemical diversity (five example molecules are shown). The bottom panel focuses on the torsion angle between two aromatic rings in the biaryl torsion benchmark~\cite{lahey2020biaryl_torsion} which contains 78 molecules (five examples are shown). }
    \label{fig:torsiondrives}
\end{figure*}

\subsubsection{TorsionNet-500}

The top panel of Figure~\ref{fig:torsiondrives} summarizes the results for the TorsionNet-500 dataset~\cite{rai2022torsionnet}. This dataset contains torsion drives of 500 different molecules, selected to cover a wide range of pharmaceutically relevant chemical space. The original dataset was reported at the B3LYP/6-31G(d) level of DFT theory. For consistency with the SPICE dataset, we recreated the torsion profiles using the DFT setting of SPICE, which uses a higher level of theory and basis set ({\bf Section S3}).

The first panel shows an example of a torsion drive, indicating the complex energy profile that the MACE models are able to capture closely, including geometries far from equilibrium. The center panel shows the mean barrier height error of a number of representative models, comparing the Sage classical empirical force field~\cite{openff-sage}, a semi-empirical quantum mechanical method GFN2-xTB~\cite{bannwarth2019gfn2}, a recent transferable machine learning force field AIMNet2~\cite{anstine2023aimnet2}, and the three MACE-OFF23 models.
Again, systematic improvements in accuracy with MACE model size are observed, with medium and large models, in particular, achieving errors of around 0.25~kcal/mol compared to the reference method. The AIMNet2 model achieves comparable accuracy to the small MACE-OFF23 model. 
A similar conclusion can be drawn from the comparison of the molecular geometries by looking at the root mean squared deviation of the atomic positions averaged over the full torsion scans, as indicated by the top right panel of Figure~\ref{fig:torsiondrives}. It shows that MACE-optimized geometries have a deviation of about 0.025~\AA, meaning they are almost indistinguishable from DFT-optimized structures. It is important to note that the other models were trained to different levels of DFT, which might also contribute to the observed differences~\cite{behara2024}.


\subsubsection{Biaryl fragments}
The biaryl torsion benchmark investigates the torsional profiles of about 100 different biaryl fragments computed at the coupled cluster level of theory~\cite{lahey2020biaryl_torsion}. In these molecules, the rotatable bond is between two aromatic rings. Such chemistry frequently occurs in drug-like molecules, and the profiles are typically challenging to model accurately using empirical classical force fields. Following previous studies, we used a subset of 78 molecules to facilitate comparisons with the ANI-1ccx model, which is only parameterized for H, C, N, and O chemical elements~\cite{lahey2020biaryl_torsion}. 

In the bottom panel of Figure~\ref{fig:torsiondrives}, we compare the results of torsion drives of empirical force fields, semi-empirical QM methods, and the ANI-1ccx machine learning force field~\cite{smith2019approaching_ani1ccx} with the MACE-OFF23 models. We also compare the DFT potential energy surfaces (using the SPICE level of theory) with the published gold standard coupled cluster data. 

The DFT torsion drives achieve a mean barrier height error of 0.2~kcal/mol with respect to coupled cluster data, and is therefore the best result theoretically possible for MACE-OFF23 models which were trained on the same DFT level of theory. 
We find that the large model comes close to this, with a mean absolute error of 0.3 kcal/mol. The medium and small MACE models have barrier height errors of 0.4 and 0.5~kcal/mol, respectively. Remarkably, the small MACE model is significantly more accurate than the next-best non-DFT methods, ANI-1ccx and GFN2-xTB, as illustrated in the bottom-center plot of Figure~\ref{fig:torsiondrives}. In particular, unlike MACE-OFF23, coupled cluster reference calculations were used to parameterize the ANI-1ccx and GFN2 models.
In the bottom right, we also show the cumulative error distributions to verify that the MACE-OFF23 barrier height errors are not only accurate on average but are also robust, with essentially no outliers.

\subsubsection{Infrared spectroscopy of a drug molecule}
\label{sec:ir_spec_paracetamol}

\begin{figure}
    \centering
    \includegraphics[width=\columnwidth]{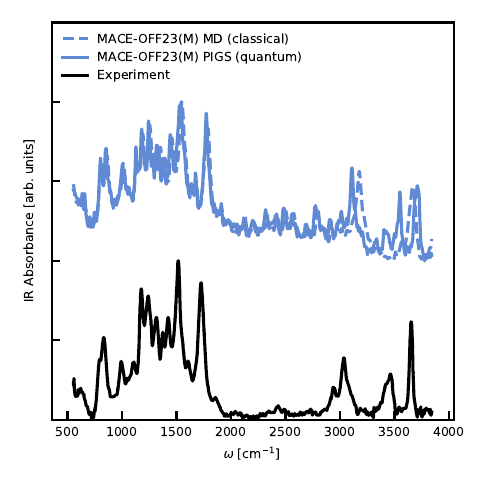}
    \caption{\textbf{Infrared (IR) spectrum of a paracetamol molecule.} The classical (blue dashed) and quantum (blue solid) IR spectrum at 293\,K using the MACE-OFF23(M) model is computed by propagating the system and estimating the time correlation function of the time derivative of the total dipole moment. The experimental data are taken from Ref.~\citenum{nist_webbook}. }
    \label{fig:IR_mol}
\end{figure}

To assess the accuracy of the MACE-OFF23(M) model for finite temperature properties of molecules, we study the infrared spectrum of an isolated paracetamol molecule at room temperature.
Molecular dipole moments are predicted using the MACE architecture, trained on the SPICE dataset, and quantum nucelear effects are incorporated using path-integral coarse-grained simulations (PIGS)~\cite{Musil2022}. 
Full details are provided in {\bf Section~S8}.
This is a stringent test, as reproducing the line positions and their relative intensities requires an accurate treatment of both intramolecular interactions and the dipole moment in out-of-equilibrium configurations.
As shown in Figure~\ref{fig:IR_mol}, we estimate the classical and approximate quantum IR spectra~\cite{Musil2022}, and compare them directly with experimental data at 293\,K~\cite{nist_webbook}.
For low and intermediate frequencies (up to around 2000 cm$^{-1}$), the classical and quantum spectra agree with each other and with the experimental data, allowing identification of all collective modes.
For the high-frequency O--H and N--H modes, we observe that the classical spectra are blue-shifted due to the absence of anharmonic zero-point fluctuations, which typically soften these bonds. 
A quantum treatment of nuclear dynamics results in good agreement with the experimental data, with the exception of a net blue shift and a slight discrepancy in the relative intensities of the high-frequency stretching with respect to the lower-frequency modes, likely due to the limitations of the reference density functional.
%
%

\subsection{Molecular crystals and organic liquids}
\label{sec:condensed_phase}

In the following, we demonstrate the ability of the MACE-OFF force fields to simulate the vibrational and thermal properties of molecular crystals.
Based on relative accuracy and computational expense, in what follows we present data only for the medium MACE-OFF models. Additional experiments, where either the higher throughput of the small model or the additional accuracy of the large model are desired, are reported in the Supporting Information.

\subsubsection{Lattice enthalphies}

We assess the ability of the MACE-OFF23(M) model to describe the stability of molecular crystals. We computed the enthalpies of sublimation of a range of 23 representative small molecular crystals~\cite{reilly2013understanding} following the protocol of Ref.~\cite{dolgonos2019revised_x23} ({\bf Section~S4}). 

\begin{figure}
    \centering
    \includegraphics[width=0.8\columnwidth]{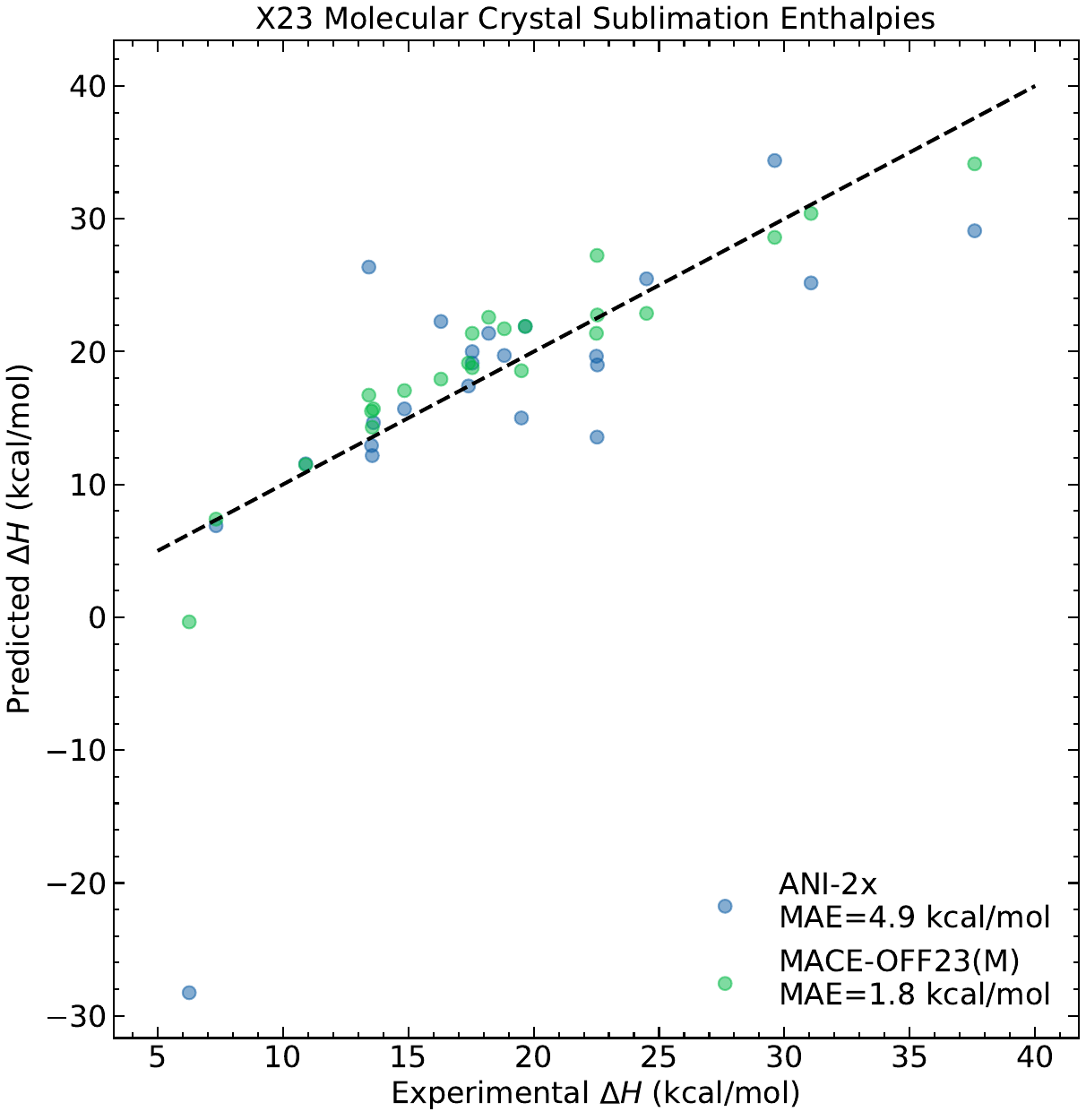}
    \caption{\textbf{Sublimation enthalphies of molecular crystals.} Comparison between predicted sublimation enthalphies of the MACE-OFF23(M) and ANI models and experiment.
    }
    \label{fig:x23}
\end{figure}

Figure~\ref{fig:x23} compares the predicted sublimation enthalpies with the experimentally measured ones. This task is often used to test various DFT functionals~\cite{reilly2013understanding}, as well as beyond-DFT methods~\cite{PhysRevLett.133.046401} and tuned machine learning potentials~\cite{kaur2025}. Since the \textomega B97M-D3(BJ) functional used to parameterize MACE-OFF23 does not have a periodic implementation, an estimate of the highest achievable accuracy is not available. The figure shows that the MACE-OFF23(M) model is able to capture trends and improve significantly compared to the ANI-2x model. The mean absolute error of 1.8~kcal/mol is comparable to the errors of several different dispersion-corrected density functionals for a fraction of the computational cost~\cite{reilly2013understanding}. The sublimation enthalpies for all MACE-OFF models are tabulated in {\bf Section~S4}, and we observe that the small model is intermediate in accuracy between the medium model and ANI-2x, while the large model does not give any significant performance improvement.

Additional analyses of the relaxed unit cell vectors using the MACE-OFF23(L) model, and the powder Raman spectrum of crystalline paracetamol are presented in {\bf Sections S5} and {\bf S8}.


\subsubsection{Condensed phase properties of organic liquids}

\begin{figure}
    \centering
    \includegraphics[width=0.8\columnwidth]{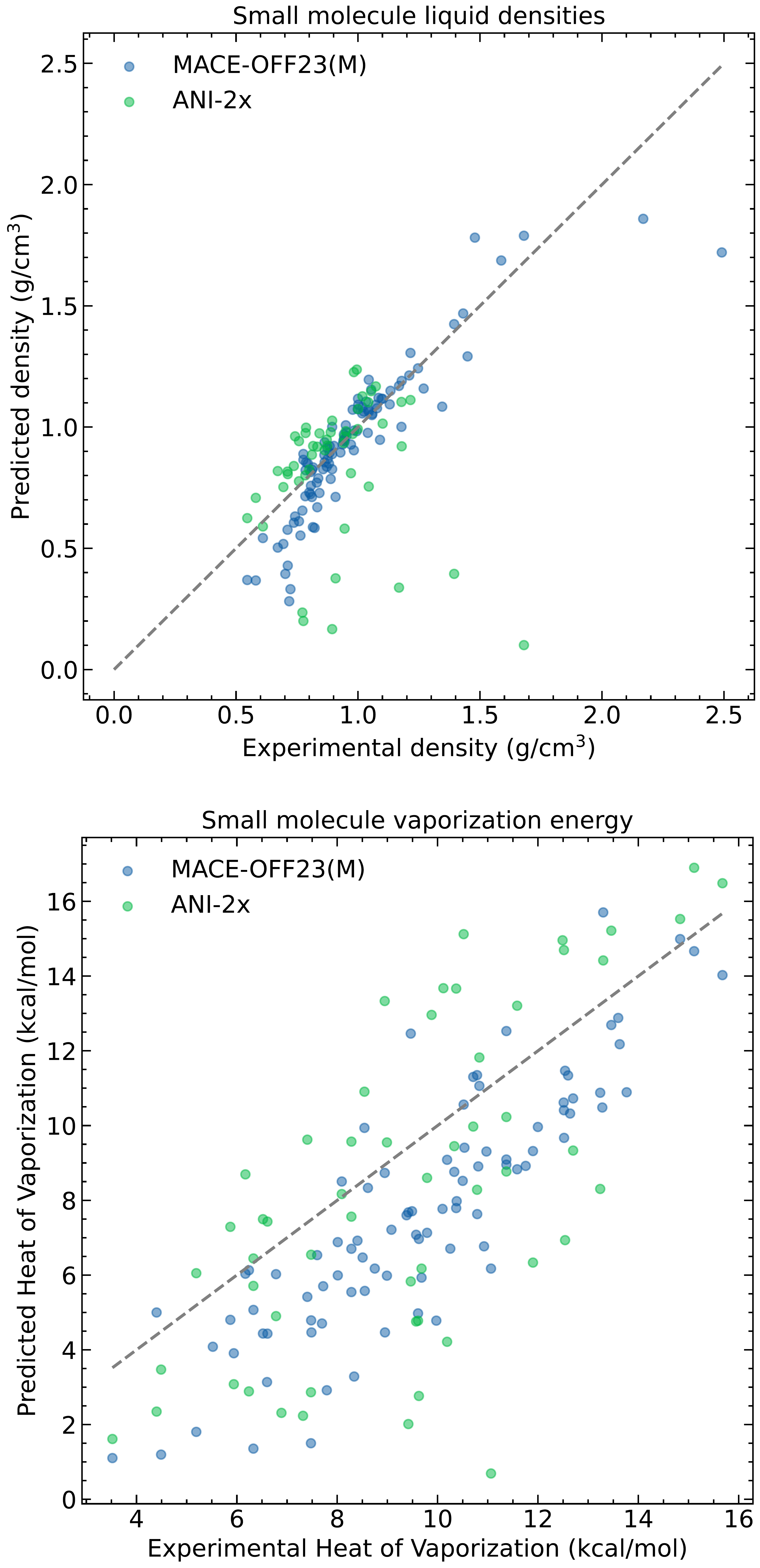}
    \caption{\textbf{Molecular liquids.} Comparison between MACE-OFF23(M) and ANI-2x with experiment for densities (top) and heats of vaporization (bottom) of condensed phase organic liquids.}
    \label{fig:mace_condensed_phase}
\end{figure}


Small molecule empirical force fields are fitted to reproduce experimental condensed phase properties including densities and heats of vaporization.  As such, errors on small molecular systems are expected to be well within chemical accuracy~\cite{caleman2012,openff-sage}.
For machine learning force fields, this task represents a greater challenge, since the potential is typically trained only using QM reference data for isolated molecules and molecular dimers. From these ab initio data, the models must learn the long-range interactions necessary to reproduce bulk properties without ever fitting to them.

To investigate the performance of the MACE-OFF23 models in the condensed phase, we selected a benchmark set of 109 molecules from Ref.~\cite{horton2019qubekit}, containing a representative set of functional groups relevant to medicinal chemistry and biology.  
%
%
For the liquid boxes containing approximately 600 atoms, MACE-OFF23(S) achieved a throughput of $2.1\times10^6$ steps/day via OpenMM on a single 80GB NVIDIA A100 GPU, whilst the medium model achieved a throughput of $1.1\times10^6$ steps/day.  Meanwhile MACE-OFF23(L) achieved a throughput of $2.8\times10^5$ steps/day on the same system, highlighting the significant overhead compared to the other two models. See section \ref{sec:performance} for further discussion of the computational performance.


First, we investigate the predictions of molecular liquid densities under ambient conditions using the medium MACE model shown on the top panel of Figure~\ref{fig:mace_condensed_phase}. MACE-OFF23(M) achieves a MAE of 0.09~g/cm\textsuperscript{3}, indicating that it can generalize from the intermolecular interactions between molecular dimers in the training set to the condensed phase, retaining a good correlation with experiment. In contrast, the ANI-2x potential has significantly higher errors, with MAE and RMSE errors twice as large as MACE-OFF23(M) ({\bf Section~S6}). 

The vast majority of the outliers in the density predictions can be assigned as either low boiling-point ether-containing compounds, polychlorinated hydrocarbons or dibromo-containing compounds.  In the first case, this may be due to a marginal under-prediction of the boiling point, since these compounds were simulated at 10~K below their tabulated boiling points.  In the latter two cases, this functional group dependence suggests these nonbonded interactions are insufficiently represented in the present training set, which contains, for example, only 84 examples of dibromo-containing dimers. This highlights the need for additional coverage of specific functional groups within SPICE (for example, the recent AIMNet2 dataset includes 20 million configurations to cover 14 chemical elements~\cite{anstine2023aimnet2}). 

We further investigated the performance of MACE-OFF23(M) on heats of vaporization. The predictions correlate strongly with the experimental data; however, there is a systematic offset of approximately 2~kcal/mol (Figure~\ref{fig:mace_condensed_phase}).  
We hypothesized that this may result from an under-prediction of the total intermolecular interactions originating from two sources: those that appear in molecular dynamics simulations outside the effective cutoff of the MACE model (compare classical simulations, which typically employ long-range corrections to the total energy~\cite{longrangecorrection}), and those that the model has not learned from the limited set of intermolecular interactions in the training set.

The effect of the choice of MACE-OFF model on condensed phase properties is further investigated in {\bf Section S6}.
In general, we observe that the small model systematically over-predicts the experimental density (MAE 0.23~g/cm\textsuperscript{3}), which is consistent with its larger intermolecular force errors ({\bf Table~S1}). 
We did not observe any significant improvements in accuracy, over the medium model, using either the large model or a model trained with an extended cutoff of 6~\AA{}. This further reinforces the hypothesis that the poorly predicted densities are a result of under-representation of the functional groups in the training set.
Nevertheless, these data indicate that, given a reasonable receptive field and a sufficiently expressive model, it is possible to fit a local force field that captures the interactions required to recover experimental properties.
Further improvements in accuracy are possible without resorting to fitting to experiment by increasing further the size of the training dataset and the explicit addition of long-range Coulomb interactions into the model, which we leave for future work.

\subsection{Biomolecular simulations}
\label{sec:biologics}
%
%
In the following section, we benchmark the performance of the MACE-OFF models on several well-studied large-scale biomolecular systems.  
This particular application area is of key interest, and state-of-the-art classical protein force fields have been carefully parameterized over many decades to reproduce key quantum mechanical and thermodynamic properties. However, this extensive parameterization process makes it difficult to extend classical force fields, and overall accuracy and transferability remain fundamentally limited by their functional form. This typically becomes apparent only when each generation of advancing hardware enables sufficiently long simulations to identify issues.

\subsubsection{Water structure and dynamics}
\label{section:water_structure}

\begin{figure*}[!ht]
    \centering
    \includegraphics[width=0.9\textwidth]{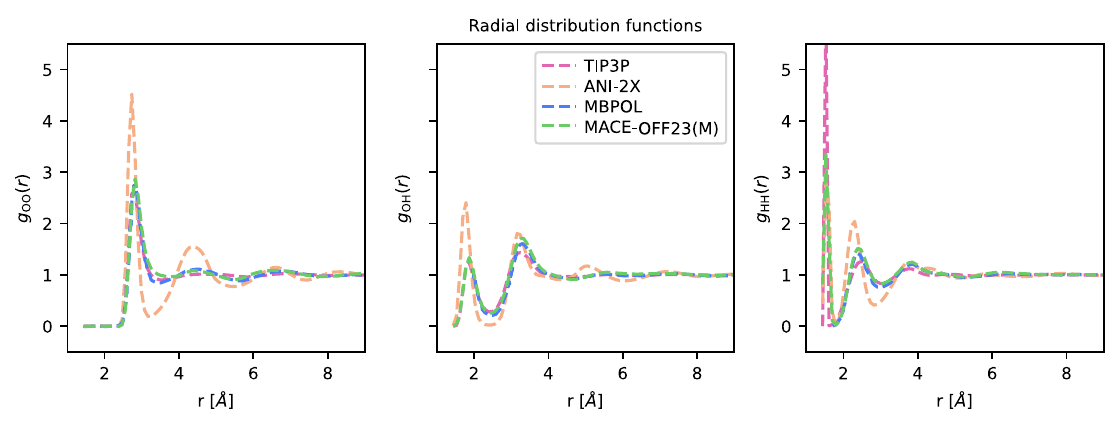}
    \caption{\textbf{Water structure.} Radial distribution functions of MACE, ANI and TIP3P models. MB-pol traces are reproduced from Ref.~\cite{meddersDevelopmentFirstPrinciplesWater2014}. }
    \label{fig:mace_ani_tip3p_rdf}
\end{figure*}

A key requirement for bio-organic force fields is an accurate description of water. 
Several empirical water models have been developed for this purpose, the most common for all-atom biomolecular simulations being TIP3P. Figure~\ref{fig:mace_ani_tip3p_rdf} shows that MACE-OFF23(M) correctly predicts the radial distribution function (RDF) of water, performing comparably with both TIP3P and the highly accurate MB-pol water model, whose RDF is indistinguishable from experiment~\cite{meddersDevelopmentFirstPrinciplesWater2014}.  
%
%
These interactions are modelled in a purely local way by MACE, having been trained on small water clusters containing up to 50 water molecules.  This result further emphasizes the ability of the model to generalize to bulk simulations, even though the training data only contained non-periodic clusters. In contrast, ANI-2X shows a significant over-structuring of the RDF, however this is likely driven primarily by the lack of dispersion correction in the underlying quantum mechanical method used to label the training set. Additional tests of the vibrational density of states of liquid water are presented in {\bf Section~S8} and show good agreement with experiment.


\begin{figure}[h]
    \centering
    \includegraphics[width=0.8\columnwidth]{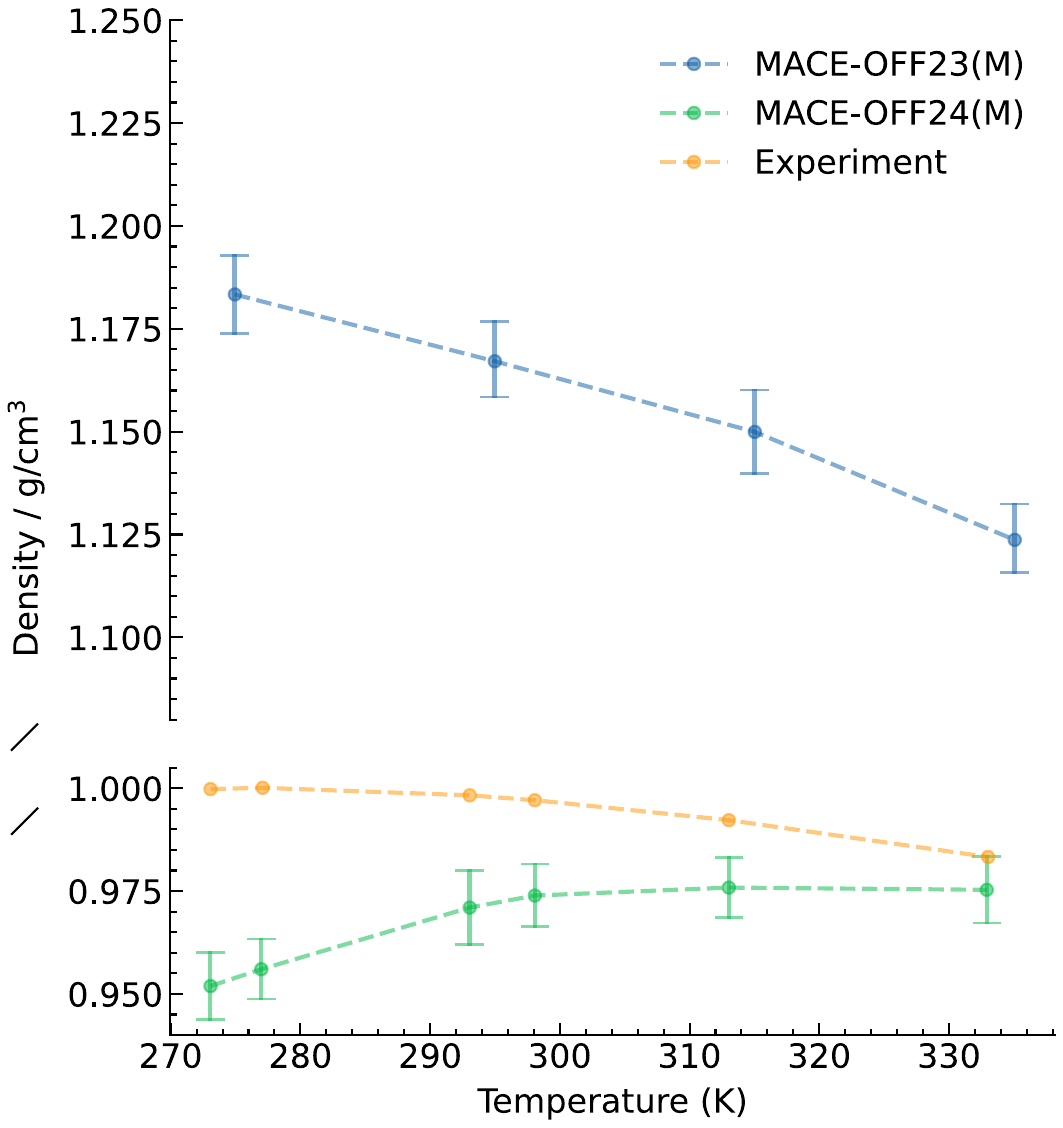}
    \caption{\textbf{Water density as a function of temperature.} Comparison between MACE-OFF23(M), and the extended cutoff MACE-OFF24(M) model.}
    \label{fig:water_dens_curves}
\end{figure}

We further calculated the temperature dependence of the water density using NPT molecular dynamics simulations, as shown in Figure~\ref{fig:water_dens_curves}. This is a highly sensitive test of the accuracy of intermolecular interactions~\cite{magduau2023machine}. 
%
%
The MACE-OFF23(M) model over-estimates the density of liquid water by approximately 20\%.  
Since there is extensive coverage of water clusters in the training set, we hypothesized that this results from a missing long-range contribution beyond the receptive field.  
To investigate this hypothesis, we fitted an additional model, which has an increased 6~\AA{} layerwise cutoff compared to 5~\AA{}. 
The larger receptive field leads to a predicted water density within 2\% of the experimental value at room temperature, and within 5\% across the full temperature range (Figure~\ref{fig:water_dens_curves}). 
We note that similar errors are reported using many dispersion-corrected DFT exchange-correlation functionals~\cite{10.1063/1.4944633}.
%
%
At this point in the model development, the second version of SPICE had been released, which included Solvated Pubchem and Amino Acid Ligand Pairs datasets~\cite{eastman_2024_10975225}.
%
%
Given the direct applicability of these training data and the importance of modelling water to biomolecular simulations, the remainder of this section uses the MACE-OFF24(M) model, which includes the additional training data and extended 6~\AA{} cutoff (Table~\ref{tab:mace_models}).
To confirm no loss of performance on incorporation of the additional training data, we confirmed that the test set errors are in agreement with MACE-OFF23(M) (\textbf{Section~S2.1}).

%


\subsubsection{Ala\textsubscript{3} Free Energy Surface}

We first investigated the ability of MACE-OFF24(M) to reproduce the free energy surface of the Ala\textsubscript{3} tripeptide. We observed computational performance of $9.6\times 10^{6}$ steps/day for the vacuum system, and $2.2\times 10^{6}$ steps/day for the solvated system on an NVIDIA 80GB A100 GPU via the OpenMM interface.  Although this remains significantly more expensive than the MM simulation (around $150\times 10^{6}$ steps/day), it nevertheless enables sufficient sampling of the free energy surface.

\begin{figure}
    \centering
    \includegraphics[width=\linewidth]{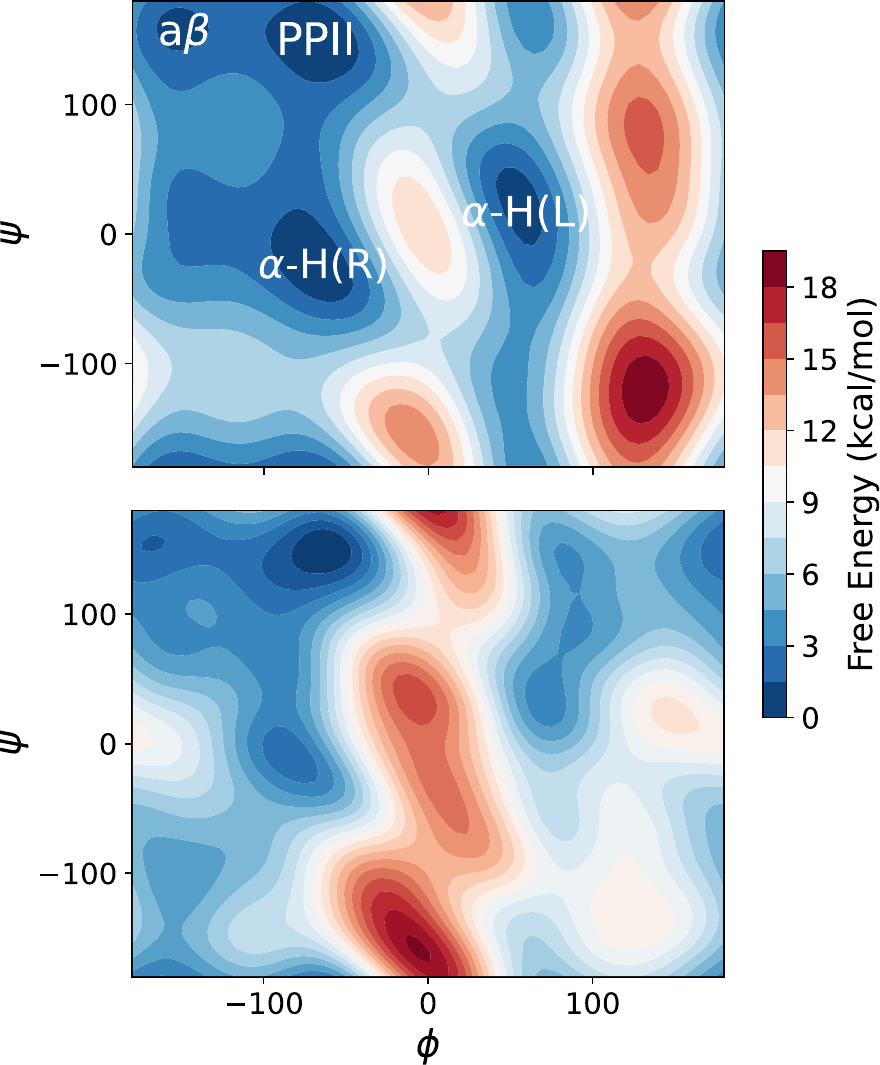}
    \caption{\textbf{Free energy surfaces of Ala\textsubscript{3} in explicit solvent.} AMBER14SB/TIP3P (top) and MACE-OFF24(M) (bottom).}
    \label{fig:mace_amber_ala3_fes}
\end{figure}

Figure~\ref{fig:mace_amber_ala3_fes} shows the free energy surface of Ala\textsubscript{3} in water, modelled by AMBER14SB/TIP3P and MACE-OFF24(M), respectively. Well-tempered metadynamics was used to enhance sampling along the central $ \phi $ and $ \psi $ backbone angles ({\bf Section~S7.1}).
%
%
The AMBER14SB simulation identifies four local minima, corresponding to the anti-parallel $\beta$-sheet ($ \phi < -120^\circ, \psi > 120^\circ $), a right-handed $\alpha$-helix ($ \phi = -60^\circ, \psi > -60^\circ $), the corresponding left-handed $\alpha$-helix ($ \phi = 60^\circ, \psi > 60^\circ $), and a polyproline II (PPII)-type structure ($ \phi = -60^\circ, \psi > 120^\circ $). Comparing the dihedral angle distribution with NMR J-coupling constants, the AMBER14SB simulation is generally in good agreement, with a slight overpopulation of both $ \alpha $-helical structures and underpopulation of the PPII mesostate~\cite{zhangMolecularDynamicsForce2020}. We therefore use this to enable visual comparison with a well-tuned classical forcefield.

The corresponding MACE simulations show a similar free energy surface to AMBER14SB/TIP3P, with the same four local minima identified. Whilst the relative depths of the \textalpha-helical conformations are in agreement, AMBER predicts the free energy of the anti-parallel \textbeta-sheet to be 1~kcal/mol higher than the PPII minimum, compared to 0.1~kcal/mol for MACE. 

We also compare the unbiased conformational ensemble from 20~ns of NPT dynamics sampled with MACE-OFF24(M) directly to experiment by calculating \textsuperscript{3}J coupling constants from the Ramachandran distribution. Computed values are shown in Table \ref{tab:coupling} in addition to literature values for AMBER14SB, ANI-2x and experiment~\cite{zhangMolecularDynamicsForce2020, rosenberger2021modeling, huDeterminationH1Angles1997a}.  MACE-OFF24(M) accurately predicts coupling constants, and is comparable with AMBER14SB. In agreement with our previous findings, both methods significantly outperform ANI-2x, highlighting the difficulties in computing accurate dynamics in the solution phase with MLPs.



\begin{table}[h!]
    \centering
    \begin{tabular}{@{}lccc@{}}
        \toprule
         & \textsuperscript{3}J(H\textsubscript{N},H\textsubscript{\textalpha}) & \textsuperscript{3}J(H\textsubscript{N},C\textsubscript{\textbeta}) & \textsuperscript{3}J(H\textsubscript{N},C$^\prime$) \\
        \toprule
        MACE-OFF24(M)   & \textbf{5.70}(0.43) & 1.64(0.18) & 1.37(0.30) \\
        ANI-2x           & 7.5 & 1.2 & 1.8 \\
        AMBER14SB        & 6.07 & \textbf{1.87} & \textbf{1.13} \\ 
        \midrule
        EXP              & 5.68(0.11) & 2.39(0.09) & 1.13(0.08) \\
        \bottomrule
    \end{tabular}
    \caption{\textbf{Summary of \textsuperscript{3}J couplings computed from dihedral distributions}. AMBER14SB and experimental values are taken from Ref.~\cite{zhangMolecularDynamicsForce2020}, and ANI-2x data from Ref.~\cite{rosenberger2021modeling}. Standard errors in the mean from block averaging are shown in parentheses for MACE-OFF, and experimental uncertainties are reproduced from Ref.~\cite{zhangMolecularDynamicsForce2020}}
        \label{tab:coupling}
\end{table}


\subsubsection{Folding dynamics of Ala\textsubscript{15}}

Having confirmed that MACE-OFF24(M) is capable of recovering a good approximation to the free energy surface of a small peptide, we next investigated the folding of the longer helical peptide Ala\textsubscript{15}.  Because this is a significantly larger system size than the dipeptide configurations on which the model was trained, this represents a nontrivial test of the extrapolation capability of the potential, including its ability to capture complex hydrogen bonding interactions required to stabilize the secondary structure. Such simulations are believed to be particularly difficult with purely local models such as those in the MACE-OFF family. 

\begin{figure}
    \centering
    \includegraphics[width=\linewidth]{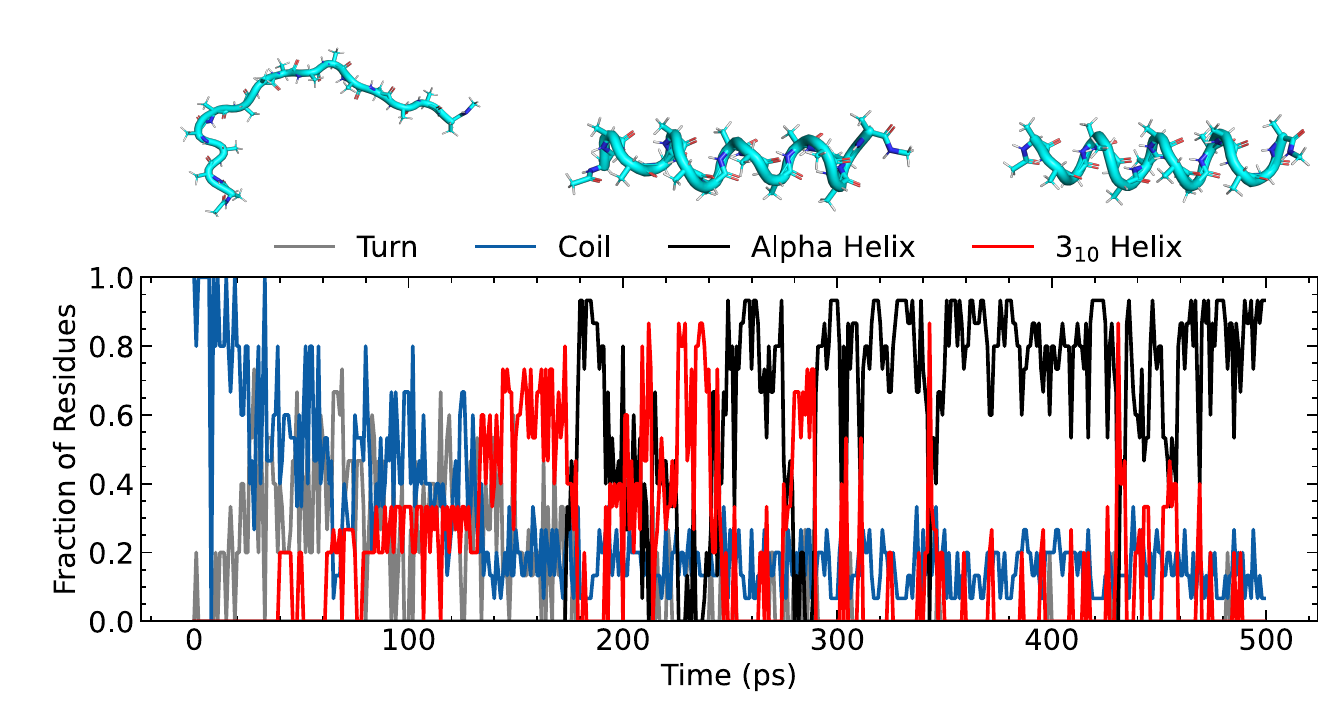}
    \caption{\textbf{Ala\textsubscript{15} folding dynamics.} Secondary structure assignment during a 500~ps trajectory generated with MACE-OFF24(M). The structure proceeds via a ``wavy'' intermediate and oscillates between the \textalpha- and 3\textsubscript{10} helices. Secondary structure assignment was performed with the STRIDE algorithm~\cite{heinigSTRIDEWebServer2004}.}
    \label{fig:ala15}
\end{figure}

As an initial test, the fully extended Ala\textsubscript{15} structure was simulated \textit{in vacuo}. Figure~\ref{fig:ala15} shows the assignment of the secondary structure during the simulation. 
The peptide folded within 200~ps, adopting initially a ``wavy'' intermediate structure, before first folding into a 3\textsubscript{10} helix. The secondary structure oscillates between the \textalpha-helix and the 3\textsubscript{10} helix for the remainder of the simulation, with the \textalpha-helix being the predominant motif. These results are in agreement with other works~\cite{kabylda2023efficient, unke-gems-sciadv}, in which polyalanine peptides are observed to fold through a ``wavy intermediate'' and oscillate between 3\textsubscript{10} and \textalpha-helices. Although, MACE-OFF24(M) predicts the 3\textsubscript{10} helix in lower ratio than SO3LR and GEMS. The oscillatory behavior has also been observed experimentally in alanine-rich polypeptides~\cite{bolin310SplitPersonality1999}.

\begin{figure}
    \centering
    \includegraphics[width=\linewidth]{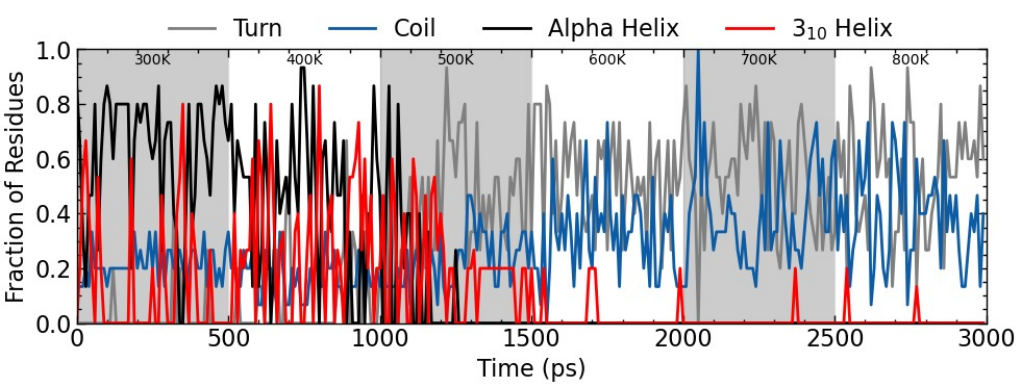}
    \caption{\textbf{Ala\textsubscript{15} Temperature dependence.} Secondary structure assignment with increasing temperature.}
    \label{fig:ala15-temp}
\end{figure}

We also measured the secondary structure of the peptide as the temperature was increased. Starting from the folded structure at 300~K, the temperature was increased in steps from 300~K to 900~K over the course of 3~ns (Figure~\ref{fig:ala15-temp}). The folded structure remains stable up to around 500~K, after which point the helix unfolds and the structure is primarily a random coil. 



\subsubsection{Protein simulation in explicit solvent}
\label{section:solvated_protein_sims}

Finally, we investigated the simulation of a fully solvated protein with MACE-OFF.  As a test case, we chose Crambin, a 42-residue threonin storage protein that contains four charged residues. This test goes significantly beyond what can typically be expected from a local force field trained on neutral species only. 

Using the MACE-OFF24(M) model, we first confirmed that the force field was indeed capable of simulating a large biomolecular system containing charged residues. In particular, it has recently been shown that the FENNIX potential fails to simulate a solvated 28 residue peptide due to unphysical proton transfers involving charged residues~\cite{Ple2023fennix}.  We confirmed that the root mean square fluctuations of the protein backbone relative to the minimized structure remained below 1~\AA{} for the duration of a 1~ns simulation, no bond breaking occurred and the secondary structure motifs remained intact ({\bf Figure S2}). We obtained a throughput of $3\times 10^5$ steps per day on a single NVIDIA A100 80GB GPU.

\begin{figure}
    \centering
    \includegraphics[width=\linewidth]{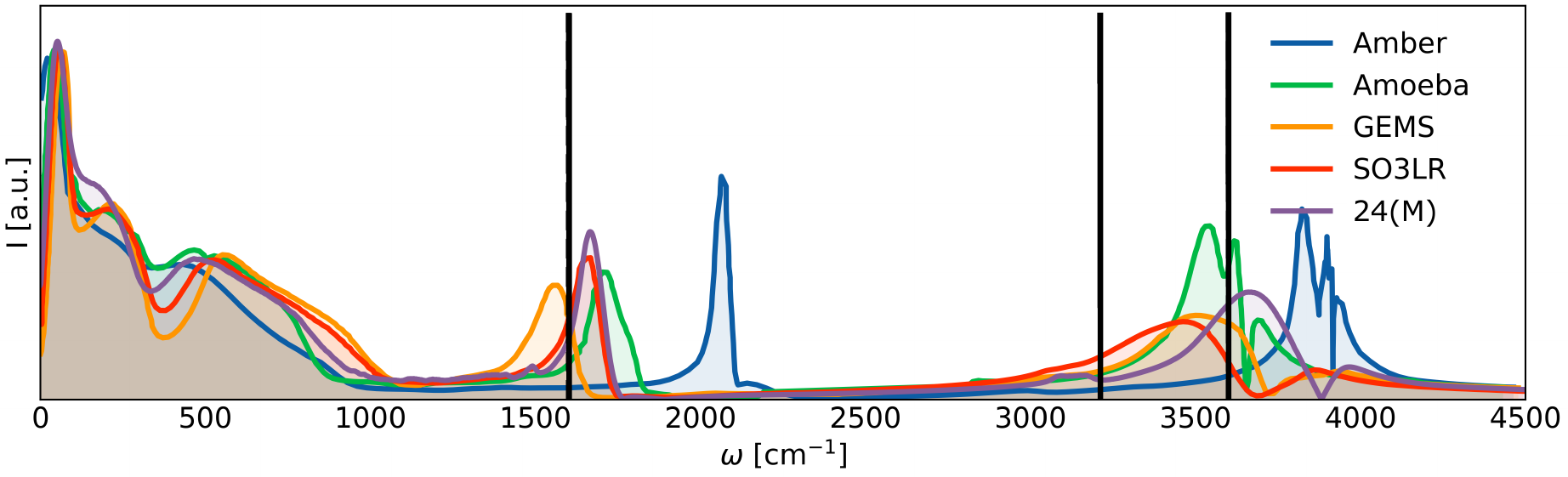}
    \caption{\textbf{Protein in explicit solvent.} Power spectrum of the high hydration state of crambin computed from the final 125~ps of a 1.6 ns simulation with 
    MACE-OFF24(M). GEMS, AMBER and SO3LR spectra are reproduced from Ref.~\cite{kabyldaMolecularSimulationsPretrained2024}. Black lines denote experimental peaks at 1640~cm\textsuperscript{-1} and the range 3200-3600~cm\textsuperscript{-1}.}
    \label{fig:crambin_power_spec_hhs}
\end{figure}

We then investigated the ability of MACE-OFF24(M) to capture the key vibrational modes of the system. To this end, we prepared a simulation box containing crambin in the high-hydration state, that is, fully solvated in explicit water, for a total system size of approximately 18,000 atoms. The power spectrum was calculated from the final 125~ps of a 1.6 ns simulation recorded at 2~fs resolution, to match the simulation length and sampling frequency employed in Ref \cite{kabyldaMolecularSimulationsPretrained2024}. 
This system has also been investigated with the AMBER classical force field and several machine learned potentials, including GEMS and, more recently, SO3LR~\cite{unke2021spookynet, unke-gems-sciadv, kabylda2023efficient}.
All three MLPs produce spectra in qualitative agreement with each other.  The water peaks at 1640~cm\textsuperscript{-1} and 3200-3600~cm\textsuperscript{-1} are reproduced reasonably accurately, especially considering that no nuclear quantum effects are taken into account (Figure~\ref{fig:crambin_power_spec_hhs}).  Furthermore, in the low frequency region, two peaks are identified in the power spectrum by GEMS, SO3LR and MACE-OFF24(M). The most obvious of these occurs at 60~cm\textsuperscript{-1} and corresponds to localized internal side chain fluctuations, while a peak at 220~cm\textsuperscript{-1} is due to specific configurations of water at the protein surface. 

We note that, whilst a 125~ps trajectory is sufficient to qualitatively identify the spectral features, a much longer simulation of at least several nanoseconds would be required to fully converge the THz region of the spectrum and enable full quantitative comparison with experiment, as demonstrated for the GEMS model in Ref.~\cite{unke-gems-sciadv}. We note that the power spectrum computed on the final 1.3~ns simulation (discarding the initial 300~ps to equilibration) does not yield a significantly different spectrum to that of the 125~ps simulation used in Figure \ref{fig:crambin_power_spec_hhs}, indicating that significantly more sampling would be required to resolve the THz lineshapes sufficiently to enable experimental comparison~\cite{woodsGlassyStateCrambin2014}.

It is also noteworthy that whilst the three MLPs compared in this section give a qualitatively similar result, their modelling choices are significantly different. Most notably, GEMS and SO3LR explicitly include long-range interactions, whilst MACE is local. Additionally, of these models, both MACE and SO3LR are universal potentials, whilst GEMS is fitted in a system-specific fashion.
We  mention that, in addition to the GEMS model, the authors benchmarked a variant which was trained only on small fragments, denoted GEMS*, instead of the large system-specific configurations used to fit GEMS~\cite{unke-gems-sciadv}.  The power spectrum produced by this model is also in qualitative agreement with the other ML potentials, however, the RMSD is reduced, indicating the protein structure is confined to small fluctuations around local minima.  This model also fails to reproduce experimental observables of other biomolecular systems, for example Ala\textsubscript{15} is not predicted to fold under the GEMS* potential, suggesting the top-down fragments are required to accurately reproduce nonbonded interactions within the GEMS family of models.

The results of sections \ref{sec:condensed_phase} and \ref{sec:biologics} highlight the most significant differences between previous transferable organic ML potentials and the MACE-OFF models.   Whilst there is still significant work to be done to comprehensively show improvement over classical forcefields for large biomolecular systems, these results demonstrate the capabilities of models fit `bottom up' from quantum mechanical data to learn subtle intermolecular interactions and correctly predict macroscopic properties. 


\subsection{Computational performance}
\label{sec:performance}
\begin{figure}
    \centering
    \includegraphics[width=\linewidth]{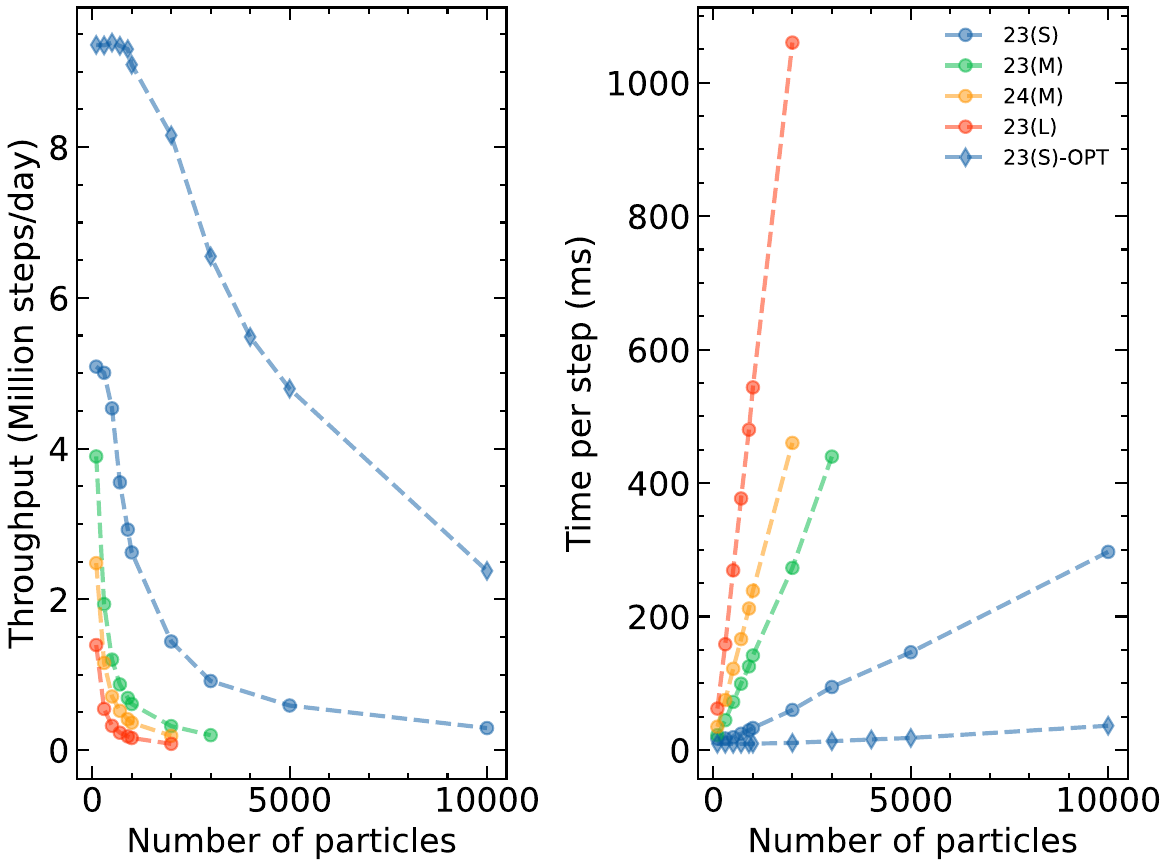}
    \caption{\textbf{Performance in OpenMM.} Performance of MACE-OFF models in  OpenMM for liquid water in the NVT ensemble at 1 g/cm\textsuperscript{3} at 300~K with a 1~fs timestep on a single NVIDIA 80GB A100 GPU. The \texttt{cuda\_mace} implementation is labeled `OPT'. Forces were evaluated in double precision, and integrated in single precision.}
    \label{fig:openmm-perf}
\end{figure}

To evaluate the computational performance of the MACE-OFF models, we performed NVT molecular dynamics for water using both OpenMM and LAMMPS, with simulation boxes ranging from $10^2$ to $10^5$ atoms. The OpenMM results shown in Figure~\ref{fig:openmm-perf} were collected with a single NVIDIA A100 80GB GPU. The LAMMPS results shown in {\bf Section S9} highlight both CPU and GPU evaluation.

In addition to the native PyTorch implementations of the MACE-OFF23 models, we also report results for optimized implementations leveraging custom C++, CUDA, and/or Kokkos code. One optimized variant incorporates custom CUDA kernels which are packaged as a standalone, inference-only library,  \texttt{cuda\_mace}~\cite{NJB2024}. Another pure-C++ implementation, as well as a Kokkos implementation, will be made available as part of the \texttt{symmetrix} library~\cite{WCW2025}.

In these optimized variants, the learnable radial function $R^{(t)}_{k \eta_1 l_1 l_2 l_3}(r_{ij})$ is evaluated with cubic splines, which efficiently map $r_{ij}$ to each of the outputs of the radial MLP. Furthermore, eqs. \ref{eq:phi-basis-t} and \ref{eq:atomic-basis-t} are implemented in unified kernels that simultaneously compute the one-particle basis and perform the summation over neighbors to generate the invariant atomic basis $A_i$. The computation of spherical harmonics, as detailed in equation \ref{eq:phi-basis-t} is accelerated using the sphericart~\cite{sphericart} library, which offers a highly optimized implementation for real spherical harmonics. All linear and equivariant linear operations have been substituted with custom kernels, and the \texttt{cuda\_mace} implementation further leverages tensor core acceleration while incorporating error-correction techniques to ensure numerical accuracy. Additionally, the computation of the product basis $\mathbf{B}_i$ in eq~\ref{eq:symmbasis_L1} is optimized using the sparsity of the generalized Clebsch-Gordon coefficients $\mathcal{C}^{LM}_{\eta_{\nu} \bm l \bm m}$, significantly accelerating the product calculations over both the body order $v$ and the coupling terms.

\section{Outlook}
In this paper, we presented the series of MACE-OFF force field models, demonstrating the broad applicability, transferability and capability of purely short-range machine learning potentials for organic (bio)molecular simulations. We have shown that the models, based on the MACE higher order equivariant message passing architecture, can improve on the pioneering ANI models~\cite{devereux2020ani_2x} that served as the only widely applicable machine learning force fields for molecules for a number of years. While a full ablation study between different ML models is not possible due to a wide range of design choices made in their construction, the MACE-OFF model architecture, training data and loss function all contribute to significant improvements in both accuracy and extrapolation, combined with high computational speed. 

We trained here so-called small, medium and large versions of the MACE-OFF models, with systematically improving accuracy when benchmarked against gas phase quantum mechanical data. We further investigated the accuracy of the medium models against a range of crystalline, liquid and biomolecular condensed phase data, as we found them to provide a good compromise between accuracy and computational expense. We have made all force fields available to the community, so that users can choose the model most suited to their application.

The lack of explicit long-range interactions limits the domain of applicability of the present model to neutral, non-radical and non-reactive systems. This is something that the recently published AIMNet2 model aims to address by extending the ANI models to include charged species and long-range interactions~\cite{anstine2023aimnet2}. We are currently working on a next-generation MACE-OFF model that will similarly include an explicit description of charges, enabling the description of amino acids with different protonation states, charged nucleic acids, and counter-ions. This will pave the way towards obtaining an accurate quantum mechanical transferable machine learning force field for simulating the full range of biologically relevant systems. 

\section*{Acknowledgements}
DPK acknowledges support from AstraZeneca and the EPSRC. DJC and JH acknowledge support from a UKRI Future Leaders Fellowship (grant MR/T019654/1). V.K. acknowledges support from the Ernest Oppenheimer Early Career Fellowship, the Sydney Harvey Junior Research Fellowship, Churchill College, University of Cambridge, and UCL's startup funds. W.C.W. acknowledges support from the EPSRC (Grant EP/V062654/1), as well as an engagement from the Virtual Institute for Scientific Software at Georgia Tech, funded by Schmidt Sciences. In particular, Ketan Bhardwaj and Dave Brownwell of the Georgia Tech Scientific Software Engineering Center contributed to the Kokkos-based MACE evaluator. JHM acknowledges support from an AstraZeneca Non-Clinical PhD studentship, and thanks Prof. Ola Engkvist, Dr Marco Kl\"ahn and Dr Graeme Robb from AstraZeneca for their helpful discussions and supervision. The authors also thank Christoph Schran for providing the carved water cluster geometries. We also thank Dr Peter Eastman and Dr Stephen Farr for their help implementing the MACE-OpenMM interface. 

We are grateful for computational support from the Swiss National Supercomputing Centre under project s1209, UCL Myriad High Performance Computing Facility, the Rocket High Performance Computing service at Newcastle University, the UK national high-performance computing service, ARCHER2, for which access was obtained via the UKCP consortium and EPSRC grant EP/P022561/1, the Cambridge Service for Data Driven Discovery (CSD3), obtained through a University of Cambridge EPSRC Core Equipment Award (EP/X034712/1), and Ångström AI. 

Elements of this work are adapted from a PhD thesis.

\section*{Data Availability}
The data used to train the models are publicly available at: \url{https://doi.org/10.17863/CAM.107498}. The torsion drive dataset is also available at: \url{https://zenodo.org/records/11385284}. The MACE-OFF series of models is available at: \url{https://github.com/ACEsuit/mace-off/tree/main}.

\section*{Supplementary Information}

The supplementary information contains the following sections:
\begin{itemize}
    \item Training details for the MACE-OFF models, including filtering of the SPICE dataset
    \item Detailed breakdown of test set errors, including intramolecular force errors
    \item Details of the torsion scans
    \item Details of lattice enthalpy calculations
    \item Crystal structure geometries
    \item Details of condensed phase simulations, including error tables
    \item Details of biological simulations
    \item Simulation details and results for dipole models
    \item Computational performance in LAMMPS
\end{itemize}

\bibliography{references}

\providecommand{\latin}[1]{#1}
\makeatletter
\providecommand{\doi}
  {\begingroup\let\do\@makeother\dospecials
  \catcode`\{=1 \catcode`\}=2 \doi@aux}
\providecommand{\doi@aux}[1]{\endgroup\texttt{#1}}
\makeatother
\providecommand*\mcitethebibliography{\thebibliography}
\csname @ifundefined\endcsname{endmcitethebibliography}  {\let\endmcitethebibliography\endthebibliography}{}
\begin{mcitethebibliography}{36}
\providecommand*\natexlab[1]{#1}
\providecommand*\mciteSetBstSublistMode[1]{}
\providecommand*\mciteSetBstMaxWidthForm[2]{}
\providecommand*\mciteBstWouldAddEndPuncttrue
  {\def\EndOfBibitem{\unskip.}}
\providecommand*\mciteBstWouldAddEndPunctfalse
  {\let\EndOfBibitem\relax}
\providecommand*\mciteSetBstMidEndSepPunct[3]{}
\providecommand*\mciteSetBstSublistLabelBeginEnd[3]{}
\providecommand*\EndOfBibitem{}
\mciteSetBstSublistMode{f}
\mciteSetBstMaxWidthForm{subitem}{(\alph{mcitesubitemcount})}
\mciteSetBstSublistLabelBeginEnd
  {\mcitemaxwidthsubitemform\space}
  {\relax}
  {\relax}

\bibitem[Magd{\u{a}}u \latin{et~al.}(2023)Magd{\u{a}}u, Arismendi-Arrieta, Smith, Grey, Hermansson, and Cs{\'a}nyi]{magduau2023machine}
Magd{\u{a}}u,~I.-B.; Arismendi-Arrieta,~D.~J.; Smith,~H.~E.; Grey,~C.~P.; Hermansson,~K.; Cs{\'a}nyi,~G. Machine learning force fields for molecular liquids: Ethylene Carbonate/Ethyl Methyl Carbonate binary solvent. \emph{npj Computational Materials} \textbf{2023}, \emph{9}, 146\relax
\mciteBstWouldAddEndPuncttrue
\mciteSetBstMidEndSepPunct{\mcitedefaultmidpunct}
{\mcitedefaultendpunct}{\mcitedefaultseppunct}\relax
\EndOfBibitem
\bibitem[Rai \latin{et~al.}(2022)Rai, Sresht, Yang, Unwalla, Tu, Mathiowetz, and Bakken]{rai2022torsionnet}
Rai,~B.~K.; Sresht,~V.; Yang,~Q.; Unwalla,~R.; Tu,~M.; Mathiowetz,~A.~M.; Bakken,~G.~A. Torsionnet: A deep neural network to rapidly predict small-molecule torsional energy profiles with the accuracy of quantum mechanics. \emph{Journal of Chemical Information and Modeling} \textbf{2022}, \emph{62}, 785--800\relax
\mciteBstWouldAddEndPuncttrue
\mciteSetBstMidEndSepPunct{\mcitedefaultmidpunct}
{\mcitedefaultendpunct}{\mcitedefaultseppunct}\relax
\EndOfBibitem
\bibitem[Lahey \latin{et~al.}(2020)Lahey, Thien~Phuc, and Rowley]{lahey2020biaryl_torsion}
Lahey,~S.-L.~J.; Thien~Phuc,~T.~N.; Rowley,~C.~N. Benchmarking force field and the ani neural network potentials for the torsional potential energy surface of biaryl drug fragments. \emph{Journal of Chemical Information and Modeling} \textbf{2020}, \emph{60}, 6258--6268\relax
\mciteBstWouldAddEndPuncttrue
\mciteSetBstMidEndSepPunct{\mcitedefaultmidpunct}
{\mcitedefaultendpunct}{\mcitedefaultseppunct}\relax
\EndOfBibitem
\bibitem[Eastman \latin{et~al.}(2023)Eastman, Behara, Dotson, Galvelis, Herr, Horton, Mao, Chodera, Pritchard, Wang, \latin{et~al.} others]{eastman2023spice}
Eastman,~P.; Behara,~P.~K.; Dotson,~D.~L.; Galvelis,~R.; Herr,~J.~E.; Horton,~J.~T.; Mao,~Y.; Chodera,~J.~D.; Pritchard,~B.~P.; Wang,~Y.; others Spice, a dataset of drug-like molecules and peptides for training machine learning potentials. \emph{Scientific Data} \textbf{2023}, \emph{10}, 11\relax
\mciteBstWouldAddEndPuncttrue
\mciteSetBstMidEndSepPunct{\mcitedefaultmidpunct}
{\mcitedefaultendpunct}{\mcitedefaultseppunct}\relax
\EndOfBibitem
\bibitem[Najibi and Goerigk(2018)Najibi, and Goerigk]{najibi2018nonlocal}
Najibi,~A.; Goerigk,~L. The nonlocal kernel in van der Waals density functionals as an additive correction: An extensive analysis with special emphasis on the B97M-V and $\omega$B97M-V approaches. \emph{Journal of Chemical Theory and Computation} \textbf{2018}, \emph{14}, 5725--5738\relax
\mciteBstWouldAddEndPuncttrue
\mciteSetBstMidEndSepPunct{\mcitedefaultmidpunct}
{\mcitedefaultendpunct}{\mcitedefaultseppunct}\relax
\EndOfBibitem
\bibitem[Weigend and Ahlrichs(2005)Weigend, and Ahlrichs]{weigend2005balanced}
Weigend,~F.; Ahlrichs,~R. Balanced basis sets of split valence, triple zeta valence and quadruple zeta valence quality for H to Rn: Design and assessment of accuracy. \emph{Physical Chemistry Chemical Physics} \textbf{2005}, \emph{7}, 3297--3305\relax
\mciteBstWouldAddEndPuncttrue
\mciteSetBstMidEndSepPunct{\mcitedefaultmidpunct}
{\mcitedefaultendpunct}{\mcitedefaultseppunct}\relax
\EndOfBibitem
\bibitem[Rappoport and Furche(2010)Rappoport, and Furche]{rappoport2010property}
Rappoport,~D.; Furche,~F. Property-optimized Gaussian basis sets for molecular response calculations. \emph{The Journal of chemical physics} \textbf{2010}, \emph{133}\relax
\mciteBstWouldAddEndPuncttrue
\mciteSetBstMidEndSepPunct{\mcitedefaultmidpunct}
{\mcitedefaultendpunct}{\mcitedefaultseppunct}\relax
\EndOfBibitem
\bibitem[Grimme \latin{et~al.}(2011)Grimme, Ehrlich, and Goerigk]{grimme2011bj_damp}
Grimme,~S.; Ehrlich,~S.; Goerigk,~L. Effect of the damping function in dispersion corrected density functional theory. \emph{Journal of computational chemistry} \textbf{2011}, \emph{32}, 1456--1465\relax
\mciteBstWouldAddEndPuncttrue
\mciteSetBstMidEndSepPunct{\mcitedefaultmidpunct}
{\mcitedefaultendpunct}{\mcitedefaultseppunct}\relax
\EndOfBibitem
\bibitem[Grimme \latin{et~al.}(2010)Grimme, Antony, Ehrlich, and Krieg]{grimme2010d3}
Grimme,~S.; Antony,~J.; Ehrlich,~S.; Krieg,~H. A consistent and accurate ab initio parametrization of density functional dispersion correction (DFT-D) for the 94 elements H-Pu. \emph{The Journal of chemical physics} \textbf{2010}, \emph{132}\relax
\mciteBstWouldAddEndPuncttrue
\mciteSetBstMidEndSepPunct{\mcitedefaultmidpunct}
{\mcitedefaultendpunct}{\mcitedefaultseppunct}\relax
\EndOfBibitem
\bibitem[Qiu \latin{et~al.}(2020)Qiu, Smith, Stern, Feng, Jang, and Wang]{qiu2020torsiondrive}
Qiu,~Y.; Smith,~D.~G.; Stern,~C.~D.; Feng,~M.; Jang,~H.; Wang,~L.-P. Driving torsion scans with wavefront propagation. \emph{The Journal of chemical physics} \textbf{2020}, \emph{152}\relax
\mciteBstWouldAddEndPuncttrue
\mciteSetBstMidEndSepPunct{\mcitedefaultmidpunct}
{\mcitedefaultendpunct}{\mcitedefaultseppunct}\relax
\EndOfBibitem
\bibitem[Smith \latin{et~al.}(2021)Smith, Lolinco, Glick, Lee, Alenaizan, Barnes, Borca, Di~Remigio, Dotson, Ehlert, \latin{et~al.} others]{smith2021qcengine}
Smith,~D.~G.; Lolinco,~A.~T.; Glick,~Z.~L.; Lee,~J.; Alenaizan,~A.; Barnes,~T.~A.; Borca,~C.~H.; Di~Remigio,~R.; Dotson,~D.~L.; Ehlert,~S.; others Quantum chemistry common driver and databases (QCDB) and quantum chemistry engine (QCEngine): Automation and interoperability among computational chemistry programs. \emph{The Journal of chemical physics} \textbf{2021}, \emph{155}\relax
\mciteBstWouldAddEndPuncttrue
\mciteSetBstMidEndSepPunct{\mcitedefaultmidpunct}
{\mcitedefaultendpunct}{\mcitedefaultseppunct}\relax
\EndOfBibitem
\bibitem[Wang and Song(2016)Wang, and Song]{wang2016geometry}
Wang,~L.-P.; Song,~C. Geometry optimization made simple with translation and rotation coordinates. \emph{The Journal of chemical physics} \textbf{2016}, \emph{144}\relax
\mciteBstWouldAddEndPuncttrue
\mciteSetBstMidEndSepPunct{\mcitedefaultmidpunct}
{\mcitedefaultendpunct}{\mcitedefaultseppunct}\relax
\EndOfBibitem
\bibitem[Smith \latin{et~al.}(2020)Smith, Burns, Simmonett, Parrish, Schieber, Galvelis, Kraus, Kruse, Di~Remigio, Alenaizan, \latin{et~al.} others]{smith2020psi4}
Smith,~D.~G.; Burns,~L.~A.; Simmonett,~A.~C.; Parrish,~R.~M.; Schieber,~M.~C.; Galvelis,~R.; Kraus,~P.; Kruse,~H.; Di~Remigio,~R.; Alenaizan,~A.; others PSI4 1.4: Open-source software for high-throughput quantum chemistry. \emph{The Journal of chemical physics} \textbf{2020}, \emph{152}\relax
\mciteBstWouldAddEndPuncttrue
\mciteSetBstMidEndSepPunct{\mcitedefaultmidpunct}
{\mcitedefaultendpunct}{\mcitedefaultseppunct}\relax
\EndOfBibitem
\bibitem[Reilly and Tkatchenko(2013)Reilly, and Tkatchenko]{reilly2013understanding}
Reilly,~A.~M.; Tkatchenko,~A. Understanding the role of vibrations, exact exchange, and many-body van der Waals interactions in the cohesive properties of molecular crystals. \emph{The Journal of chemical physics} \textbf{2013}, \emph{139}, 024705\relax
\mciteBstWouldAddEndPuncttrue
\mciteSetBstMidEndSepPunct{\mcitedefaultmidpunct}
{\mcitedefaultendpunct}{\mcitedefaultseppunct}\relax
\EndOfBibitem
\bibitem[Dolgonos \latin{et~al.}(2019)Dolgonos, Hoja, and Boese]{dolgonos2019revised_x23}
Dolgonos,~G.~A.; Hoja,~J.; Boese,~A.~D. Revised values for the X23 benchmark set of molecular crystals. \emph{Physical Chemistry Chemical Physics} \textbf{2019}, \emph{21}, 24333--24344\relax
\mciteBstWouldAddEndPuncttrue
\mciteSetBstMidEndSepPunct{\mcitedefaultmidpunct}
{\mcitedefaultendpunct}{\mcitedefaultseppunct}\relax
\EndOfBibitem
\bibitem[Horton \latin{et~al.}(2019)Horton, Allen, Dodda, and Cole]{horton2019qubekit}
Horton,~J.~T.; Allen,~A.~E.; Dodda,~L.~S.; Cole,~D.~J. QUBEKit: Automating the derivation of force field parameters from quantum mechanics. \emph{Journal of chemical information and modeling} \textbf{2019}, \emph{59}, 1366--1381\relax
\mciteBstWouldAddEndPuncttrue
\mciteSetBstMidEndSepPunct{\mcitedefaultmidpunct}
{\mcitedefaultendpunct}{\mcitedefaultseppunct}\relax
\EndOfBibitem
\bibitem[Mart{\'i}nez \latin{et~al.}(2009)Mart{\'i}nez, Andrade, Birgin, and Mart{\'i}nez]{martinezPACKMOLPackageBuilding2009a}
Mart{\'i}nez,~L.; Andrade,~R.; Birgin,~E.~G.; Mart{\'i}nez,~J.~M. {PACKMOL: A package for building initial configurations for molecular dynamics simulations}. \emph{Journal of Computational Chemistry} \textbf{2009}, \emph{30}, 2157--2164\relax
\mciteBstWouldAddEndPuncttrue
\mciteSetBstMidEndSepPunct{\mcitedefaultmidpunct}
{\mcitedefaultendpunct}{\mcitedefaultseppunct}\relax
\EndOfBibitem
\bibitem[Eastman \latin{et~al.}(2017)Eastman, Swails, Chodera, McGibbon, Zhao, Beauchamp, Wang, Simmonett, Harrigan, Stern, Wiewiora, Brooks, and Pande]{eastmanOpenMMRapidDevelopment2017}
Eastman,~P.; Swails,~J.; Chodera,~J.~D.; McGibbon,~R.~T.; Zhao,~Y.; Beauchamp,~K.~A.; Wang,~L.-P.; Simmonett,~A.~C.; Harrigan,~M.~P.; Stern,~C.~D.; Wiewiora,~R.~P.; Brooks,~B.~R.; Pande,~V.~S. {{OpenMM}} 7: {{Rapid}} Development of High Performance Algorithms for Molecular Dynamics. \emph{PLOS Computational Biology} \textbf{2017}, \emph{13}, e1005659\relax
\mciteBstWouldAddEndPuncttrue
\mciteSetBstMidEndSepPunct{\mcitedefaultmidpunct}
{\mcitedefaultendpunct}{\mcitedefaultseppunct}\relax
\EndOfBibitem
\bibitem[Brehm and Kirchner(2011)Brehm, and Kirchner]{brehmTRAVISFreeAnalyzer2011}
Brehm,~M.; Kirchner,~B. {{TRAVIS}} - {{A Free Analyzer}} and {{Visualizer}} for {{Monte Carlo}} and {{Molecular Dynamics Trajectories}}. \emph{Journal of Chemical Information and Modeling} \textbf{2011}, \emph{51}, 2007--2023\relax
\mciteBstWouldAddEndPuncttrue
\mciteSetBstMidEndSepPunct{\mcitedefaultmidpunct}
{\mcitedefaultendpunct}{\mcitedefaultseppunct}\relax
\EndOfBibitem
\bibitem[Brehm \latin{et~al.}(2020)Brehm, Thomas, Gehrke, and Kirchner]{brehmTRAVISFreeAnalyzer2020a}
Brehm,~M.; Thomas,~M.; Gehrke,~S.; Kirchner,~B. {{TRAVIS}}{\textemdash}{{A}} Free Analyzer for Trajectories from Molecular Simulation. \emph{The Journal of Chemical Physics} \textbf{2020}, \emph{152}, 164105\relax
\mciteBstWouldAddEndPuncttrue
\mciteSetBstMidEndSepPunct{\mcitedefaultmidpunct}
{\mcitedefaultendpunct}{\mcitedefaultseppunct}\relax
\EndOfBibitem
\bibitem[Thomas \latin{et~al.}(2013)Thomas, Brehm, Fligg, V{\"o}hringer, and Kirchner]{thomasComputingVibrationalSpectra2013a}
Thomas,~M.; Brehm,~M.; Fligg,~R.; V{\"o}hringer,~P.; Kirchner,~B. Computing Vibrational Spectra from Ab Initio Molecular Dynamics. \emph{Physical Chemistry Chemical Physics} \textbf{2013}, \emph{15}, 6608--6622\relax
\mciteBstWouldAddEndPuncttrue
\mciteSetBstMidEndSepPunct{\mcitedefaultmidpunct}
{\mcitedefaultendpunct}{\mcitedefaultseppunct}\relax
\EndOfBibitem
\bibitem[Thomas \latin{et~al.}(2015)Thomas, Brehm, and Kirchner]{thomasVoronoiDipoleMoments2015}
Thomas,~M.; Brehm,~M.; Kirchner,~B. Voronoi Dipole Moments for the Simulation of Bulk Phase Vibrational Spectra. \emph{Physical Chemistry Chemical Physics} \textbf{2015}, \emph{17}, 3207--3213\relax
\mciteBstWouldAddEndPuncttrue
\mciteSetBstMidEndSepPunct{\mcitedefaultmidpunct}
{\mcitedefaultendpunct}{\mcitedefaultseppunct}\relax
\EndOfBibitem
\bibitem[Pracht \latin{et~al.}(2024)Pracht, Pillai, Kapil, Csányi, G\"{o}nnheimer, Vondrák, Margraf, and Wales]{Pracht2024}
Pracht,~P.; Pillai,~Y.; Kapil,~V.; Csányi,~G.; G\"{o}nnheimer,~N.; Vondrák,~M.; Margraf,~J.~T.; Wales,~D.~J. Efficient Composite Infrared Spectroscopy: Combining the Double-Harmonic Approximation with Machine Learning Potentials. \emph{Journal of Chemical Theory and Computation} \textbf{2024}, \emph{20}, 10986–11004\relax
\mciteBstWouldAddEndPuncttrue
\mciteSetBstMidEndSepPunct{\mcitedefaultmidpunct}
{\mcitedefaultendpunct}{\mcitedefaultseppunct}\relax
\EndOfBibitem
\bibitem[Kapil \latin{et~al.}(2023)Kapil, Kovács, Csányi, and Michaelides]{Kapil2023}
Kapil,~V.; Kovács,~D.~P.; Csányi,~G.; Michaelides,~A. First-principles spectroscopy of aqueous interfaces using machine-learned electronic and quantum nuclear effects. \emph{Faraday Discussions} \textbf{2023}, \relax
\mciteBstWouldAddEndPunctfalse
\mciteSetBstMidEndSepPunct{\mcitedefaultmidpunct}
{}{\mcitedefaultseppunct}\relax
\EndOfBibitem
\bibitem[Musil \latin{et~al.}(2022)Musil, Zaporozhets, Noé, Clementi, and Kapil]{Musil2022}
Musil,~F.; Zaporozhets,~I.; Noé,~F.; Clementi,~C.; Kapil,~V. Quantum dynamics using path integral coarse-graining. \emph{The Journal of Chemical Physics} \textbf{2022}, \emph{157}\relax
\mciteBstWouldAddEndPuncttrue
\mciteSetBstMidEndSepPunct{\mcitedefaultmidpunct}
{\mcitedefaultendpunct}{\mcitedefaultseppunct}\relax
\EndOfBibitem
\bibitem[Kapil \latin{et~al.}(2019)Kapil, Wieme, Vandenbrande, Lamaire, Van~Speybroeck, and Ceriotti]{Kapil2019}
Kapil,~V.; Wieme,~J.; Vandenbrande,~S.; Lamaire,~A.; Van~Speybroeck,~V.; Ceriotti,~M. Modeling the Structural and Thermal Properties of Loaded Metal–Organic Frameworks. An Interplay of Quantum and Anharmonic Fluctuations. \emph{Journal of Chemical Theory and Computation} \textbf{2019}, \emph{15}, 3237–3249\relax
\mciteBstWouldAddEndPuncttrue
\mciteSetBstMidEndSepPunct{\mcitedefaultmidpunct}
{\mcitedefaultendpunct}{\mcitedefaultseppunct}\relax
\EndOfBibitem
\bibitem[Ceriotti \latin{et~al.}(2010)Ceriotti, Parrinello, Markland, and Manolopoulos]{Ceriotti2010}
Ceriotti,~M.; Parrinello,~M.; Markland,~T.~E.; Manolopoulos,~D.~E. Efficient stochastic thermostatting of path integral molecular dynamics. \emph{The Journal of Chemical Physics} \textbf{2010}, \emph{133}\relax
\mciteBstWouldAddEndPuncttrue
\mciteSetBstMidEndSepPunct{\mcitedefaultmidpunct}
{\mcitedefaultendpunct}{\mcitedefaultseppunct}\relax
\EndOfBibitem
\bibitem[Kapil \latin{et~al.}(2019)Kapil, Rossi, Marsalek, Petraglia, Litman, Spura, Cheng, Cuzzocrea, Meißner, Wilkins, Helfrecht, Juda, Bienvenue, Fang, Kessler, Poltavsky, Vandenbrande, Wieme, Corminboeuf, K\"{u}hne, Manolopoulos, Markland, Richardson, Tkatchenko, Tribello, Van~Speybroeck, and Ceriotti]{ipi}
Kapil,~V. \latin{et~al.}  i-PI 2.0: A universal force engine for advanced molecular simulations. \emph{Computer Physics Communications} \textbf{2019}, \emph{236}, 214–223\relax
\mciteBstWouldAddEndPuncttrue
\mciteSetBstMidEndSepPunct{\mcitedefaultmidpunct}
{\mcitedefaultendpunct}{\mcitedefaultseppunct}\relax
\EndOfBibitem
\bibitem[Raimbault \latin{et~al.}(2019)Raimbault, Athavale, and Rossi]{raimbault2019parac_raman_AbInitio}
Raimbault,~N.; Athavale,~V.; Rossi,~M. Anharmonic effects in the low-frequency vibrational modes of aspirin and paracetamol crystals. \emph{Physical Review Materials} \textbf{2019}, \emph{3}, 053605\relax
\mciteBstWouldAddEndPuncttrue
\mciteSetBstMidEndSepPunct{\mcitedefaultmidpunct}
{\mcitedefaultendpunct}{\mcitedefaultseppunct}\relax
\EndOfBibitem
\bibitem[Raimbault \latin{et~al.}(2019)Raimbault, Grisafi, Ceriotti, and Rossi]{Raimbault2019Using}
Raimbault,~N.; Grisafi,~A.; Ceriotti,~M.; Rossi,~M. Using Gaussian process regression to simulate the vibrational Raman spectra of molecular crystals. \emph{New Journal of Physics} \textbf{2019}, \emph{21}, 105001\relax
\mciteBstWouldAddEndPuncttrue
\mciteSetBstMidEndSepPunct{\mcitedefaultmidpunct}
{\mcitedefaultendpunct}{\mcitedefaultseppunct}\relax
\EndOfBibitem
\bibitem[Raimbault \latin{et~al.}(2019)Raimbault, Grisafi, Ceriotti, and Rossi]{raimbault2019paracetamol_SAGPR}
Raimbault,~N.; Grisafi,~A.; Ceriotti,~M.; Rossi,~M. Using Gaussian process regression to simulate the vibrational Raman spectra of molecular crystals. \emph{New Journal of Physics} \textbf{2019}, \emph{21}, 105001\relax
\mciteBstWouldAddEndPuncttrue
\mciteSetBstMidEndSepPunct{\mcitedefaultmidpunct}
{\mcitedefaultendpunct}{\mcitedefaultseppunct}\relax
\EndOfBibitem
\bibitem[Kolesov \latin{et~al.}(2011)Kolesov, Mikhailenko, and Boldyreva]{Kolesov2011}
Kolesov,~B.~A.; Mikhailenko,~M.~A.; Boldyreva,~E.~V. Dynamics of the intermolecular hydrogen bonds in the polymorphs of paracetamol in relation to crystal packing and conformational transitions: a variable-temperature polarized Raman spectroscopy study. \emph{Physical Chemistry Chemical Physics} \textbf{2011}, \emph{13}, 14243\relax
\mciteBstWouldAddEndPuncttrue
\mciteSetBstMidEndSepPunct{\mcitedefaultmidpunct}
{\mcitedefaultendpunct}{\mcitedefaultseppunct}\relax
\EndOfBibitem
\bibitem[Raimbault \latin{et~al.}(2019)Raimbault, Athavale, and Rossi]{Raimbault2019Anharmonic}
Raimbault,~N.; Athavale,~V.; Rossi,~M. Anharmonic effects in the low-frequency vibrational modes of aspirin and paracetamol crystals. \emph{Physical Review Materials} \textbf{2019}, \emph{3}\relax
\mciteBstWouldAddEndPuncttrue
\mciteSetBstMidEndSepPunct{\mcitedefaultmidpunct}
{\mcitedefaultendpunct}{\mcitedefaultseppunct}\relax
\EndOfBibitem
\bibitem[Ceriotti \latin{et~al.}(2016)Ceriotti, Fang, Kusalik, McKenzie, Michaelides, Morales, and Markland]{Ceriotti2016}
Ceriotti,~M.; Fang,~W.; Kusalik,~P.~G.; McKenzie,~R.~H.; Michaelides,~A.; Morales,~M.~A.; Markland,~T.~E. Nuclear Quantum Effects in Water and Aqueous Systems: Experiment, Theory, and Current Challenges. \emph{Chemical Reviews} \textbf{2016}, \emph{116}, 7529–7550\relax
\mciteBstWouldAddEndPuncttrue
\mciteSetBstMidEndSepPunct{\mcitedefaultmidpunct}
{\mcitedefaultendpunct}{\mcitedefaultseppunct}\relax
\EndOfBibitem
\bibitem[Morawietz \latin{et~al.}(2018)Morawietz, Marsalek, Pattenaude, Streacker, Ben-Amotz, and Markland]{morawietz2018}
Morawietz,~T.; Marsalek,~O.; Pattenaude,~S.~R.; Streacker,~L.~M.; Ben-Amotz,~D.; Markland,~T.~E. The Interplay of Structure and Dynamics in the Raman Spectrum of Liquid Water over the Full Frequency and Temperature Range. \emph{The Journal of Physical Chemistry Letters} \textbf{2018}, \emph{9}, 851--857\relax
\mciteBstWouldAddEndPuncttrue
\mciteSetBstMidEndSepPunct{\mcitedefaultmidpunct}
{\mcitedefaultendpunct}{\mcitedefaultseppunct}\relax
\EndOfBibitem
\end{mcitethebibliography}


\begin{thebibliography}{90}%
\makeatletter
\providecommand \@ifxundefined [1]{%
 \@ifx{#1\undefined}
}%
\providecommand \@ifnum [1]{%
 \ifnum #1\expandafter \@firstoftwo
 \else \expandafter \@secondoftwo
 \fi
}%
\providecommand \@ifx [1]{%
 \ifx #1\expandafter \@firstoftwo
 \else \expandafter \@secondoftwo
 \fi
}%
\providecommand \natexlab [1]{#1}%
\providecommand \enquote  [1]{``#1''}%
\providecommand \bibnamefont  [1]{#1}%
\providecommand \bibfnamefont [1]{#1}%
\providecommand \citenamefont [1]{#1}%
\providecommand \href@noop [0]{\@secondoftwo}%
\providecommand \href [0]{\begingroup \@sanitize@url \@href}%
\providecommand \@href[1]{\@@startlink{#1}\@@href}%
\providecommand \@@href[1]{\endgroup#1\@@endlink}%
\providecommand \@sanitize@url [0]{\catcode `\\12\catcode `\$12\catcode `\&12\catcode `\#12\catcode `\^12\catcode `\_12\catcode `\%12\relax}%
\providecommand \@@startlink[1]{}%
\providecommand \@@endlink[0]{}%
\providecommand \url  [0]{\begingroup\@sanitize@url \@url }%
\providecommand \@url [1]{\endgroup\@href {#1}{\urlprefix }}%
\providecommand \urlprefix  [0]{URL }%
\providecommand \Eprint [0]{\href }%
\providecommand \doibase [0]{https://doi.org/}%
\providecommand \selectlanguage [0]{\@gobble}%
\providecommand \bibinfo  [0]{\@secondoftwo}%
\providecommand \bibfield  [0]{\@secondoftwo}%
\providecommand \translation [1]{[#1]}%
\providecommand \BibitemOpen [0]{}%
\providecommand \bibitemStop [0]{}%
\providecommand \bibitemNoStop [0]{.\EOS\space}%
\providecommand \EOS [0]{\spacefactor3000\relax}%
\providecommand \BibitemShut  [1]{\csname bibitem#1\endcsname}%
\let\auto@bib@innerbib\@empty
\bibitem [{\citenamefont {Lysogorskiy}\ \emph {et~al.}(2021)\citenamefont {Lysogorskiy}, \citenamefont {Oord}, \citenamefont {Bochkarev}, \citenamefont {Menon}, \citenamefont {Rinaldi}, \citenamefont {Hammerschmidt}, \citenamefont {Mrovec}, \citenamefont {Thompson}, \citenamefont {Cs{\'a}nyi}, \citenamefont {Ortner} \emph {et~al.}}]{lysogorskiy2021performant}%
  \BibitemOpen
  \bibfield  {author} {\bibinfo {author} {\bibfnamefont {Y.}~\bibnamefont {Lysogorskiy}}, \bibinfo {author} {\bibfnamefont {C.~v.~d.}\ \bibnamefont {Oord}}, \bibinfo {author} {\bibfnamefont {A.}~\bibnamefont {Bochkarev}}, \bibinfo {author} {\bibfnamefont {S.}~\bibnamefont {Menon}}, \bibinfo {author} {\bibfnamefont {M.}~\bibnamefont {Rinaldi}}, \bibinfo {author} {\bibfnamefont {T.}~\bibnamefont {Hammerschmidt}}, \bibinfo {author} {\bibfnamefont {M.}~\bibnamefont {Mrovec}}, \bibinfo {author} {\bibfnamefont {A.}~\bibnamefont {Thompson}}, \bibinfo {author} {\bibfnamefont {G.}~\bibnamefont {Cs{\'a}nyi}}, \bibinfo {author} {\bibfnamefont {C.}~\bibnamefont {Ortner}}, \emph {et~al.},\ }\bibfield  {title} {\bibinfo {title} {Performant implementation of the atomic cluster expansion (pace) and application to copper and silicon},\ }\href@noop {} {\bibfield  {journal} {\bibinfo  {journal} {npj computational materials}\ }\textbf {\bibinfo {volume} {7}},\ \bibinfo {pages} {97} (\bibinfo {year} {2021})}\BibitemShut {NoStop}%
\bibitem [{\citenamefont {Batatia}\ \emph {et~al.}(2022)\citenamefont {Batatia}, \citenamefont {Kovacs}, \citenamefont {Simm}, \citenamefont {Ortner},\ and\ \citenamefont {Cs{\'a}nyi}}]{batatia2022mace}%
  \BibitemOpen
  \bibfield  {author} {\bibinfo {author} {\bibfnamefont {I.}~\bibnamefont {Batatia}}, \bibinfo {author} {\bibfnamefont {D.~P.}\ \bibnamefont {Kovacs}}, \bibinfo {author} {\bibfnamefont {G.}~\bibnamefont {Simm}}, \bibinfo {author} {\bibfnamefont {C.}~\bibnamefont {Ortner}},\ and\ \bibinfo {author} {\bibfnamefont {G.}~\bibnamefont {Cs{\'a}nyi}},\ }\bibfield  {title} {\bibinfo {title} {Mace: Higher order equivariant message passing neural networks for fast and accurate force fields},\ }\href@noop {} {\bibfield  {journal} {\bibinfo  {journal} {Advances in Neural Information Processing Systems}\ }\textbf {\bibinfo {volume} {35}},\ \bibinfo {pages} {11423} (\bibinfo {year} {2022})}\BibitemShut {NoStop}%
\bibitem [{\citenamefont {Kovács}\ \emph {et~al.}(2023)\citenamefont {Kovács}, \citenamefont {Batatia}, \citenamefont {Arany},\ and\ \citenamefont {Csányi}}]{kovacs2023mace_eval}%
  \BibitemOpen
  \bibfield  {author} {\bibinfo {author} {\bibfnamefont {D.~P.}\ \bibnamefont {Kovács}}, \bibinfo {author} {\bibfnamefont {I.}~\bibnamefont {Batatia}}, \bibinfo {author} {\bibfnamefont {E.~S.}\ \bibnamefont {Arany}},\ and\ \bibinfo {author} {\bibfnamefont {G.}~\bibnamefont {Csányi}},\ }\bibfield  {title} {\bibinfo {title} {{Evaluation of the MACE force field architecture: From medicinal chemistry to materials science}},\ }\href {https://doi.org/10.1063/5.0155322} {\bibfield  {journal} {\bibinfo  {journal} {The Journal of Chemical Physics}\ }\textbf {\bibinfo {volume} {159}},\ \bibinfo {pages} {044118} (\bibinfo {year} {2023})}\BibitemShut {NoStop}%
\bibitem [{\citenamefont {Batzner}\ \emph {et~al.}(2022)\citenamefont {Batzner}, \citenamefont {Musaelian}, \citenamefont {Sun}, \citenamefont {Geiger}, \citenamefont {Mailoa}, \citenamefont {Kornbluth}, \citenamefont {Molinari}, \citenamefont {Smidt},\ and\ \citenamefont {Kozinsky}}]{batzner2023nequip}%
  \BibitemOpen
  \bibfield  {author} {\bibinfo {author} {\bibfnamefont {S.}~\bibnamefont {Batzner}}, \bibinfo {author} {\bibfnamefont {A.}~\bibnamefont {Musaelian}}, \bibinfo {author} {\bibfnamefont {L.}~\bibnamefont {Sun}}, \bibinfo {author} {\bibfnamefont {M.}~\bibnamefont {Geiger}}, \bibinfo {author} {\bibfnamefont {J.~P.}\ \bibnamefont {Mailoa}}, \bibinfo {author} {\bibfnamefont {M.}~\bibnamefont {Kornbluth}}, \bibinfo {author} {\bibfnamefont {N.}~\bibnamefont {Molinari}}, \bibinfo {author} {\bibfnamefont {T.~E.}\ \bibnamefont {Smidt}},\ and\ \bibinfo {author} {\bibfnamefont {B.}~\bibnamefont {Kozinsky}},\ }\bibfield  {title} {\bibinfo {title} {E (3)-equivariant graph neural networks for data-efficient and accurate interatomic potentials},\ }\href@noop {} {\bibfield  {journal} {\bibinfo  {journal} {Nature communications}\ }\textbf {\bibinfo {volume} {13}},\ \bibinfo {pages} {2453} (\bibinfo {year} {2022})}\BibitemShut {NoStop}%
\bibitem [{\citenamefont {Zaverkin}\ \emph {et~al.}(2023)\citenamefont {Zaverkin}, \citenamefont {Holzm{\"u}ller}, \citenamefont {Bonfirraro},\ and\ \citenamefont {K{\"a}stner}}]{zaverkin2023transfer}%
  \BibitemOpen
  \bibfield  {author} {\bibinfo {author} {\bibfnamefont {V.}~\bibnamefont {Zaverkin}}, \bibinfo {author} {\bibfnamefont {D.}~\bibnamefont {Holzm{\"u}ller}}, \bibinfo {author} {\bibfnamefont {L.}~\bibnamefont {Bonfirraro}},\ and\ \bibinfo {author} {\bibfnamefont {J.}~\bibnamefont {K{\"a}stner}},\ }\bibfield  {title} {\bibinfo {title} {Transfer learning for chemically accurate interatomic neural network potentials},\ }\href@noop {} {\bibfield  {journal} {\bibinfo  {journal} {Physical Chemistry Chemical Physics}\ }\textbf {\bibinfo {volume} {25}},\ \bibinfo {pages} {5383} (\bibinfo {year} {2023})}\BibitemShut {NoStop}%
\bibitem [{\citenamefont {Haghighatlari}\ \emph {et~al.}(2022)\citenamefont {Haghighatlari}, \citenamefont {Li}, \citenamefont {Guan}, \citenamefont {Zhang}, \citenamefont {Das}, \citenamefont {Stein}, \citenamefont {Heidar-Zadeh}, \citenamefont {Liu}, \citenamefont {Head-Gordon}, \citenamefont {Bertels} \emph {et~al.}}]{haghighatlari2022newtonnet}%
  \BibitemOpen
  \bibfield  {author} {\bibinfo {author} {\bibfnamefont {M.}~\bibnamefont {Haghighatlari}}, \bibinfo {author} {\bibfnamefont {J.}~\bibnamefont {Li}}, \bibinfo {author} {\bibfnamefont {X.}~\bibnamefont {Guan}}, \bibinfo {author} {\bibfnamefont {O.}~\bibnamefont {Zhang}}, \bibinfo {author} {\bibfnamefont {A.}~\bibnamefont {Das}}, \bibinfo {author} {\bibfnamefont {C.~J.}\ \bibnamefont {Stein}}, \bibinfo {author} {\bibfnamefont {F.}~\bibnamefont {Heidar-Zadeh}}, \bibinfo {author} {\bibfnamefont {M.}~\bibnamefont {Liu}}, \bibinfo {author} {\bibfnamefont {M.}~\bibnamefont {Head-Gordon}}, \bibinfo {author} {\bibfnamefont {L.}~\bibnamefont {Bertels}}, \emph {et~al.},\ }\bibfield  {title} {\bibinfo {title} {Newtonnet: A newtonian message passing network for deep learning of interatomic potentials and forces},\ }\href@noop {} {\bibfield  {journal} {\bibinfo  {journal} {Digital Discovery}\ }\textbf {\bibinfo {volume} {1}},\ \bibinfo {pages} {333} (\bibinfo {year} {2022})}\BibitemShut {NoStop}%
\bibitem [{\citenamefont {Shapeev}(2016)}]{shapeev2016moment}%
  \BibitemOpen
  \bibfield  {author} {\bibinfo {author} {\bibfnamefont {A.~V.}\ \bibnamefont {Shapeev}},\ }\bibfield  {title} {\bibinfo {title} {Moment tensor potentials: A class of systematically improvable interatomic potentials},\ }\href@noop {} {\bibfield  {journal} {\bibinfo  {journal} {Multiscale Modeling \& Simulation}\ }\textbf {\bibinfo {volume} {14}},\ \bibinfo {pages} {1153} (\bibinfo {year} {2016})}\BibitemShut {NoStop}%
\bibitem [{\citenamefont {Anstine}\ \emph {et~al.}()\citenamefont {Anstine}, \citenamefont {Zubatyuk},\ and\ \citenamefont {Isayev}}]{anstine2023aimnet2}%
  \BibitemOpen
  \bibfield  {author} {\bibinfo {author} {\bibfnamefont {D.}~\bibnamefont {Anstine}}, \bibinfo {author} {\bibfnamefont {R.}~\bibnamefont {Zubatyuk}},\ and\ \bibinfo {author} {\bibfnamefont {O.}~\bibnamefont {Isayev}},\ }\bibfield  {title} {\bibinfo {title} {Aimnet2: A neural network potential to meet your neutral, charged, organic, and elemental-organic needs},\ }\href@noop {} {\bibfield  {journal} {\bibinfo  {journal} {ChemRxiv}\ }}\bibinfo {note} {Sumbitted 20-12-2024, accessed 2025-04-15, \url{10.26434/chemrxiv-2023-296ch-v3}}\BibitemShut {NoStop}%
\bibitem [{\citenamefont {Kabylda}\ \emph {et~al.}(2023)\citenamefont {Kabylda}, \citenamefont {Vassilev-Galindo}, \citenamefont {Chmiela}, \citenamefont {Poltavsky},\ and\ \citenamefont {Tkatchenko}}]{kabylda2023efficient}%
  \BibitemOpen
  \bibfield  {author} {\bibinfo {author} {\bibfnamefont {A.}~\bibnamefont {Kabylda}}, \bibinfo {author} {\bibfnamefont {V.}~\bibnamefont {Vassilev-Galindo}}, \bibinfo {author} {\bibfnamefont {S.}~\bibnamefont {Chmiela}}, \bibinfo {author} {\bibfnamefont {I.}~\bibnamefont {Poltavsky}},\ and\ \bibinfo {author} {\bibfnamefont {A.}~\bibnamefont {Tkatchenko}},\ }\bibfield  {title} {\bibinfo {title} {Efficient interatomic descriptors for accurate machine learning force fields of extended molecules},\ }\href@noop {} {\bibfield  {journal} {\bibinfo  {journal} {Nature Communications}\ }\textbf {\bibinfo {volume} {14}},\ \bibinfo {pages} {3562} (\bibinfo {year} {2023})}\BibitemShut {NoStop}%
\bibitem [{\citenamefont {Behler}(2021)}]{behler2021four}%
  \BibitemOpen
  \bibfield  {author} {\bibinfo {author} {\bibfnamefont {J.}~\bibnamefont {Behler}},\ }\bibfield  {title} {\bibinfo {title} {Four generations of high-dimensional neural network potentials},\ }\href@noop {} {\bibfield  {journal} {\bibinfo  {journal} {Chemical Reviews}\ }\textbf {\bibinfo {volume} {121}},\ \bibinfo {pages} {10037} (\bibinfo {year} {2021})}\BibitemShut {NoStop}%
\bibitem [{\citenamefont {Musil}\ \emph {et~al.}(2021)\citenamefont {Musil}, \citenamefont {Grisafi}, \citenamefont {Bart{\'o}k}, \citenamefont {Ortner}, \citenamefont {Cs{\'a}nyi},\ and\ \citenamefont {Ceriotti}}]{musil2021physics}%
  \BibitemOpen
  \bibfield  {author} {\bibinfo {author} {\bibfnamefont {F.}~\bibnamefont {Musil}}, \bibinfo {author} {\bibfnamefont {A.}~\bibnamefont {Grisafi}}, \bibinfo {author} {\bibfnamefont {A.~P.}\ \bibnamefont {Bart{\'o}k}}, \bibinfo {author} {\bibfnamefont {C.}~\bibnamefont {Ortner}}, \bibinfo {author} {\bibfnamefont {G.}~\bibnamefont {Cs{\'a}nyi}},\ and\ \bibinfo {author} {\bibfnamefont {M.}~\bibnamefont {Ceriotti}},\ }\bibfield  {title} {\bibinfo {title} {Physics-inspired structural representations for molecules and materials},\ }\href@noop {} {\bibfield  {journal} {\bibinfo  {journal} {Chemical Reviews}\ }\textbf {\bibinfo {volume} {121}},\ \bibinfo {pages} {9759} (\bibinfo {year} {2021})}\BibitemShut {NoStop}%
\bibitem [{\citenamefont {Deringer}\ \emph {et~al.}(2021{\natexlab{a}})\citenamefont {Deringer}, \citenamefont {Bart{\'o}k}, \citenamefont {Bernstein}, \citenamefont {Wilkins}, \citenamefont {Ceriotti},\ and\ \citenamefont {Cs{\'a}nyi}}]{deringer2021gaussian}%
  \BibitemOpen
  \bibfield  {author} {\bibinfo {author} {\bibfnamefont {V.~L.}\ \bibnamefont {Deringer}}, \bibinfo {author} {\bibfnamefont {A.~P.}\ \bibnamefont {Bart{\'o}k}}, \bibinfo {author} {\bibfnamefont {N.}~\bibnamefont {Bernstein}}, \bibinfo {author} {\bibfnamefont {D.~M.}\ \bibnamefont {Wilkins}}, \bibinfo {author} {\bibfnamefont {M.}~\bibnamefont {Ceriotti}},\ and\ \bibinfo {author} {\bibfnamefont {G.}~\bibnamefont {Cs{\'a}nyi}},\ }\bibfield  {title} {\bibinfo {title} {Gaussian process regression for materials and molecules},\ }\href@noop {} {\bibfield  {journal} {\bibinfo  {journal} {Chemical Reviews}\ }\textbf {\bibinfo {volume} {121}},\ \bibinfo {pages} {10073} (\bibinfo {year} {2021}{\natexlab{a}})}\BibitemShut {NoStop}%
\bibitem [{\citenamefont {Huang}\ and\ \citenamefont {Von~Lilienfeld}(2021)}]{huang2021ab}%
  \BibitemOpen
  \bibfield  {author} {\bibinfo {author} {\bibfnamefont {B.}~\bibnamefont {Huang}}\ and\ \bibinfo {author} {\bibfnamefont {O.~A.}\ \bibnamefont {Von~Lilienfeld}},\ }\bibfield  {title} {\bibinfo {title} {Ab initio machine learning in chemical compound space},\ }\href@noop {} {\bibfield  {journal} {\bibinfo  {journal} {Chemical reviews}\ }\textbf {\bibinfo {volume} {121}},\ \bibinfo {pages} {10001} (\bibinfo {year} {2021})}\BibitemShut {NoStop}%
\bibitem [{\citenamefont {Daw}\ and\ \citenamefont {Baskes}(1984)}]{daw1984embedded}%
  \BibitemOpen
  \bibfield  {author} {\bibinfo {author} {\bibfnamefont {M.~S.}\ \bibnamefont {Daw}}\ and\ \bibinfo {author} {\bibfnamefont {M.~I.}\ \bibnamefont {Baskes}},\ }\bibfield  {title} {\bibinfo {title} {Embedded-atom method: Derivation and application to impurities, surfaces, and other defects in metals},\ }\href@noop {} {\bibfield  {journal} {\bibinfo  {journal} {Physical Review B}\ }\textbf {\bibinfo {volume} {29}},\ \bibinfo {pages} {6443} (\bibinfo {year} {1984})}\BibitemShut {NoStop}%
\bibitem [{\citenamefont {Deringer}\ \emph {et~al.}(2021{\natexlab{b}})\citenamefont {Deringer}, \citenamefont {Bernstein}, \citenamefont {Cs{\'a}nyi}, \citenamefont {Ben~Mahmoud}, \citenamefont {Ceriotti}, \citenamefont {Wilson}, \citenamefont {Drabold},\ and\ \citenamefont {Elliott}}]{deringer2021origins}%
  \BibitemOpen
  \bibfield  {author} {\bibinfo {author} {\bibfnamefont {V.~L.}\ \bibnamefont {Deringer}}, \bibinfo {author} {\bibfnamefont {N.}~\bibnamefont {Bernstein}}, \bibinfo {author} {\bibfnamefont {G.}~\bibnamefont {Cs{\'a}nyi}}, \bibinfo {author} {\bibfnamefont {C.}~\bibnamefont {Ben~Mahmoud}}, \bibinfo {author} {\bibfnamefont {M.}~\bibnamefont {Ceriotti}}, \bibinfo {author} {\bibfnamefont {M.}~\bibnamefont {Wilson}}, \bibinfo {author} {\bibfnamefont {D.~A.}\ \bibnamefont {Drabold}},\ and\ \bibinfo {author} {\bibfnamefont {S.~R.}\ \bibnamefont {Elliott}},\ }\bibfield  {title} {\bibinfo {title} {Origins of structural and electronic transitions in disordered silicon},\ }\href@noop {} {\bibfield  {journal} {\bibinfo  {journal} {Nature}\ }\textbf {\bibinfo {volume} {589}},\ \bibinfo {pages} {59} (\bibinfo {year} {2021}{\natexlab{b}})}\BibitemShut {NoStop}%
\bibitem [{\citenamefont {Baldwin}\ \emph {et~al.}(2023)\citenamefont {Baldwin}, \citenamefont {Liang}, \citenamefont {Klarbring}, \citenamefont {Dubajic}, \citenamefont {Dell'Angelo}, \citenamefont {Sutton}, \citenamefont {Caddeo}, \citenamefont {Stranks}, \citenamefont {Mattoni}, \citenamefont {Walsh} \emph {et~al.}}]{baldwin2023dynamic}%
  \BibitemOpen
  \bibfield  {author} {\bibinfo {author} {\bibfnamefont {W.~J.}\ \bibnamefont {Baldwin}}, \bibinfo {author} {\bibfnamefont {X.}~\bibnamefont {Liang}}, \bibinfo {author} {\bibfnamefont {J.}~\bibnamefont {Klarbring}}, \bibinfo {author} {\bibfnamefont {M.}~\bibnamefont {Dubajic}}, \bibinfo {author} {\bibfnamefont {D.}~\bibnamefont {Dell'Angelo}}, \bibinfo {author} {\bibfnamefont {C.}~\bibnamefont {Sutton}}, \bibinfo {author} {\bibfnamefont {C.}~\bibnamefont {Caddeo}}, \bibinfo {author} {\bibfnamefont {S.~D.}\ \bibnamefont {Stranks}}, \bibinfo {author} {\bibfnamefont {A.}~\bibnamefont {Mattoni}}, \bibinfo {author} {\bibfnamefont {A.}~\bibnamefont {Walsh}}, \emph {et~al.},\ }\bibfield  {title} {\bibinfo {title} {Dynamic local structure in caesium lead iodide: Spatial correlation and transient domains},\ }\href@noop {} {\bibfield  {journal} {\bibinfo  {journal} {Small}\ ,\ \bibinfo {pages} {2303565}} (\bibinfo {year} {2023})}\BibitemShut {NoStop}%
\bibitem [{\citenamefont {Rosenbrock}\ \emph {et~al.}(2021)\citenamefont {Rosenbrock}, \citenamefont {Gubaev}, \citenamefont {Shapeev}, \citenamefont {P{\'a}rtay}, \citenamefont {Bernstein}, \citenamefont {Cs{\'a}nyi},\ and\ \citenamefont {Hart}}]{rosenbrock2021machine}%
  \BibitemOpen
  \bibfield  {author} {\bibinfo {author} {\bibfnamefont {C.~W.}\ \bibnamefont {Rosenbrock}}, \bibinfo {author} {\bibfnamefont {K.}~\bibnamefont {Gubaev}}, \bibinfo {author} {\bibfnamefont {A.~V.}\ \bibnamefont {Shapeev}}, \bibinfo {author} {\bibfnamefont {L.~B.}\ \bibnamefont {P{\'a}rtay}}, \bibinfo {author} {\bibfnamefont {N.}~\bibnamefont {Bernstein}}, \bibinfo {author} {\bibfnamefont {G.}~\bibnamefont {Cs{\'a}nyi}},\ and\ \bibinfo {author} {\bibfnamefont {G.~L.}\ \bibnamefont {Hart}},\ }\bibfield  {title} {\bibinfo {title} {Machine-learned interatomic potentials for alloys and alloy phase diagrams},\ }\href@noop {} {\bibfield  {journal} {\bibinfo  {journal} {npj Computational Materials}\ }\textbf {\bibinfo {volume} {7}},\ \bibinfo {pages} {24} (\bibinfo {year} {2021})}\BibitemShut {NoStop}%
\bibitem [{\citenamefont {Zhou}\ \emph {et~al.}(2023)\citenamefont {Zhou}, \citenamefont {Zhang}, \citenamefont {Ma},\ and\ \citenamefont {Deringer}}]{zhou2023device}%
  \BibitemOpen
  \bibfield  {author} {\bibinfo {author} {\bibfnamefont {Y.}~\bibnamefont {Zhou}}, \bibinfo {author} {\bibfnamefont {W.}~\bibnamefont {Zhang}}, \bibinfo {author} {\bibfnamefont {E.}~\bibnamefont {Ma}},\ and\ \bibinfo {author} {\bibfnamefont {V.~L.}\ \bibnamefont {Deringer}},\ }\bibfield  {title} {\bibinfo {title} {Device-scale atomistic modelling of phase-change memory materials},\ }\href@noop {} {\bibfield  {journal} {\bibinfo  {journal} {Nature Electronics}\ ,\ \bibinfo {pages} {1}} (\bibinfo {year} {2023})}\BibitemShut {NoStop}%
\bibitem [{\citenamefont {Dauber-Osguthorpe}\ and\ \citenamefont {Hagler}(2019)}]{hagler-1}%
  \BibitemOpen
  \bibfield  {author} {\bibinfo {author} {\bibfnamefont {P.}~\bibnamefont {Dauber-Osguthorpe}}\ and\ \bibinfo {author} {\bibfnamefont {A.~T.}\ \bibnamefont {Hagler}},\ }\bibfield  {title} {\bibinfo {title} {Biomolecular force fields: where have we been, where are we now, where do we need to go and how do we get there?},\ }\href {https://doi.org/10.1007/s10822-018-0111-4} {\bibfield  {journal} {\bibinfo  {journal} {J Comput Aided Mol Des}\ }\textbf {\bibinfo {volume} {33}},\ \bibinfo {pages} {133–203} (\bibinfo {year} {2019})}\BibitemShut {NoStop}%
\bibitem [{\citenamefont {Hagler}(2019)}]{hagler-2}%
  \BibitemOpen
  \bibfield  {author} {\bibinfo {author} {\bibfnamefont {A.~T.}\ \bibnamefont {Hagler}},\ }\bibfield  {title} {\bibinfo {title} {Force field development phase ii: Relaxation of physics-based criteria... or inclusion of more rigorous physics into the representation of molecular energetics},\ }\href {https://doi.org/10.1007/s10822-018-0134-x} {\bibfield  {journal} {\bibinfo  {journal} {J Comput Aided Mol Des}\ }\textbf {\bibinfo {volume} {33}},\ \bibinfo {pages} {205–264} (\bibinfo {year} {2019})}\BibitemShut {NoStop}%
\bibitem [{\citenamefont {Boothroyd}\ \emph {et~al.}(2023)\citenamefont {Boothroyd}, \citenamefont {Behara}, \citenamefont {Madin}, \citenamefont {Hahn}, \citenamefont {Jang}, \citenamefont {Gapsys}, \citenamefont {Wagner}, \citenamefont {Horton}, \citenamefont {Dotson}, \citenamefont {Thompson}, \citenamefont {Maat}, \citenamefont {Gokey}, \citenamefont {Wang}, \citenamefont {Cole}, \citenamefont {Gilson}, \citenamefont {Chodera}, \citenamefont {Bayly}, \citenamefont {Shirts},\ and\ \citenamefont {Mobley}}]{openff-sage}%
  \BibitemOpen
  \bibfield  {author} {\bibinfo {author} {\bibfnamefont {S.}~\bibnamefont {Boothroyd}}, \bibinfo {author} {\bibfnamefont {P.~K.}\ \bibnamefont {Behara}}, \bibinfo {author} {\bibfnamefont {O.~C.}\ \bibnamefont {Madin}}, \bibinfo {author} {\bibfnamefont {D.~F.}\ \bibnamefont {Hahn}}, \bibinfo {author} {\bibfnamefont {H.}~\bibnamefont {Jang}}, \bibinfo {author} {\bibfnamefont {V.}~\bibnamefont {Gapsys}}, \bibinfo {author} {\bibfnamefont {J.~R.}\ \bibnamefont {Wagner}}, \bibinfo {author} {\bibfnamefont {J.~T.}\ \bibnamefont {Horton}}, \bibinfo {author} {\bibfnamefont {D.~L.}\ \bibnamefont {Dotson}}, \bibinfo {author} {\bibfnamefont {M.~W.}\ \bibnamefont {Thompson}}, \bibinfo {author} {\bibfnamefont {J.}~\bibnamefont {Maat}}, \bibinfo {author} {\bibfnamefont {T.}~\bibnamefont {Gokey}}, \bibinfo {author} {\bibfnamefont {L.-P.}\ \bibnamefont {Wang}}, \bibinfo {author} {\bibfnamefont {D.~J.}\ \bibnamefont {Cole}}, \bibinfo {author} {\bibfnamefont {M.~K.}\ \bibnamefont {Gilson}}, \bibinfo {author} {\bibfnamefont
  {J.~D.}\ \bibnamefont {Chodera}}, \bibinfo {author} {\bibfnamefont {C.~I.}\ \bibnamefont {Bayly}}, \bibinfo {author} {\bibfnamefont {M.~R.}\ \bibnamefont {Shirts}},\ and\ \bibinfo {author} {\bibfnamefont {D.~L.}\ \bibnamefont {Mobley}},\ }\bibfield  {title} {\bibinfo {title} {Development and benchmarking of open force field 2.0.0: The sage small molecule force field},\ }\href {https://doi.org/10.1021/acs.jctc.3c00039} {\bibfield  {journal} {\bibinfo  {journal} {J. Chem. Theory Comput.}\ }\textbf {\bibinfo {volume} {0}},\ \bibinfo {pages} {null} (\bibinfo {year} {2023})},\ \bibinfo {note} {pMID: 37167319}\BibitemShut {NoStop}%
\bibitem [{\citenamefont {Gapsys}\ \emph {et~al.}(2020)\citenamefont {Gapsys}, \citenamefont {Pérez-Benito}, \citenamefont {Aldeghi}, \citenamefont {Seeliger}, \citenamefont {Van~Vlijmen}, \citenamefont {Tresadern},\ and\ \citenamefont {De~Groot}}]{Gapsys2020LargeAlchemy}%
  \BibitemOpen
  \bibfield  {author} {\bibinfo {author} {\bibfnamefont {V.}~\bibnamefont {Gapsys}}, \bibinfo {author} {\bibfnamefont {L.}~\bibnamefont {Pérez-Benito}}, \bibinfo {author} {\bibfnamefont {M.}~\bibnamefont {Aldeghi}}, \bibinfo {author} {\bibfnamefont {D.}~\bibnamefont {Seeliger}}, \bibinfo {author} {\bibfnamefont {H.}~\bibnamefont {Van~Vlijmen}}, \bibinfo {author} {\bibfnamefont {G.}~\bibnamefont {Tresadern}},\ and\ \bibinfo {author} {\bibfnamefont {B.~L.}\ \bibnamefont {De~Groot}},\ }\bibfield  {title} {\bibinfo {title} {{Large scale relative protein ligand binding affinities using non-equilibrium alchemy}},\ }\href {https://doi.org/10.1039/c9sc03754c} {\bibfield  {journal} {\bibinfo  {journal} {Chem. Sci.}\ }\textbf {\bibinfo {volume} {11}},\ \bibinfo {pages} {1140} (\bibinfo {year} {2020})}\BibitemShut {NoStop}%
\bibitem [{\citenamefont {Bannwarth}\ \emph {et~al.}(2019)\citenamefont {Bannwarth}, \citenamefont {Ehlert},\ and\ \citenamefont {Grimme}}]{bannwarth2019gfn2}%
  \BibitemOpen
  \bibfield  {author} {\bibinfo {author} {\bibfnamefont {C.}~\bibnamefont {Bannwarth}}, \bibinfo {author} {\bibfnamefont {S.}~\bibnamefont {Ehlert}},\ and\ \bibinfo {author} {\bibfnamefont {S.}~\bibnamefont {Grimme}},\ }\bibfield  {title} {\bibinfo {title} {Gfn2-xtb—an accurate and broadly parametrized self-consistent tight-binding quantum chemical method with multipole electrostatics and density-dependent dispersion contributions},\ }\href@noop {} {\bibfield  {journal} {\bibinfo  {journal} {Journal of chemical theory and computation}\ }\textbf {\bibinfo {volume} {15}},\ \bibinfo {pages} {1652} (\bibinfo {year} {2019})}\BibitemShut {NoStop}%
\bibitem [{\citenamefont {Smith}\ \emph {et~al.}(2017{\natexlab{a}})\citenamefont {Smith}, \citenamefont {Isayev},\ and\ \citenamefont {Roitberg}}]{smith2017ani}%
  \BibitemOpen
  \bibfield  {author} {\bibinfo {author} {\bibfnamefont {J.~S.}\ \bibnamefont {Smith}}, \bibinfo {author} {\bibfnamefont {O.}~\bibnamefont {Isayev}},\ and\ \bibinfo {author} {\bibfnamefont {A.~E.}\ \bibnamefont {Roitberg}},\ }\bibfield  {title} {\bibinfo {title} {Ani-1: an extensible neural network potential with dft accuracy at force field computational cost},\ }\href@noop {} {\bibfield  {journal} {\bibinfo  {journal} {Chemical science}\ }\textbf {\bibinfo {volume} {8}},\ \bibinfo {pages} {3192} (\bibinfo {year} {2017}{\natexlab{a}})}\BibitemShut {NoStop}%
\bibitem [{\citenamefont {Smith}\ \emph {et~al.}(2018)\citenamefont {Smith}, \citenamefont {Nebgen}, \citenamefont {Lubbers}, \citenamefont {Isayev},\ and\ \citenamefont {Roitberg}}]{smith2018ani_1x}%
  \BibitemOpen
  \bibfield  {author} {\bibinfo {author} {\bibfnamefont {J.~S.}\ \bibnamefont {Smith}}, \bibinfo {author} {\bibfnamefont {B.}~\bibnamefont {Nebgen}}, \bibinfo {author} {\bibfnamefont {N.}~\bibnamefont {Lubbers}}, \bibinfo {author} {\bibfnamefont {O.}~\bibnamefont {Isayev}},\ and\ \bibinfo {author} {\bibfnamefont {A.~E.}\ \bibnamefont {Roitberg}},\ }\bibfield  {title} {\bibinfo {title} {Less is more: Sampling chemical space with active learning},\ }\href@noop {} {\bibfield  {journal} {\bibinfo  {journal} {The Journal of chemical physics}\ }\textbf {\bibinfo {volume} {148}},\ \bibinfo {pages} {241733} (\bibinfo {year} {2018})}\BibitemShut {NoStop}%
\bibitem [{\citenamefont {Smith}\ \emph {et~al.}(2019)\citenamefont {Smith}, \citenamefont {Nebgen}, \citenamefont {Zubatyuk}, \citenamefont {Lubbers}, \citenamefont {Devereux}, \citenamefont {Barros}, \citenamefont {Tretiak}, \citenamefont {Isayev},\ and\ \citenamefont {Roitberg}}]{smith2019approaching_ani1ccx}%
  \BibitemOpen
  \bibfield  {author} {\bibinfo {author} {\bibfnamefont {J.~S.}\ \bibnamefont {Smith}}, \bibinfo {author} {\bibfnamefont {B.~T.}\ \bibnamefont {Nebgen}}, \bibinfo {author} {\bibfnamefont {R.}~\bibnamefont {Zubatyuk}}, \bibinfo {author} {\bibfnamefont {N.}~\bibnamefont {Lubbers}}, \bibinfo {author} {\bibfnamefont {C.}~\bibnamefont {Devereux}}, \bibinfo {author} {\bibfnamefont {K.}~\bibnamefont {Barros}}, \bibinfo {author} {\bibfnamefont {S.}~\bibnamefont {Tretiak}}, \bibinfo {author} {\bibfnamefont {O.}~\bibnamefont {Isayev}},\ and\ \bibinfo {author} {\bibfnamefont {A.~E.}\ \bibnamefont {Roitberg}},\ }\bibfield  {title} {\bibinfo {title} {Approaching coupled cluster accuracy with a general-purpose neural network potential through transfer learning},\ }\href@noop {} {\bibfield  {journal} {\bibinfo  {journal} {Nature communications}\ }\textbf {\bibinfo {volume} {10}},\ \bibinfo {pages} {1} (\bibinfo {year} {2019})}\BibitemShut {NoStop}%
\bibitem [{\citenamefont {Devereux}\ \emph {et~al.}(2020)\citenamefont {Devereux}, \citenamefont {Smith}, \citenamefont {Huddleston}, \citenamefont {Barros}, \citenamefont {Zubatyuk}, \citenamefont {Isayev},\ and\ \citenamefont {Roitberg}}]{devereux2020ani_2x}%
  \BibitemOpen
  \bibfield  {author} {\bibinfo {author} {\bibfnamefont {C.}~\bibnamefont {Devereux}}, \bibinfo {author} {\bibfnamefont {J.~S.}\ \bibnamefont {Smith}}, \bibinfo {author} {\bibfnamefont {K.~K.}\ \bibnamefont {Huddleston}}, \bibinfo {author} {\bibfnamefont {K.}~\bibnamefont {Barros}}, \bibinfo {author} {\bibfnamefont {R.}~\bibnamefont {Zubatyuk}}, \bibinfo {author} {\bibfnamefont {O.}~\bibnamefont {Isayev}},\ and\ \bibinfo {author} {\bibfnamefont {A.~E.}\ \bibnamefont {Roitberg}},\ }\bibfield  {title} {\bibinfo {title} {Extending the applicability of the ani deep learning molecular potential to sulfur and halogens},\ }\href@noop {} {\bibfield  {journal} {\bibinfo  {journal} {Journal of Chemical Theory and Computation}\ }\textbf {\bibinfo {volume} {16}},\ \bibinfo {pages} {4192} (\bibinfo {year} {2020})}\BibitemShut {NoStop}%
\bibitem [{\citenamefont {Zubatyuk}\ \emph {et~al.}(2019)\citenamefont {Zubatyuk}, \citenamefont {Smith}, \citenamefont {Leszczynski},\ and\ \citenamefont {Isayev}}]{Zubatyuk2019aimnet}%
  \BibitemOpen
  \bibfield  {author} {\bibinfo {author} {\bibfnamefont {R.}~\bibnamefont {Zubatyuk}}, \bibinfo {author} {\bibfnamefont {J.~S.}\ \bibnamefont {Smith}}, \bibinfo {author} {\bibfnamefont {J.}~\bibnamefont {Leszczynski}},\ and\ \bibinfo {author} {\bibfnamefont {O.}~\bibnamefont {Isayev}},\ }\bibfield  {title} {\bibinfo {title} {Accurate and transferable multitask prediction of chemical properties with an atoms-in-molecules neural network},\ }\href@noop {} {\bibfield  {journal} {\bibinfo  {journal} {Science advances}\ }\textbf {\bibinfo {volume} {5}},\ \bibinfo {pages} {eaav6490} (\bibinfo {year} {2019})}\BibitemShut {NoStop}%
\bibitem [{\citenamefont {Zubatyuk}\ \emph {et~al.}(2021)\citenamefont {Zubatyuk}, \citenamefont {Smith}, \citenamefont {Nebgen}, \citenamefont {Tretiak},\ and\ \citenamefont {Isayev}}]{zubatyuk2021aimnet_nse}%
  \BibitemOpen
  \bibfield  {author} {\bibinfo {author} {\bibfnamefont {R.}~\bibnamefont {Zubatyuk}}, \bibinfo {author} {\bibfnamefont {J.~S.}\ \bibnamefont {Smith}}, \bibinfo {author} {\bibfnamefont {B.~T.}\ \bibnamefont {Nebgen}}, \bibinfo {author} {\bibfnamefont {S.}~\bibnamefont {Tretiak}},\ and\ \bibinfo {author} {\bibfnamefont {O.}~\bibnamefont {Isayev}},\ }\bibfield  {title} {\bibinfo {title} {Teaching a neural network to attach and detach electrons from molecules},\ }\href@noop {} {\bibfield  {journal} {\bibinfo  {journal} {Nature Communications}\ }\textbf {\bibinfo {volume} {12}},\ \bibinfo {pages} {4870} (\bibinfo {year} {2021})}\BibitemShut {NoStop}%
\bibitem [{\citenamefont {Behler}\ and\ \citenamefont {Parrinello}(2007)}]{behler2007generalized}%
  \BibitemOpen
  \bibfield  {author} {\bibinfo {author} {\bibfnamefont {J.}~\bibnamefont {Behler}}\ and\ \bibinfo {author} {\bibfnamefont {M.}~\bibnamefont {Parrinello}},\ }\bibfield  {title} {\bibinfo {title} {Generalized neural-network representation of high-dimensional potential-energy surfaces},\ }\href@noop {} {\bibfield  {journal} {\bibinfo  {journal} {Physical review letters}\ }\textbf {\bibinfo {volume} {98}},\ \bibinfo {pages} {146401} (\bibinfo {year} {2007})}\BibitemShut {NoStop}%
\bibitem [{\citenamefont {Smith}\ \emph {et~al.}(2017{\natexlab{b}})\citenamefont {Smith}, \citenamefont {Isayev},\ and\ \citenamefont {Roitberg}}]{smith2017ani_data}%
  \BibitemOpen
  \bibfield  {author} {\bibinfo {author} {\bibfnamefont {J.~S.}\ \bibnamefont {Smith}}, \bibinfo {author} {\bibfnamefont {O.}~\bibnamefont {Isayev}},\ and\ \bibinfo {author} {\bibfnamefont {A.~E.}\ \bibnamefont {Roitberg}},\ }\bibfield  {title} {\bibinfo {title} {Ani-1, a data set of 20 million calculated off-equilibrium conformations for organic molecules},\ }\href@noop {} {\bibfield  {journal} {\bibinfo  {journal} {Scientific data}\ }\textbf {\bibinfo {volume} {4}},\ \bibinfo {pages} {1} (\bibinfo {year} {2017}{\natexlab{b}})}\BibitemShut {NoStop}%
\bibitem [{\citenamefont {Smith}\ \emph {et~al.}(2020{\natexlab{a}})\citenamefont {Smith}, \citenamefont {Zubatyuk}, \citenamefont {Nebgen}, \citenamefont {Lubbers}, \citenamefont {Barros}, \citenamefont {Roitberg}, \citenamefont {Isayev},\ and\ \citenamefont {Tretiak}}]{smith2020ani_cc_data}%
  \BibitemOpen
  \bibfield  {author} {\bibinfo {author} {\bibfnamefont {J.~S.}\ \bibnamefont {Smith}}, \bibinfo {author} {\bibfnamefont {R.}~\bibnamefont {Zubatyuk}}, \bibinfo {author} {\bibfnamefont {B.}~\bibnamefont {Nebgen}}, \bibinfo {author} {\bibfnamefont {N.}~\bibnamefont {Lubbers}}, \bibinfo {author} {\bibfnamefont {K.}~\bibnamefont {Barros}}, \bibinfo {author} {\bibfnamefont {A.~E.}\ \bibnamefont {Roitberg}}, \bibinfo {author} {\bibfnamefont {O.}~\bibnamefont {Isayev}},\ and\ \bibinfo {author} {\bibfnamefont {S.}~\bibnamefont {Tretiak}},\ }\bibfield  {title} {\bibinfo {title} {The ani-1ccx and ani-1x data sets, coupled-cluster and density functional theory properties for molecules},\ }\href@noop {} {\bibfield  {journal} {\bibinfo  {journal} {Scientific data}\ }\textbf {\bibinfo {volume} {7}},\ \bibinfo {pages} {134} (\bibinfo {year} {2020}{\natexlab{a}})}\BibitemShut {NoStop}%
\bibitem [{\citenamefont {Inizan}\ \emph {et~al.}(2023)\citenamefont {Inizan}, \citenamefont {Pl{\'e}}, \citenamefont {Adjoua}, \citenamefont {Ren}, \citenamefont {G{\"o}kcan}, \citenamefont {Isayev}, \citenamefont {Lagard{\`e}re},\ and\ \citenamefont {Piquemal}}]{inizan2023scalable}%
  \BibitemOpen
  \bibfield  {author} {\bibinfo {author} {\bibfnamefont {T.~J.}\ \bibnamefont {Inizan}}, \bibinfo {author} {\bibfnamefont {T.}~\bibnamefont {Pl{\'e}}}, \bibinfo {author} {\bibfnamefont {O.}~\bibnamefont {Adjoua}}, \bibinfo {author} {\bibfnamefont {P.}~\bibnamefont {Ren}}, \bibinfo {author} {\bibfnamefont {H.}~\bibnamefont {G{\"o}kcan}}, \bibinfo {author} {\bibfnamefont {O.}~\bibnamefont {Isayev}}, \bibinfo {author} {\bibfnamefont {L.}~\bibnamefont {Lagard{\`e}re}},\ and\ \bibinfo {author} {\bibfnamefont {J.-P.}\ \bibnamefont {Piquemal}},\ }\bibfield  {title} {\bibinfo {title} {Scalable hybrid deep neural networks/polarizable potentials biomolecular simulations including long-range effects},\ }\href@noop {} {\bibfield  {journal} {\bibinfo  {journal} {Chemical Science}\ }\textbf {\bibinfo {volume} {14}},\ \bibinfo {pages} {5438} (\bibinfo {year} {2023})}\BibitemShut {NoStop}%
\bibitem [{\citenamefont {Tu}\ \emph {et~al.}(2023)\citenamefont {Tu}, \citenamefont {Rezajooei}, \citenamefont {Johnson},\ and\ \citenamefont {Rowley}}]{tu2023neural}%
  \BibitemOpen
  \bibfield  {author} {\bibinfo {author} {\bibfnamefont {N.~T.~P.}\ \bibnamefont {Tu}}, \bibinfo {author} {\bibfnamefont {N.}~\bibnamefont {Rezajooei}}, \bibinfo {author} {\bibfnamefont {E.~R.}\ \bibnamefont {Johnson}},\ and\ \bibinfo {author} {\bibfnamefont {C.}~\bibnamefont {Rowley}},\ }\bibfield  {title} {\bibinfo {title} {A neural network potential with rigorous treatment of long-range dispersion},\ }\href@noop {} {\bibfield  {journal} {\bibinfo  {journal} {Digital Discovery}\ } (\bibinfo {year} {2023})}\BibitemShut {NoStop}%
\bibitem [{\citenamefont {Gilmer}\ \emph {et~al.}(2017)\citenamefont {Gilmer}, \citenamefont {Schoenholz}, \citenamefont {Riley}, \citenamefont {Vinyals},\ and\ \citenamefont {Dahl}}]{gilmer2017neural}%
  \BibitemOpen
  \bibfield  {author} {\bibinfo {author} {\bibfnamefont {J.}~\bibnamefont {Gilmer}}, \bibinfo {author} {\bibfnamefont {S.~S.}\ \bibnamefont {Schoenholz}}, \bibinfo {author} {\bibfnamefont {P.~F.}\ \bibnamefont {Riley}}, \bibinfo {author} {\bibfnamefont {O.}~\bibnamefont {Vinyals}},\ and\ \bibinfo {author} {\bibfnamefont {G.~E.}\ \bibnamefont {Dahl}},\ }\bibfield  {title} {\bibinfo {title} {Neural message passing for quantum chemistry},\ }in\ \href@noop {} {\emph {\bibinfo {booktitle} {International conference on machine learning}}}\ (\bibinfo {organization} {PMLR},\ \bibinfo {year} {2017})\ pp.\ \bibinfo {pages} {1263--1272}\BibitemShut {NoStop}%
\bibitem [{\citenamefont {Unke}\ and\ \citenamefont {Meuwly}(2019)}]{unke2019physnet}%
  \BibitemOpen
  \bibfield  {author} {\bibinfo {author} {\bibfnamefont {O.~T.}\ \bibnamefont {Unke}}\ and\ \bibinfo {author} {\bibfnamefont {M.}~\bibnamefont {Meuwly}},\ }\bibfield  {title} {\bibinfo {title} {Physnet: A neural network for predicting energies, forces, dipole moments, and partial charges},\ }\href@noop {} {\bibfield  {journal} {\bibinfo  {journal} {Journal of chemical theory and computation}\ }\textbf {\bibinfo {volume} {15}},\ \bibinfo {pages} {3678} (\bibinfo {year} {2019})}\BibitemShut {NoStop}%
\bibitem [{\citenamefont {Pl{\'{e} }}\ \emph {et~al.}(2023)\citenamefont {Pl{\'{e} }}, \citenamefont {Lagard{\`{e}}re},\ and\ \citenamefont {Piquemal}}]{Ple2023fennix}%
  \BibitemOpen
  \bibfield  {author} {\bibinfo {author} {\bibfnamefont {T.}~\bibnamefont {Pl{\'{e} }}}, \bibinfo {author} {\bibfnamefont {L.}~\bibnamefont {Lagard{\`{e}}re}},\ and\ \bibinfo {author} {\bibfnamefont {J.-P.}\ \bibnamefont {Piquemal}},\ }\bibfield  {title} {\bibinfo {title} {Force-field-enhanced neural network interactions: from local equivariant embedding to atom-in-molecule properties and long-range effects},\ }\bibfield  {journal} {\bibinfo  {journal} {Chemical Science}\ }\href {https://doi.org/10.1039/d3sc02581k} {10.1039/d3sc02581k} (\bibinfo {year} {2023})\BibitemShut {NoStop}%
\bibitem [{\citenamefont {Th{\"u}rlemann}\ and\ \citenamefont {Riniker}(2023)}]{ana2b}%
  \BibitemOpen
  \bibfield  {author} {\bibinfo {author} {\bibfnamefont {M.}~\bibnamefont {Th{\"u}rlemann}}\ and\ \bibinfo {author} {\bibfnamefont {S.}~\bibnamefont {Riniker}},\ }\bibfield  {title} {\bibinfo {title} {Hybrid classical/machine-learning force fields for the accurate description of molecular condensed-phase systems},\ }\href@noop {} {\bibfield  {journal} {\bibinfo  {journal} {Chemical Science}\ }\textbf {\bibinfo {volume} {14}},\ \bibinfo {pages} {12661} (\bibinfo {year} {2023})}\BibitemShut {NoStop}%
\bibitem [{\citenamefont {Unke}\ \emph {et~al.}(2024)\citenamefont {Unke}, \citenamefont {Stöhr}, \citenamefont {Ganscha}, \citenamefont {Unterthiner}, \citenamefont {Maennel}, \citenamefont {Kashubin}, \citenamefont {Ahlin}, \citenamefont {Gastegger}, \citenamefont {Sandonas}, \citenamefont {Berryman}, \citenamefont {Tkatchenko},\ and\ \citenamefont {Müller}}]{unke-gems-sciadv}%
  \BibitemOpen
  \bibfield  {author} {\bibinfo {author} {\bibfnamefont {O.~T.}\ \bibnamefont {Unke}}, \bibinfo {author} {\bibfnamefont {M.}~\bibnamefont {Stöhr}}, \bibinfo {author} {\bibfnamefont {S.}~\bibnamefont {Ganscha}}, \bibinfo {author} {\bibfnamefont {T.}~\bibnamefont {Unterthiner}}, \bibinfo {author} {\bibfnamefont {H.}~\bibnamefont {Maennel}}, \bibinfo {author} {\bibfnamefont {S.}~\bibnamefont {Kashubin}}, \bibinfo {author} {\bibfnamefont {D.}~\bibnamefont {Ahlin}}, \bibinfo {author} {\bibfnamefont {M.}~\bibnamefont {Gastegger}}, \bibinfo {author} {\bibfnamefont {L.~M.}\ \bibnamefont {Sandonas}}, \bibinfo {author} {\bibfnamefont {J.~T.}\ \bibnamefont {Berryman}}, \bibinfo {author} {\bibfnamefont {A.}~\bibnamefont {Tkatchenko}},\ and\ \bibinfo {author} {\bibfnamefont {K.-R.}\ \bibnamefont {Müller}},\ }\bibfield  {title} {\bibinfo {title} {Biomolecular dynamics with machine-learned quantum-mechanical force fields trained on diverse chemical fragments},\ }\href {https://doi.org/10.1126/sciadv.adn4397} {\bibfield
  {journal} {\bibinfo  {journal} {Science Advances}\ }\textbf {\bibinfo {volume} {10}},\ \bibinfo {pages} {eadn4397} (\bibinfo {year} {2024})}\BibitemShut {NoStop}%
\bibitem [{\citenamefont {Unke}\ \emph {et~al.}(2021)\citenamefont {Unke}, \citenamefont {Chmiela}, \citenamefont {Gastegger}, \citenamefont {Sch{\"u}tt}, \citenamefont {Sauceda},\ and\ \citenamefont {M{\"u}ller}}]{unke2021spookynet}%
  \BibitemOpen
  \bibfield  {author} {\bibinfo {author} {\bibfnamefont {O.~T.}\ \bibnamefont {Unke}}, \bibinfo {author} {\bibfnamefont {S.}~\bibnamefont {Chmiela}}, \bibinfo {author} {\bibfnamefont {M.}~\bibnamefont {Gastegger}}, \bibinfo {author} {\bibfnamefont {K.~T.}\ \bibnamefont {Sch{\"u}tt}}, \bibinfo {author} {\bibfnamefont {H.~E.}\ \bibnamefont {Sauceda}},\ and\ \bibinfo {author} {\bibfnamefont {K.-R.}\ \bibnamefont {M{\"u}ller}},\ }\bibfield  {title} {\bibinfo {title} {Spookynet: Learning force fields with electronic degrees of freedom and nonlocal effects},\ }\href@noop {} {\bibfield  {journal} {\bibinfo  {journal} {Nature communications}\ }\textbf {\bibinfo {volume} {12}},\ \bibinfo {pages} {7273} (\bibinfo {year} {2021})}\BibitemShut {NoStop}%
\bibitem [{\citenamefont {Kabylda}\ \emph {et~al.}(2024)\citenamefont {Kabylda}, \citenamefont {Frank}, \citenamefont {Dou}, \citenamefont {Khabibrakhmanov}, \citenamefont {Sandonas}, \citenamefont {Unke}, \citenamefont {Chmiela}, \citenamefont {Muller},\ and\ \citenamefont {Tkatchenko}}]{kabyldaMolecularSimulationsPretrained2024}%
  \BibitemOpen
  \bibfield  {author} {\bibinfo {author} {\bibfnamefont {A.}~\bibnamefont {Kabylda}}, \bibinfo {author} {\bibfnamefont {J.~T.}\ \bibnamefont {Frank}}, \bibinfo {author} {\bibfnamefont {S.~S.}\ \bibnamefont {Dou}}, \bibinfo {author} {\bibfnamefont {A.}~\bibnamefont {Khabibrakhmanov}}, \bibinfo {author} {\bibfnamefont {L.~M.}\ \bibnamefont {Sandonas}}, \bibinfo {author} {\bibfnamefont {O.~T.}\ \bibnamefont {Unke}}, \bibinfo {author} {\bibfnamefont {S.}~\bibnamefont {Chmiela}}, \bibinfo {author} {\bibfnamefont {K.-R.}\ \bibnamefont {Muller}},\ and\ \bibinfo {author} {\bibfnamefont {A.}~\bibnamefont {Tkatchenko}},\ }\href {https://doi.org/10.26434/chemrxiv-2024-bdfr0} {\bibinfo {title} {Molecular {{Simulations}} with a {{Pretrained Neural Network}} and {{Universal Pairwise Force Fields}}}} (\bibinfo {year} {2024})\BibitemShut {NoStop}%
\bibitem [{\citenamefont {Sch{\"u}tt}\ \emph {et~al.}(2017)\citenamefont {Sch{\"u}tt}, \citenamefont {Kindermans}, \citenamefont {Sauceda~Felix}, \citenamefont {Chmiela}, \citenamefont {Tkatchenko},\ and\ \citenamefont {M{\"u}ller}}]{schutt2017schnet}%
  \BibitemOpen
  \bibfield  {author} {\bibinfo {author} {\bibfnamefont {K.}~\bibnamefont {Sch{\"u}tt}}, \bibinfo {author} {\bibfnamefont {P.-J.}\ \bibnamefont {Kindermans}}, \bibinfo {author} {\bibfnamefont {H.~E.}\ \bibnamefont {Sauceda~Felix}}, \bibinfo {author} {\bibfnamefont {S.}~\bibnamefont {Chmiela}}, \bibinfo {author} {\bibfnamefont {A.}~\bibnamefont {Tkatchenko}},\ and\ \bibinfo {author} {\bibfnamefont {K.-R.}\ \bibnamefont {M{\"u}ller}},\ }\bibfield  {title} {\bibinfo {title} {Schnet: A continuous-filter convolutional neural network for modeling quantum interactions},\ }\href@noop {} {\bibfield  {journal} {\bibinfo  {journal} {Advances in neural information processing systems}\ }\textbf {\bibinfo {volume} {30}} (\bibinfo {year} {2017})}\BibitemShut {NoStop}%
\bibitem [{\citenamefont {Sch{\"u}tt}\ \emph {et~al.}(2021)\citenamefont {Sch{\"u}tt}, \citenamefont {Unke},\ and\ \citenamefont {Gastegger}}]{schutt2021PAINN}%
  \BibitemOpen
  \bibfield  {author} {\bibinfo {author} {\bibfnamefont {K.}~\bibnamefont {Sch{\"u}tt}}, \bibinfo {author} {\bibfnamefont {O.}~\bibnamefont {Unke}},\ and\ \bibinfo {author} {\bibfnamefont {M.}~\bibnamefont {Gastegger}},\ }\bibfield  {title} {\bibinfo {title} {Equivariant message passing for the prediction of tensorial properties and molecular spectra},\ }in\ \href@noop {} {\emph {\bibinfo {booktitle} {International Conference on Machine Learning}}}\ (\bibinfo {organization} {PMLR},\ \bibinfo {year} {2021})\ pp.\ \bibinfo {pages} {9377--9388}\BibitemShut {NoStop}%
\bibitem [{\citenamefont {Gasteiger}\ \emph {et~al.}(2021)\citenamefont {Gasteiger}, \citenamefont {Becker},\ and\ \citenamefont {G{\"u}nnemann}}]{gasteiger2021gemnet}%
  \BibitemOpen
  \bibfield  {author} {\bibinfo {author} {\bibfnamefont {J.}~\bibnamefont {Gasteiger}}, \bibinfo {author} {\bibfnamefont {F.}~\bibnamefont {Becker}},\ and\ \bibinfo {author} {\bibfnamefont {S.}~\bibnamefont {G{\"u}nnemann}},\ }\bibfield  {title} {\bibinfo {title} {Gemnet: Universal directional graph neural networks for molecules},\ }\href@noop {} {\bibfield  {journal} {\bibinfo  {journal} {Advances in Neural Information Processing Systems}\ }\textbf {\bibinfo {volume} {34}},\ \bibinfo {pages} {6790} (\bibinfo {year} {2021})}\BibitemShut {NoStop}%
\bibitem [{\citenamefont {Willatt}\ \emph {et~al.}(2018)\citenamefont {Willatt}, \citenamefont {Musil},\ and\ \citenamefont {Ceriotti}}]{willatt2018feature}%
  \BibitemOpen
  \bibfield  {author} {\bibinfo {author} {\bibfnamefont {M.~J.}\ \bibnamefont {Willatt}}, \bibinfo {author} {\bibfnamefont {F.}~\bibnamefont {Musil}},\ and\ \bibinfo {author} {\bibfnamefont {M.}~\bibnamefont {Ceriotti}},\ }\bibfield  {title} {\bibinfo {title} {Feature optimization for atomistic machine learning yields a data-driven construction of the periodic table of the elements},\ }\href@noop {} {\bibfield  {journal} {\bibinfo  {journal} {Physical Chemistry Chemical Physics}\ }\textbf {\bibinfo {volume} {20}},\ \bibinfo {pages} {29661} (\bibinfo {year} {2018})}\BibitemShut {NoStop}%
\bibitem [{\citenamefont {Darby}\ \emph {et~al.}(2023{\natexlab{a}})\citenamefont {Darby}, \citenamefont {Kov{\'a}cs}, \citenamefont {Batatia}, \citenamefont {Caro}, \citenamefont {Hart}, \citenamefont {Ortner},\ and\ \citenamefont {Cs{\'a}nyi}}]{darby2023tensor}%
  \BibitemOpen
  \bibfield  {author} {\bibinfo {author} {\bibfnamefont {J.~P.}\ \bibnamefont {Darby}}, \bibinfo {author} {\bibfnamefont {D.~P.}\ \bibnamefont {Kov{\'a}cs}}, \bibinfo {author} {\bibfnamefont {I.}~\bibnamefont {Batatia}}, \bibinfo {author} {\bibfnamefont {M.~A.}\ \bibnamefont {Caro}}, \bibinfo {author} {\bibfnamefont {G.~L.}\ \bibnamefont {Hart}}, \bibinfo {author} {\bibfnamefont {C.}~\bibnamefont {Ortner}},\ and\ \bibinfo {author} {\bibfnamefont {G.}~\bibnamefont {Cs{\'a}nyi}},\ }\bibfield  {title} {\bibinfo {title} {Tensor-reduced atomic density representations},\ }\href@noop {} {\bibfield  {journal} {\bibinfo  {journal} {Physical Review Letters}\ }\textbf {\bibinfo {volume} {131}},\ \bibinfo {pages} {028001} (\bibinfo {year} {2023}{\natexlab{a}})}\BibitemShut {NoStop}%
\bibitem [{\citenamefont {Wigner}(2012)}]{wigner2012group}%
  \BibitemOpen
  \bibfield  {author} {\bibinfo {author} {\bibfnamefont {E.}~\bibnamefont {Wigner}},\ }\href@noop {} {\emph {\bibinfo {title} {Group theory: and its application to the quantum mechanics of atomic spectra}}},\ Vol.~\bibinfo {volume} {5}\ (\bibinfo  {publisher} {Elsevier},\ \bibinfo {year} {2012})\BibitemShut {NoStop}%
\bibitem [{\citenamefont {Dusson}\ \emph {et~al.}(2022)\citenamefont {Dusson}, \citenamefont {Bachmayr}, \citenamefont {Cs{\'a}nyi}, \citenamefont {Drautz}, \citenamefont {Etter}, \citenamefont {van~der Oord},\ and\ \citenamefont {Ortner}}]{dusson2022atomic}%
  \BibitemOpen
  \bibfield  {author} {\bibinfo {author} {\bibfnamefont {G.}~\bibnamefont {Dusson}}, \bibinfo {author} {\bibfnamefont {M.}~\bibnamefont {Bachmayr}}, \bibinfo {author} {\bibfnamefont {G.}~\bibnamefont {Cs{\'a}nyi}}, \bibinfo {author} {\bibfnamefont {R.}~\bibnamefont {Drautz}}, \bibinfo {author} {\bibfnamefont {S.}~\bibnamefont {Etter}}, \bibinfo {author} {\bibfnamefont {C.}~\bibnamefont {van~der Oord}},\ and\ \bibinfo {author} {\bibfnamefont {C.}~\bibnamefont {Ortner}},\ }\bibfield  {title} {\bibinfo {title} {Atomic cluster expansion: Completeness, efficiency and stability},\ }\href@noop {} {\bibfield  {journal} {\bibinfo  {journal} {Journal of Computational Physics}\ }\textbf {\bibinfo {volume} {454}},\ \bibinfo {pages} {110946} (\bibinfo {year} {2022})}\BibitemShut {NoStop}%
\bibitem [{\citenamefont {Batatia}\ \emph {et~al.}(2025)\citenamefont {Batatia}, \citenamefont {Batzner}, \citenamefont {Kov{\'a}cs}, \citenamefont {Musaelian}, \citenamefont {Simm}, \citenamefont {Drautz}, \citenamefont {Ortner}, \citenamefont {Kozinsky},\ and\ \citenamefont {Cs{\'a}nyi}}]{batatiaDesignSpaceE3equivariant2025}%
  \BibitemOpen
  \bibfield  {author} {\bibinfo {author} {\bibfnamefont {I.}~\bibnamefont {Batatia}}, \bibinfo {author} {\bibfnamefont {S.}~\bibnamefont {Batzner}}, \bibinfo {author} {\bibfnamefont {D.~P.}\ \bibnamefont {Kov{\'a}cs}}, \bibinfo {author} {\bibfnamefont {A.}~\bibnamefont {Musaelian}}, \bibinfo {author} {\bibfnamefont {G.~N.~C.}\ \bibnamefont {Simm}}, \bibinfo {author} {\bibfnamefont {R.}~\bibnamefont {Drautz}}, \bibinfo {author} {\bibfnamefont {C.}~\bibnamefont {Ortner}}, \bibinfo {author} {\bibfnamefont {B.}~\bibnamefont {Kozinsky}},\ and\ \bibinfo {author} {\bibfnamefont {G.}~\bibnamefont {Cs{\'a}nyi}},\ }\bibfield  {title} {\bibinfo {title} {The design space of {{E}}(3)-equivariant atom-centred interatomic potentials},\ }\href {https://doi.org/10.1038/s42256-024-00956-x} {\bibfield  {journal} {\bibinfo  {journal} {Nature Machine Intelligence}\ }\textbf {\bibinfo {volume} {7}},\ \bibinfo {pages} {56} (\bibinfo {year} {2025})}\BibitemShut {NoStop}%
\bibitem [{\citenamefont {Kov{\'a}cs}\ \emph {et~al.}(2021)\citenamefont {Kov{\'a}cs}, \citenamefont {Oord}, \citenamefont {Kucera}, \citenamefont {Allen}, \citenamefont {Cole}, \citenamefont {Ortner},\ and\ \citenamefont {Cs{\'a}nyi}}]{kovacs2021linear}%
  \BibitemOpen
  \bibfield  {author} {\bibinfo {author} {\bibfnamefont {D.~P.}\ \bibnamefont {Kov{\'a}cs}}, \bibinfo {author} {\bibfnamefont {C.~v.~d.}\ \bibnamefont {Oord}}, \bibinfo {author} {\bibfnamefont {J.}~\bibnamefont {Kucera}}, \bibinfo {author} {\bibfnamefont {A.~E.}\ \bibnamefont {Allen}}, \bibinfo {author} {\bibfnamefont {D.~J.}\ \bibnamefont {Cole}}, \bibinfo {author} {\bibfnamefont {C.}~\bibnamefont {Ortner}},\ and\ \bibinfo {author} {\bibfnamefont {G.}~\bibnamefont {Cs{\'a}nyi}},\ }\bibfield  {title} {\bibinfo {title} {Linear atomic cluster expansion force fields for organic molecules: beyond rmse},\ }\href@noop {} {\bibfield  {journal} {\bibinfo  {journal} {Journal of chemical theory and computation}\ }\textbf {\bibinfo {volume} {17}},\ \bibinfo {pages} {7696} (\bibinfo {year} {2021})}\BibitemShut {NoStop}%
\bibitem [{\citenamefont {Eastman}\ \emph {et~al.}(2023)\citenamefont {Eastman}, \citenamefont {Behara}, \citenamefont {Dotson}, \citenamefont {Galvelis}, \citenamefont {Herr}, \citenamefont {Horton}, \citenamefont {Mao}, \citenamefont {Chodera}, \citenamefont {Pritchard}, \citenamefont {Wang} \emph {et~al.}}]{eastman2023spice}%
  \BibitemOpen
  \bibfield  {author} {\bibinfo {author} {\bibfnamefont {P.}~\bibnamefont {Eastman}}, \bibinfo {author} {\bibfnamefont {P.~K.}\ \bibnamefont {Behara}}, \bibinfo {author} {\bibfnamefont {D.~L.}\ \bibnamefont {Dotson}}, \bibinfo {author} {\bibfnamefont {R.}~\bibnamefont {Galvelis}}, \bibinfo {author} {\bibfnamefont {J.~E.}\ \bibnamefont {Herr}}, \bibinfo {author} {\bibfnamefont {J.~T.}\ \bibnamefont {Horton}}, \bibinfo {author} {\bibfnamefont {Y.}~\bibnamefont {Mao}}, \bibinfo {author} {\bibfnamefont {J.~D.}\ \bibnamefont {Chodera}}, \bibinfo {author} {\bibfnamefont {B.~P.}\ \bibnamefont {Pritchard}}, \bibinfo {author} {\bibfnamefont {Y.}~\bibnamefont {Wang}}, \emph {et~al.},\ }\bibfield  {title} {\bibinfo {title} {Spice, a dataset of drug-like molecules and peptides for training machine learning potentials},\ }\href@noop {} {\bibfield  {journal} {\bibinfo  {journal} {Scientific Data}\ }\textbf {\bibinfo {volume} {10}},\ \bibinfo {pages} {11} (\bibinfo {year} {2023})}\BibitemShut {NoStop}%
\bibitem [{\citenamefont {Najibi}\ and\ \citenamefont {Goerigk}(2018)}]{najibi2018nonlocal}%
  \BibitemOpen
  \bibfield  {author} {\bibinfo {author} {\bibfnamefont {A.}~\bibnamefont {Najibi}}\ and\ \bibinfo {author} {\bibfnamefont {L.}~\bibnamefont {Goerigk}},\ }\bibfield  {title} {\bibinfo {title} {The nonlocal kernel in van der waals density functionals as an additive correction: An extensive analysis with special emphasis on the b97m-v and $\omega$b97m-v approaches},\ }\href@noop {} {\bibfield  {journal} {\bibinfo  {journal} {Journal of Chemical Theory and Computation}\ }\textbf {\bibinfo {volume} {14}},\ \bibinfo {pages} {5725} (\bibinfo {year} {2018})}\BibitemShut {NoStop}%
\bibitem [{\citenamefont {Weigend}\ and\ \citenamefont {Ahlrichs}(2005)}]{weigend2005balanced}%
  \BibitemOpen
  \bibfield  {author} {\bibinfo {author} {\bibfnamefont {F.}~\bibnamefont {Weigend}}\ and\ \bibinfo {author} {\bibfnamefont {R.}~\bibnamefont {Ahlrichs}},\ }\bibfield  {title} {\bibinfo {title} {Balanced basis sets of split valence, triple zeta valence and quadruple zeta valence quality for h to rn: Design and assessment of accuracy},\ }\href@noop {} {\bibfield  {journal} {\bibinfo  {journal} {Physical Chemistry Chemical Physics}\ }\textbf {\bibinfo {volume} {7}},\ \bibinfo {pages} {3297} (\bibinfo {year} {2005})}\BibitemShut {NoStop}%
\bibitem [{\citenamefont {Rappoport}\ and\ \citenamefont {Furche}(2010)}]{rappoport2010property}%
  \BibitemOpen
  \bibfield  {author} {\bibinfo {author} {\bibfnamefont {D.}~\bibnamefont {Rappoport}}\ and\ \bibinfo {author} {\bibfnamefont {F.}~\bibnamefont {Furche}},\ }\bibfield  {title} {\bibinfo {title} {Property-optimized gaussian basis sets for molecular response calculations},\ }\href@noop {} {\bibfield  {journal} {\bibinfo  {journal} {The Journal of chemical physics}\ }\textbf {\bibinfo {volume} {133}} (\bibinfo {year} {2010})}\BibitemShut {NoStop}%
\bibitem [{\citenamefont {Grimme}\ \emph {et~al.}(2011)\citenamefont {Grimme}, \citenamefont {Ehrlich},\ and\ \citenamefont {Goerigk}}]{grimme2011bj_damp}%
  \BibitemOpen
  \bibfield  {author} {\bibinfo {author} {\bibfnamefont {S.}~\bibnamefont {Grimme}}, \bibinfo {author} {\bibfnamefont {S.}~\bibnamefont {Ehrlich}},\ and\ \bibinfo {author} {\bibfnamefont {L.}~\bibnamefont {Goerigk}},\ }\bibfield  {title} {\bibinfo {title} {Effect of the damping function in dispersion corrected density functional theory},\ }\href@noop {} {\bibfield  {journal} {\bibinfo  {journal} {Journal of computational chemistry}\ }\textbf {\bibinfo {volume} {32}},\ \bibinfo {pages} {1456} (\bibinfo {year} {2011})}\BibitemShut {NoStop}%
\bibitem [{\citenamefont {Grimme}\ \emph {et~al.}(2010)\citenamefont {Grimme}, \citenamefont {Antony}, \citenamefont {Ehrlich},\ and\ \citenamefont {Krieg}}]{grimme2010d3}%
  \BibitemOpen
  \bibfield  {author} {\bibinfo {author} {\bibfnamefont {S.}~\bibnamefont {Grimme}}, \bibinfo {author} {\bibfnamefont {J.}~\bibnamefont {Antony}}, \bibinfo {author} {\bibfnamefont {S.}~\bibnamefont {Ehrlich}},\ and\ \bibinfo {author} {\bibfnamefont {H.}~\bibnamefont {Krieg}},\ }\bibfield  {title} {\bibinfo {title} {A consistent and accurate ab initio parametrization of density functional dispersion correction (dft-d) for the 94 elements h-pu},\ }\href@noop {} {\bibfield  {journal} {\bibinfo  {journal} {The Journal of chemical physics}\ }\textbf {\bibinfo {volume} {132}} (\bibinfo {year} {2010})}\BibitemShut {NoStop}%
\bibitem [{\citenamefont {Smith}\ \emph {et~al.}(2020{\natexlab{b}})\citenamefont {Smith}, \citenamefont {Burns}, \citenamefont {Simmonett}, \citenamefont {Parrish}, \citenamefont {Schieber}, \citenamefont {Galvelis}, \citenamefont {Kraus}, \citenamefont {Kruse}, \citenamefont {Di~Remigio}, \citenamefont {Alenaizan} \emph {et~al.}}]{smith2020psi4}%
  \BibitemOpen
  \bibfield  {author} {\bibinfo {author} {\bibfnamefont {D.~G.}\ \bibnamefont {Smith}}, \bibinfo {author} {\bibfnamefont {L.~A.}\ \bibnamefont {Burns}}, \bibinfo {author} {\bibfnamefont {A.~C.}\ \bibnamefont {Simmonett}}, \bibinfo {author} {\bibfnamefont {R.~M.}\ \bibnamefont {Parrish}}, \bibinfo {author} {\bibfnamefont {M.~C.}\ \bibnamefont {Schieber}}, \bibinfo {author} {\bibfnamefont {R.}~\bibnamefont {Galvelis}}, \bibinfo {author} {\bibfnamefont {P.}~\bibnamefont {Kraus}}, \bibinfo {author} {\bibfnamefont {H.}~\bibnamefont {Kruse}}, \bibinfo {author} {\bibfnamefont {R.}~\bibnamefont {Di~Remigio}}, \bibinfo {author} {\bibfnamefont {A.}~\bibnamefont {Alenaizan}}, \emph {et~al.},\ }\bibfield  {title} {\bibinfo {title} {Psi4 1.4: Open-source software for high-throughput quantum chemistry},\ }\href@noop {} {\bibfield  {journal} {\bibinfo  {journal} {The Journal of chemical physics}\ }\textbf {\bibinfo {volume} {152}} (\bibinfo {year} {2020}{\natexlab{b}})}\BibitemShut {NoStop}%
\bibitem [{\citenamefont {Maier}\ \emph {et~al.}(2015)\citenamefont {Maier}, \citenamefont {Martinez}, \citenamefont {Kasavajhala}, \citenamefont {Wickstrom}, \citenamefont {Hauser},\ and\ \citenamefont {Simmerling}}]{maier2015ff14sb}%
  \BibitemOpen
  \bibfield  {author} {\bibinfo {author} {\bibfnamefont {J.~A.}\ \bibnamefont {Maier}}, \bibinfo {author} {\bibfnamefont {C.}~\bibnamefont {Martinez}}, \bibinfo {author} {\bibfnamefont {K.}~\bibnamefont {Kasavajhala}}, \bibinfo {author} {\bibfnamefont {L.}~\bibnamefont {Wickstrom}}, \bibinfo {author} {\bibfnamefont {K.~E.}\ \bibnamefont {Hauser}},\ and\ \bibinfo {author} {\bibfnamefont {C.}~\bibnamefont {Simmerling}},\ }\bibfield  {title} {\bibinfo {title} {ff14sb: improving the accuracy of protein side chain and backbone parameters from ff99sb},\ }\href@noop {} {\bibfield  {journal} {\bibinfo  {journal} {Journal of chemical theory and computation}\ }\textbf {\bibinfo {volume} {11}},\ \bibinfo {pages} {3696} (\bibinfo {year} {2015})}\BibitemShut {NoStop}%
\bibitem [{\citenamefont {Donchev}\ \emph {et~al.}(2021)\citenamefont {Donchev}, \citenamefont {Taube}, \citenamefont {Decolvenaere}, \citenamefont {Hargus}, \citenamefont {McGibbon}, \citenamefont {Law}, \citenamefont {Gregersen}, \citenamefont {Li}, \citenamefont {Palmo}, \citenamefont {Siva} \emph {et~al.}}]{donchev2021quantum}%
  \BibitemOpen
  \bibfield  {author} {\bibinfo {author} {\bibfnamefont {A.~G.}\ \bibnamefont {Donchev}}, \bibinfo {author} {\bibfnamefont {A.~G.}\ \bibnamefont {Taube}}, \bibinfo {author} {\bibfnamefont {E.}~\bibnamefont {Decolvenaere}}, \bibinfo {author} {\bibfnamefont {C.}~\bibnamefont {Hargus}}, \bibinfo {author} {\bibfnamefont {R.~T.}\ \bibnamefont {McGibbon}}, \bibinfo {author} {\bibfnamefont {K.-H.}\ \bibnamefont {Law}}, \bibinfo {author} {\bibfnamefont {B.~A.}\ \bibnamefont {Gregersen}}, \bibinfo {author} {\bibfnamefont {J.-L.}\ \bibnamefont {Li}}, \bibinfo {author} {\bibfnamefont {K.}~\bibnamefont {Palmo}}, \bibinfo {author} {\bibfnamefont {K.}~\bibnamefont {Siva}}, \emph {et~al.},\ }\bibfield  {title} {\bibinfo {title} {Quantum chemical benchmark databases of gold-standard dimer interaction energies},\ }\href@noop {} {\bibfield  {journal} {\bibinfo  {journal} {Scientific data}\ }\textbf {\bibinfo {volume} {8}},\ \bibinfo {pages} {55} (\bibinfo {year} {2021})}\BibitemShut {NoStop}%
\bibitem [{\citenamefont {Isert}\ \emph {et~al.}(2022)\citenamefont {Isert}, \citenamefont {Atz}, \citenamefont {Jim{\'e}nez-Luna},\ and\ \citenamefont {Schneider}}]{isert2022qmugs}%
  \BibitemOpen
  \bibfield  {author} {\bibinfo {author} {\bibfnamefont {C.}~\bibnamefont {Isert}}, \bibinfo {author} {\bibfnamefont {K.}~\bibnamefont {Atz}}, \bibinfo {author} {\bibfnamefont {J.}~\bibnamefont {Jim{\'e}nez-Luna}},\ and\ \bibinfo {author} {\bibfnamefont {G.}~\bibnamefont {Schneider}},\ }\bibfield  {title} {\bibinfo {title} {Qmugs, quantum mechanical properties of drug-like molecules},\ }\href@noop {} {\bibfield  {journal} {\bibinfo  {journal} {Scientific Data}\ }\textbf {\bibinfo {volume} {9}},\ \bibinfo {pages} {273} (\bibinfo {year} {2022})}\BibitemShut {NoStop}%
\bibitem [{\citenamefont {Darby}\ \emph {et~al.}(2023{\natexlab{b}})\citenamefont {Darby}, \citenamefont {Kov{\'a}cs}, \citenamefont {Batatia}, \citenamefont {Caro}, \citenamefont {Hart}, \citenamefont {Ortner},\ and\ \citenamefont {Cs{\'a}nyi}}]{darby2023TrAce}%
  \BibitemOpen
  \bibfield  {author} {\bibinfo {author} {\bibfnamefont {J.~P.}\ \bibnamefont {Darby}}, \bibinfo {author} {\bibfnamefont {D.~P.}\ \bibnamefont {Kov{\'a}cs}}, \bibinfo {author} {\bibfnamefont {I.}~\bibnamefont {Batatia}}, \bibinfo {author} {\bibfnamefont {M.~A.}\ \bibnamefont {Caro}}, \bibinfo {author} {\bibfnamefont {G.~L.}\ \bibnamefont {Hart}}, \bibinfo {author} {\bibfnamefont {C.}~\bibnamefont {Ortner}},\ and\ \bibinfo {author} {\bibfnamefont {G.}~\bibnamefont {Cs{\'a}nyi}},\ }\bibfield  {title} {\bibinfo {title} {Tensor-reduced atomic density representations},\ }\href@noop {} {\bibfield  {journal} {\bibinfo  {journal} {Physical Review Letters}\ }\textbf {\bibinfo {volume} {131}},\ \bibinfo {pages} {028001} (\bibinfo {year} {2023}{\natexlab{b}})}\BibitemShut {NoStop}%
\bibitem [{\citenamefont {Schran}\ \emph {et~al.}(2021)\citenamefont {Schran}, \citenamefont {Thiemann}, \citenamefont {Rowe}, \citenamefont {M{\"u}ller}, \citenamefont {Marsalek},\ and\ \citenamefont {Michaelides}}]{schran2021machine}%
  \BibitemOpen
  \bibfield  {author} {\bibinfo {author} {\bibfnamefont {C.}~\bibnamefont {Schran}}, \bibinfo {author} {\bibfnamefont {F.~L.}\ \bibnamefont {Thiemann}}, \bibinfo {author} {\bibfnamefont {P.}~\bibnamefont {Rowe}}, \bibinfo {author} {\bibfnamefont {E.~A.}\ \bibnamefont {M{\"u}ller}}, \bibinfo {author} {\bibfnamefont {O.}~\bibnamefont {Marsalek}},\ and\ \bibinfo {author} {\bibfnamefont {A.}~\bibnamefont {Michaelides}},\ }\bibfield  {title} {\bibinfo {title} {Machine learning potentials for complex aqueous systems made simple},\ }\href@noop {} {\bibfield  {journal} {\bibinfo  {journal} {Proceedings of the National Academy of Sciences}\ }\textbf {\bibinfo {volume} {118}},\ \bibinfo {pages} {e2110077118} (\bibinfo {year} {2021})}\BibitemShut {NoStop}%
\bibitem [{\citenamefont {Eastman}\ \emph {et~al.}(2024)\citenamefont {Eastman}, \citenamefont {Behara}, \citenamefont {Dotson}, \citenamefont {Galvelis}, \citenamefont {Herr}, \citenamefont {Horton}, \citenamefont {Mao}, \citenamefont {Chodera}, \citenamefont {Pritchard}, \citenamefont {Wang}, \citenamefont {De~Fabritiis},\ and\ \citenamefont {Markland}}]{eastman_2024_10975225}%
  \BibitemOpen
  \bibfield  {author} {\bibinfo {author} {\bibfnamefont {P.}~\bibnamefont {Eastman}}, \bibinfo {author} {\bibfnamefont {P.~K.}\ \bibnamefont {Behara}}, \bibinfo {author} {\bibfnamefont {D.}~\bibnamefont {Dotson}}, \bibinfo {author} {\bibfnamefont {R.}~\bibnamefont {Galvelis}}, \bibinfo {author} {\bibfnamefont {J.}~\bibnamefont {Herr}}, \bibinfo {author} {\bibfnamefont {J.}~\bibnamefont {Horton}}, \bibinfo {author} {\bibfnamefont {Y.}~\bibnamefont {Mao}}, \bibinfo {author} {\bibfnamefont {J.}~\bibnamefont {Chodera}}, \bibinfo {author} {\bibfnamefont {B.}~\bibnamefont {Pritchard}}, \bibinfo {author} {\bibfnamefont {Y.}~\bibnamefont {Wang}}, \bibinfo {author} {\bibfnamefont {G.}~\bibnamefont {De~Fabritiis}},\ and\ \bibinfo {author} {\bibfnamefont {T.}~\bibnamefont {Markland}},\ }\bibfield  {title} {\bibinfo {title} {Spice 2.0.1},\ }\href {https://doi.org/10.5281/zenodo.10975225} {10.5281/zenodo.10975225} (\bibinfo {year} {2024})\BibitemShut {NoStop}%
\bibitem [{\citenamefont {Paszke}\ \emph {et~al.}(2019)\citenamefont {Paszke}, \citenamefont {Gross}, \citenamefont {Massa}, \citenamefont {Lerer}, \citenamefont {Bradbury}, \citenamefont {Chanan}, \citenamefont {Killeen}, \citenamefont {Lin}, \citenamefont {Gimelshein}, \citenamefont {Antiga}, \citenamefont {Desmaison}, \citenamefont {K{\"o}pf}, \citenamefont {Yang}, \citenamefont {DeVito}, \citenamefont {Raison}, \citenamefont {Tejani}, \citenamefont {Chilamkurthy}, \citenamefont {Steiner}, \citenamefont {Fang}, \citenamefont {Bai},\ and\ \citenamefont {Chintala}}]{Paszke2019PyTorchAI}%
  \BibitemOpen
  \bibfield  {author} {\bibinfo {author} {\bibfnamefont {A.}~\bibnamefont {Paszke}}, \bibinfo {author} {\bibfnamefont {S.}~\bibnamefont {Gross}}, \bibinfo {author} {\bibfnamefont {F.}~\bibnamefont {Massa}}, \bibinfo {author} {\bibfnamefont {A.}~\bibnamefont {Lerer}}, \bibinfo {author} {\bibfnamefont {J.}~\bibnamefont {Bradbury}}, \bibinfo {author} {\bibfnamefont {G.}~\bibnamefont {Chanan}}, \bibinfo {author} {\bibfnamefont {T.}~\bibnamefont {Killeen}}, \bibinfo {author} {\bibfnamefont {Z.}~\bibnamefont {Lin}}, \bibinfo {author} {\bibfnamefont {N.}~\bibnamefont {Gimelshein}}, \bibinfo {author} {\bibfnamefont {L.}~\bibnamefont {Antiga}}, \bibinfo {author} {\bibfnamefont {A.}~\bibnamefont {Desmaison}}, \bibinfo {author} {\bibfnamefont {A.}~\bibnamefont {K{\"o}pf}}, \bibinfo {author} {\bibfnamefont {E.}~\bibnamefont {Yang}}, \bibinfo {author} {\bibfnamefont {Z.}~\bibnamefont {DeVito}}, \bibinfo {author} {\bibfnamefont {M.}~\bibnamefont {Raison}}, \bibinfo {author} {\bibfnamefont {A.}~\bibnamefont {Tejani}},
  \bibinfo {author} {\bibfnamefont {S.}~\bibnamefont {Chilamkurthy}}, \bibinfo {author} {\bibfnamefont {B.}~\bibnamefont {Steiner}}, \bibinfo {author} {\bibfnamefont {L.}~\bibnamefont {Fang}}, \bibinfo {author} {\bibfnamefont {J.}~\bibnamefont {Bai}},\ and\ \bibinfo {author} {\bibfnamefont {S.}~\bibnamefont {Chintala}},\ }\bibfield  {title} {\bibinfo {title} {Pytorch: An imperative style, high-performance deep learning library},\ }in\ \href {https://api.semanticscholar.org/CorpusID:202786778} {\emph {\bibinfo {booktitle} {Neural Information Processing Systems}}}\ (\bibinfo {year} {2019})\BibitemShut {NoStop}%
\bibitem [{\citenamefont {Magd{\u{a}}u}\ \emph {et~al.}(2023)\citenamefont {Magd{\u{a}}u}, \citenamefont {Arismendi-Arrieta}, \citenamefont {Smith}, \citenamefont {Grey}, \citenamefont {Hermansson},\ and\ \citenamefont {Cs{\'a}nyi}}]{magduau2023machine}%
  \BibitemOpen
  \bibfield  {author} {\bibinfo {author} {\bibfnamefont {I.-B.}\ \bibnamefont {Magd{\u{a}}u}}, \bibinfo {author} {\bibfnamefont {D.~J.}\ \bibnamefont {Arismendi-Arrieta}}, \bibinfo {author} {\bibfnamefont {H.~E.}\ \bibnamefont {Smith}}, \bibinfo {author} {\bibfnamefont {C.~P.}\ \bibnamefont {Grey}}, \bibinfo {author} {\bibfnamefont {K.}~\bibnamefont {Hermansson}},\ and\ \bibinfo {author} {\bibfnamefont {G.}~\bibnamefont {Cs{\'a}nyi}},\ }\bibfield  {title} {\bibinfo {title} {Machine learning force fields for molecular liquids: Ethylene carbonate/ethyl methyl carbonate binary solvent},\ }\href@noop {} {\bibfield  {journal} {\bibinfo  {journal} {npj Computational Materials}\ }\textbf {\bibinfo {volume} {9}},\ \bibinfo {pages} {146} (\bibinfo {year} {2023})}\BibitemShut {NoStop}%
\bibitem [{\citenamefont {Horton}\ \emph {et~al.}(2022)\citenamefont {Horton}, \citenamefont {Boothroyd}, \citenamefont {Wagner}, \citenamefont {Mitchell}, \citenamefont {Gokey}, \citenamefont {Dotson}, \citenamefont {Behara}, \citenamefont {Ramaswamy}, \citenamefont {Mackey}, \citenamefont {Chodera} \emph {et~al.}}]{horton2022open}%
  \BibitemOpen
  \bibfield  {author} {\bibinfo {author} {\bibfnamefont {J.~T.}\ \bibnamefont {Horton}}, \bibinfo {author} {\bibfnamefont {S.}~\bibnamefont {Boothroyd}}, \bibinfo {author} {\bibfnamefont {J.}~\bibnamefont {Wagner}}, \bibinfo {author} {\bibfnamefont {J.~A.}\ \bibnamefont {Mitchell}}, \bibinfo {author} {\bibfnamefont {T.}~\bibnamefont {Gokey}}, \bibinfo {author} {\bibfnamefont {D.~L.}\ \bibnamefont {Dotson}}, \bibinfo {author} {\bibfnamefont {P.~K.}\ \bibnamefont {Behara}}, \bibinfo {author} {\bibfnamefont {V.~K.}\ \bibnamefont {Ramaswamy}}, \bibinfo {author} {\bibfnamefont {M.}~\bibnamefont {Mackey}}, \bibinfo {author} {\bibfnamefont {J.~D.}\ \bibnamefont {Chodera}}, \emph {et~al.},\ }\bibfield  {title} {\bibinfo {title} {Open force field bespokefit: Automating bespoke torsion parametrization at scale},\ }\href@noop {} {\bibfield  {journal} {\bibinfo  {journal} {Journal of Chemical Information and Modeling}\ }\textbf {\bibinfo {volume} {62}},\ \bibinfo {pages} {5622} (\bibinfo {year} {2022})}\BibitemShut
  {NoStop}%
\bibitem [{\citenamefont {Hao}\ \emph {et~al.}(2022)\citenamefont {Hao}, \citenamefont {He}, \citenamefont {Roitberg}, \citenamefont {Zhang},\ and\ \citenamefont {Wang}}]{hao2022ani_geom_opt}%
  \BibitemOpen
  \bibfield  {author} {\bibinfo {author} {\bibfnamefont {D.}~\bibnamefont {Hao}}, \bibinfo {author} {\bibfnamefont {X.}~\bibnamefont {He}}, \bibinfo {author} {\bibfnamefont {A.~E.}\ \bibnamefont {Roitberg}}, \bibinfo {author} {\bibfnamefont {S.}~\bibnamefont {Zhang}},\ and\ \bibinfo {author} {\bibfnamefont {J.}~\bibnamefont {Wang}},\ }\bibfield  {title} {\bibinfo {title} {Development and evaluation of geometry optimization algorithms in conjunction with ani potentials},\ }\href@noop {} {\bibfield  {journal} {\bibinfo  {journal} {Journal of Chemical Theory and Computation}\ }\textbf {\bibinfo {volume} {18}},\ \bibinfo {pages} {978} (\bibinfo {year} {2022})}\BibitemShut {NoStop}%
\bibitem [{\citenamefont {Lahey}\ \emph {et~al.}(2020)\citenamefont {Lahey}, \citenamefont {Thien~Phuc},\ and\ \citenamefont {Rowley}}]{lahey2020biaryl_torsion}%
  \BibitemOpen
  \bibfield  {author} {\bibinfo {author} {\bibfnamefont {S.-L.~J.}\ \bibnamefont {Lahey}}, \bibinfo {author} {\bibfnamefont {T.~N.}\ \bibnamefont {Thien~Phuc}},\ and\ \bibinfo {author} {\bibfnamefont {C.~N.}\ \bibnamefont {Rowley}},\ }\bibfield  {title} {\bibinfo {title} {Benchmarking force field and the ani neural network potentials for the torsional potential energy surface of biaryl drug fragments},\ }\href@noop {} {\bibfield  {journal} {\bibinfo  {journal} {Journal of Chemical Information and Modeling}\ }\textbf {\bibinfo {volume} {60}},\ \bibinfo {pages} {6258} (\bibinfo {year} {2020})}\BibitemShut {NoStop}%
\bibitem [{\citenamefont {Rai}\ \emph {et~al.}(2022)\citenamefont {Rai}, \citenamefont {Sresht}, \citenamefont {Yang}, \citenamefont {Unwalla}, \citenamefont {Tu}, \citenamefont {Mathiowetz},\ and\ \citenamefont {Bakken}}]{rai2022torsionnet}%
  \BibitemOpen
  \bibfield  {author} {\bibinfo {author} {\bibfnamefont {B.~K.}\ \bibnamefont {Rai}}, \bibinfo {author} {\bibfnamefont {V.}~\bibnamefont {Sresht}}, \bibinfo {author} {\bibfnamefont {Q.}~\bibnamefont {Yang}}, \bibinfo {author} {\bibfnamefont {R.}~\bibnamefont {Unwalla}}, \bibinfo {author} {\bibfnamefont {M.}~\bibnamefont {Tu}}, \bibinfo {author} {\bibfnamefont {A.~M.}\ \bibnamefont {Mathiowetz}},\ and\ \bibinfo {author} {\bibfnamefont {G.~A.}\ \bibnamefont {Bakken}},\ }\bibfield  {title} {\bibinfo {title} {Torsionnet: A deep neural network to rapidly predict small-molecule torsional energy profiles with the accuracy of quantum mechanics},\ }\href@noop {} {\bibfield  {journal} {\bibinfo  {journal} {Journal of Chemical Information and Modeling}\ }\textbf {\bibinfo {volume} {62}},\ \bibinfo {pages} {785} (\bibinfo {year} {2022})}\BibitemShut {NoStop}%
\bibitem [{\citenamefont {Behara}\ \emph {et~al.}(2024)\citenamefont {Behara}, \citenamefont {Jang}, \citenamefont {Horton}, \citenamefont {Gokey}, \citenamefont {Dotson}, \citenamefont {Boothroyd}, \citenamefont {Bayly}, \citenamefont {Cole}, \citenamefont {Wang},\ and\ \citenamefont {Mobley}}]{behara2024}%
  \BibitemOpen
  \bibfield  {author} {\bibinfo {author} {\bibfnamefont {P.~K.}\ \bibnamefont {Behara}}, \bibinfo {author} {\bibfnamefont {H.}~\bibnamefont {Jang}}, \bibinfo {author} {\bibfnamefont {J.~T.}\ \bibnamefont {Horton}}, \bibinfo {author} {\bibfnamefont {T.}~\bibnamefont {Gokey}}, \bibinfo {author} {\bibfnamefont {D.~L.}\ \bibnamefont {Dotson}}, \bibinfo {author} {\bibfnamefont {S.}~\bibnamefont {Boothroyd}}, \bibinfo {author} {\bibfnamefont {C.~I.}\ \bibnamefont {Bayly}}, \bibinfo {author} {\bibfnamefont {D.~J.}\ \bibnamefont {Cole}}, \bibinfo {author} {\bibfnamefont {L.-P.}\ \bibnamefont {Wang}},\ and\ \bibinfo {author} {\bibfnamefont {D.~L.}\ \bibnamefont {Mobley}},\ }\bibfield  {title} {\bibinfo {title} {Benchmarking quantum mechanical levels of theory for valence parametrization in force fields},\ }\href {https://doi.org/10.1021/acs.jpcb.4c03167} {\bibfield  {journal} {\bibinfo  {journal} {The Journal of Physical Chemistry B}\ }\textbf {\bibinfo {volume} {128}},\ \bibinfo {pages} {7888} (\bibinfo {year}
  {2024})}\BibitemShut {NoStop}%
\bibitem [{\citenamefont {Musil}\ \emph {et~al.}(2022)\citenamefont {Musil}, \citenamefont {Zaporozhets}, \citenamefont {Noé}, \citenamefont {Clementi},\ and\ \citenamefont {Kapil}}]{Musil2022}%
  \BibitemOpen
  \bibfield  {author} {\bibinfo {author} {\bibfnamefont {F.}~\bibnamefont {Musil}}, \bibinfo {author} {\bibfnamefont {I.}~\bibnamefont {Zaporozhets}}, \bibinfo {author} {\bibfnamefont {F.}~\bibnamefont {Noé}}, \bibinfo {author} {\bibfnamefont {C.}~\bibnamefont {Clementi}},\ and\ \bibinfo {author} {\bibfnamefont {V.}~\bibnamefont {Kapil}},\ }\bibfield  {title} {\bibinfo {title} {Quantum dynamics using path integral coarse-graining},\ }\bibfield  {journal} {\bibinfo  {journal} {The Journal of Chemical Physics}\ }\textbf {\bibinfo {volume} {157}},\ \href {https://doi.org/10.1063/5.0120386} {10.1063/5.0120386} (\bibinfo {year} {2022})\BibitemShut {NoStop}%
\bibitem [{\citenamefont {NIST Mass Spectrometry Data~Center}(2025)}]{nist_webbook}%
  \BibitemOpen
  \bibfield  {author} {\bibinfo {author} {\bibfnamefont {d.}~\bibnamefont {NIST Mass Spectrometry Data~Center}, \bibfnamefont {William E.~Wallace}},\ }\href {https://doi.org/10.18434/T4D303} {\bibinfo {title} {Infrared spectra}} (\bibinfo {year} {2025}),\ \bibinfo {note} {retrieved January 20, 2025}\BibitemShut {NoStop}%
\bibitem [{\citenamefont {Reilly}\ and\ \citenamefont {Tkatchenko}(2013)}]{reilly2013understanding}%
  \BibitemOpen
  \bibfield  {author} {\bibinfo {author} {\bibfnamefont {A.~M.}\ \bibnamefont {Reilly}}\ and\ \bibinfo {author} {\bibfnamefont {A.}~\bibnamefont {Tkatchenko}},\ }\bibfield  {title} {\bibinfo {title} {Understanding the role of vibrations, exact exchange, and many-body van der waals interactions in the cohesive properties of molecular crystals},\ }\href@noop {} {\bibfield  {journal} {\bibinfo  {journal} {The Journal of chemical physics}\ }\textbf {\bibinfo {volume} {139}},\ \bibinfo {pages} {024705} (\bibinfo {year} {2013})}\BibitemShut {NoStop}%
\bibitem [{\citenamefont {Dolgonos}\ \emph {et~al.}(2019)\citenamefont {Dolgonos}, \citenamefont {Hoja},\ and\ \citenamefont {Boese}}]{dolgonos2019revised_x23}%
  \BibitemOpen
  \bibfield  {author} {\bibinfo {author} {\bibfnamefont {G.~A.}\ \bibnamefont {Dolgonos}}, \bibinfo {author} {\bibfnamefont {J.}~\bibnamefont {Hoja}},\ and\ \bibinfo {author} {\bibfnamefont {A.~D.}\ \bibnamefont {Boese}},\ }\bibfield  {title} {\bibinfo {title} {Revised values for the x23 benchmark set of molecular crystals},\ }\href@noop {} {\bibfield  {journal} {\bibinfo  {journal} {Physical Chemistry Chemical Physics}\ }\textbf {\bibinfo {volume} {21}},\ \bibinfo {pages} {24333} (\bibinfo {year} {2019})}\BibitemShut {NoStop}%
\bibitem [{\citenamefont {Della~Pia}\ \emph {et~al.}(2024)\citenamefont {Della~Pia}, \citenamefont {Zen}, \citenamefont {Alf\`e},\ and\ \citenamefont {Michaelides}}]{PhysRevLett.133.046401}%
  \BibitemOpen
  \bibfield  {author} {\bibinfo {author} {\bibfnamefont {F.}~\bibnamefont {Della~Pia}}, \bibinfo {author} {\bibfnamefont {A.}~\bibnamefont {Zen}}, \bibinfo {author} {\bibfnamefont {D.}~\bibnamefont {Alf\`e}},\ and\ \bibinfo {author} {\bibfnamefont {A.}~\bibnamefont {Michaelides}},\ }\bibfield  {title} {\bibinfo {title} {How accurate are simulations and experiments for the lattice energies of molecular crystals?},\ }\href {https://doi.org/10.1103/PhysRevLett.133.046401} {\bibfield  {journal} {\bibinfo  {journal} {Phys. Rev. Lett.}\ }\textbf {\bibinfo {volume} {133}},\ \bibinfo {pages} {046401} (\bibinfo {year} {2024})}\BibitemShut {NoStop}%
\bibitem [{\citenamefont {Kaur}\ \emph {et~al.}(2025)\citenamefont {Kaur}, \citenamefont {Della~Pia}, \citenamefont {Batatia}, \citenamefont {Advincula}, \citenamefont {Shi}, \citenamefont {Lan}, \citenamefont {Csányi}, \citenamefont {Michaelides},\ and\ \citenamefont {Kapil}}]{kaur2025}%
  \BibitemOpen
  \bibfield  {author} {\bibinfo {author} {\bibfnamefont {H.}~\bibnamefont {Kaur}}, \bibinfo {author} {\bibfnamefont {F.}~\bibnamefont {Della~Pia}}, \bibinfo {author} {\bibfnamefont {I.}~\bibnamefont {Batatia}}, \bibinfo {author} {\bibfnamefont {X.~R.}\ \bibnamefont {Advincula}}, \bibinfo {author} {\bibfnamefont {B.~X.}\ \bibnamefont {Shi}}, \bibinfo {author} {\bibfnamefont {J.}~\bibnamefont {Lan}}, \bibinfo {author} {\bibfnamefont {G.}~\bibnamefont {Csányi}}, \bibinfo {author} {\bibfnamefont {A.}~\bibnamefont {Michaelides}},\ and\ \bibinfo {author} {\bibfnamefont {V.}~\bibnamefont {Kapil}},\ }\bibfield  {title} {\bibinfo {title} {Data-efficient fine-tuning of foundational models for first-principles quality sublimation enthalpies},\ }\href {https://doi.org/10.1039/D4FD00107A} {\bibfield  {journal} {\bibinfo  {journal} {Faraday Discuss.}\ }\textbf {\bibinfo {volume} {256}},\ \bibinfo {pages} {120} (\bibinfo {year} {2025})}\BibitemShut {NoStop}%
\bibitem [{\citenamefont {Caleman}\ \emph {et~al.}(2012)\citenamefont {Caleman}, \citenamefont {van Maaren}, \citenamefont {Hong}, \citenamefont {Hub}, \citenamefont {Costa},\ and\ \citenamefont {van~der Spoel}}]{caleman2012}%
  \BibitemOpen
  \bibfield  {author} {\bibinfo {author} {\bibfnamefont {C.}~\bibnamefont {Caleman}}, \bibinfo {author} {\bibfnamefont {P.~J.}\ \bibnamefont {van Maaren}}, \bibinfo {author} {\bibfnamefont {M.}~\bibnamefont {Hong}}, \bibinfo {author} {\bibfnamefont {J.~S.}\ \bibnamefont {Hub}}, \bibinfo {author} {\bibfnamefont {L.~T.}\ \bibnamefont {Costa}},\ and\ \bibinfo {author} {\bibfnamefont {D.}~\bibnamefont {van~der Spoel}},\ }\bibfield  {title} {\bibinfo {title} {Force field benchmark of organic liquids: Density, enthalpy of vaporization, heat capacities, surface tension, isothermal compressibility, volumetric expansion coefficient, and dielectric constant},\ }\href {https://doi.org/10.1021/ct200731v} {\bibfield  {journal} {\bibinfo  {journal} {Journal of Chemical Theory and Computation}\ }\textbf {\bibinfo {volume} {8}},\ \bibinfo {pages} {61} (\bibinfo {year} {2012})}\BibitemShut {NoStop}%
\bibitem [{\citenamefont {Horton}\ \emph {et~al.}(2019)\citenamefont {Horton}, \citenamefont {Allen}, \citenamefont {Dodda},\ and\ \citenamefont {Cole}}]{horton2019qubekit}%
  \BibitemOpen
  \bibfield  {author} {\bibinfo {author} {\bibfnamefont {J.~T.}\ \bibnamefont {Horton}}, \bibinfo {author} {\bibfnamefont {A.~E.}\ \bibnamefont {Allen}}, \bibinfo {author} {\bibfnamefont {L.~S.}\ \bibnamefont {Dodda}},\ and\ \bibinfo {author} {\bibfnamefont {D.~J.}\ \bibnamefont {Cole}},\ }\bibfield  {title} {\bibinfo {title} {Qubekit: Automating the derivation of force field parameters from quantum mechanics},\ }\href@noop {} {\bibfield  {journal} {\bibinfo  {journal} {Journal of chemical information and modeling}\ }\textbf {\bibinfo {volume} {59}},\ \bibinfo {pages} {1366} (\bibinfo {year} {2019})}\BibitemShut {NoStop}%
\bibitem [{\citenamefont {Shirts}\ \emph {et~al.}(2007)\citenamefont {Shirts}, \citenamefont {Mobley}, \citenamefont {Chodera},\ and\ \citenamefont {Pande}}]{longrangecorrection}%
  \BibitemOpen
  \bibfield  {author} {\bibinfo {author} {\bibfnamefont {M.~R.}\ \bibnamefont {Shirts}}, \bibinfo {author} {\bibfnamefont {D.~L.}\ \bibnamefont {Mobley}}, \bibinfo {author} {\bibfnamefont {J.~D.}\ \bibnamefont {Chodera}},\ and\ \bibinfo {author} {\bibfnamefont {V.~S.}\ \bibnamefont {Pande}},\ }\bibfield  {title} {\bibinfo {title} {Accurate and efficient corrections for missing dispersion interactions in molecular simulations},\ }\href {https://doi.org/10.1021/jp0735987} {\bibfield  {journal} {\bibinfo  {journal} {The Journal of Physical Chemistry B}\ }\textbf {\bibinfo {volume} {111}},\ \bibinfo {pages} {13052} (\bibinfo {year} {2007})}\BibitemShut {NoStop}%
\bibitem [{\citenamefont {Medders}\ \emph {et~al.}(2014)\citenamefont {Medders}, \citenamefont {Babin},\ and\ \citenamefont {Paesani}}]{meddersDevelopmentFirstPrinciplesWater2014}%
  \BibitemOpen
  \bibfield  {author} {\bibinfo {author} {\bibfnamefont {G.~R.}\ \bibnamefont {Medders}}, \bibinfo {author} {\bibfnamefont {V.}~\bibnamefont {Babin}},\ and\ \bibinfo {author} {\bibfnamefont {F.}~\bibnamefont {Paesani}},\ }\bibfield  {title} {\bibinfo {title} {Development of a ``{{First-Principles}}'' {{Water Potential}} with {{Flexible Monomers}}. {{III}}. {{Liquid Phase Properties}}},\ }\href {https://doi.org/10.1021/ct5004115} {\bibfield  {journal} {\bibinfo  {journal} {Journal of Chemical Theory and Computation}\ }\textbf {\bibinfo {volume} {10}},\ \bibinfo {pages} {2906} (\bibinfo {year} {2014})}\BibitemShut {NoStop}%
\bibitem [{\citenamefont {Gillan}\ \emph {et~al.}(2016)\citenamefont {Gillan}, \citenamefont {Alfè},\ and\ \citenamefont {Michaelides}}]{10.1063/1.4944633}%
  \BibitemOpen
  \bibfield  {author} {\bibinfo {author} {\bibfnamefont {M.~J.}\ \bibnamefont {Gillan}}, \bibinfo {author} {\bibfnamefont {D.}~\bibnamefont {Alfè}},\ and\ \bibinfo {author} {\bibfnamefont {A.}~\bibnamefont {Michaelides}},\ }\bibfield  {title} {\bibinfo {title} {Perspective: How good is dft for water?},\ }\href {https://doi.org/10.1063/1.4944633} {\bibfield  {journal} {\bibinfo  {journal} {The Journal of Chemical Physics}\ }\textbf {\bibinfo {volume} {144}},\ \bibinfo {pages} {130901} (\bibinfo {year} {2016})}\BibitemShut {NoStop}%
\bibitem [{\citenamefont {Zhang}\ \emph {et~al.}(2020)\citenamefont {Zhang}, \citenamefont {{Schweitzer-Stenner}},\ and\ \citenamefont {Urbanc}}]{zhangMolecularDynamicsForce2020}%
  \BibitemOpen
  \bibfield  {author} {\bibinfo {author} {\bibfnamefont {S.}~\bibnamefont {Zhang}}, \bibinfo {author} {\bibfnamefont {R.}~\bibnamefont {{Schweitzer-Stenner}}},\ and\ \bibinfo {author} {\bibfnamefont {B.}~\bibnamefont {Urbanc}},\ }\bibfield  {title} {\bibinfo {title} {Do {{Molecular Dynamics Force Fields Capture Conformational Dynamics}} of {{Alanine}} in {{Water}}?},\ }\href {https://doi.org/10.1021/acs.jctc.9b00588} {\bibfield  {journal} {\bibinfo  {journal} {Journal of Chemical Theory and Computation}\ }\textbf {\bibinfo {volume} {16}},\ \bibinfo {pages} {510} (\bibinfo {year} {2020})}\BibitemShut {NoStop}%
\bibitem [{\citenamefont {Rosenberger}\ \emph {et~al.}(2021)\citenamefont {Rosenberger}, \citenamefont {Smith},\ and\ \citenamefont {Garcia}}]{rosenberger2021modeling}%
  \BibitemOpen
  \bibfield  {author} {\bibinfo {author} {\bibfnamefont {D.}~\bibnamefont {Rosenberger}}, \bibinfo {author} {\bibfnamefont {J.~S.}\ \bibnamefont {Smith}},\ and\ \bibinfo {author} {\bibfnamefont {A.~E.}\ \bibnamefont {Garcia}},\ }\bibfield  {title} {\bibinfo {title} {Modeling of peptides with classical and novel machine learning force fields: A comparison},\ }\href@noop {} {\bibfield  {journal} {\bibinfo  {journal} {The Journal of Physical Chemistry B}\ }\textbf {\bibinfo {volume} {125}},\ \bibinfo {pages} {3598} (\bibinfo {year} {2021})}\BibitemShut {NoStop}%
\bibitem [{\citenamefont {Hu}\ and\ \citenamefont {Bax}(1997)}]{huDeterminationH1Angles1997a}%
  \BibitemOpen
  \bibfield  {author} {\bibinfo {author} {\bibfnamefont {J.-S.}\ \bibnamefont {Hu}}\ and\ \bibinfo {author} {\bibfnamefont {A.}~\bibnamefont {Bax}},\ }\bibfield  {title} {\bibinfo {title} {Determination of {$\varphi$} and {$X$}1 {{Angles}} in {{Proteins}} from {{13C}}-{{13C Three-Bond J Couplings Measured}} by {{Three-Dimensional Heteronuclear NMR}}. {{How Planar Is}} the {{Peptide Bond}}?},\ }\href {https://doi.org/10.1021/ja970067v} {\bibfield  {journal} {\bibinfo  {journal} {Journal of the American Chemical Society}\ }\textbf {\bibinfo {volume} {119}},\ \bibinfo {pages} {6360} (\bibinfo {year} {1997})}\BibitemShut {NoStop}%
\bibitem [{\citenamefont {Bolin}\ and\ \citenamefont {Millhauser}(1999)}]{bolin310SplitPersonality1999}%
  \BibitemOpen
  \bibfield  {author} {\bibinfo {author} {\bibfnamefont {K.~A.}\ \bibnamefont {Bolin}}\ and\ \bibinfo {author} {\bibfnamefont {G.~L.}\ \bibnamefont {Millhauser}},\ }\bibfield  {title} {\bibinfo {title} {{$\alpha$} and 310:\, {{The Split Personality}} of {{Polypeptide Helices}}},\ }\href {https://doi.org/10.1021/ar980065v} {\bibfield  {journal} {\bibinfo  {journal} {Accounts of Chemical Research}\ }\textbf {\bibinfo {volume} {32}},\ \bibinfo {pages} {1027} (\bibinfo {year} {1999})}\BibitemShut {NoStop}%
\bibitem [{\citenamefont {Heinig}\ and\ \citenamefont {Frishman}(2004)}]{heinigSTRIDEWebServer2004}%
  \BibitemOpen
  \bibfield  {author} {\bibinfo {author} {\bibfnamefont {M.}~\bibnamefont {Heinig}}\ and\ \bibinfo {author} {\bibfnamefont {D.}~\bibnamefont {Frishman}},\ }\bibfield  {title} {\bibinfo {title} {{{STRIDE}}: A web server for secondary structure assignment from known atomic coordinates of proteins},\ }\href {https://doi.org/10.1093/nar/gkh429} {\bibfield  {journal} {\bibinfo  {journal} {Nucleic Acids Research}\ }\textbf {\bibinfo {volume} {32}},\ \bibinfo {pages} {W500} (\bibinfo {year} {2004})}\BibitemShut {NoStop}%
\bibitem [{\citenamefont {Woods}(2014)}]{woodsGlassyStateCrambin2014}%
  \BibitemOpen
  \bibfield  {author} {\bibinfo {author} {\bibfnamefont {K.~N.}\ \bibnamefont {Woods}},\ }\bibfield  {title} {\bibinfo {title} {The glassy state of crambin and the {{THz}} time scale protein-solvent fluctuations possibly related to protein function},\ }\href {https://doi.org/10.1186/s13628-014-0008-0} {\bibfield  {journal} {\bibinfo  {journal} {BMC Biophysics}\ }\textbf {\bibinfo {volume} {7}},\ \bibinfo {pages} {8} (\bibinfo {year} {2014})}\BibitemShut {NoStop}%
\bibitem [{\citenamefont {Browining}(2024)}]{NJB2024}%
  \BibitemOpen
  \bibfield  {author} {\bibinfo {author} {\bibfnamefont {N.~J.}\ \bibnamefont {Browining}},\ }\href@noop {} {\bibinfo {title} {Cuda mace}},\ \bibinfo {howpublished} {\url{https://github.com/nickjbrowning/cuda_mace}} (\bibinfo {year} {2024}),\ \bibinfo {note} {accessed 2025-04-15}\BibitemShut {NoStop}%
\bibitem [{\citenamefont {Witt}(2025)}]{WCW2025}%
  \BibitemOpen
  \bibfield  {author} {\bibinfo {author} {\bibfnamefont {W.~C.}\ \bibnamefont {Witt}},\ }\href@noop {} {\bibinfo {title} {symmetrix}},\ \bibinfo {howpublished} {\url{https://github.com/wcwitt/symmetrix}} (\bibinfo {year} {2025}),\ \bibinfo {note} {accessed 2025-04-15}\BibitemShut {NoStop}%
\bibitem [{\citenamefont {Bigi}\ \emph {et~al.}(2023)\citenamefont {Bigi}, \citenamefont {Fraux}, \citenamefont {Browning},\ and\ \citenamefont {Ceriotti}}]{sphericart}%
  \BibitemOpen
  \bibfield  {author} {\bibinfo {author} {\bibfnamefont {F.}~\bibnamefont {Bigi}}, \bibinfo {author} {\bibfnamefont {G.}~\bibnamefont {Fraux}}, \bibinfo {author} {\bibfnamefont {N.~J.}\ \bibnamefont {Browning}},\ and\ \bibinfo {author} {\bibfnamefont {M.}~\bibnamefont {Ceriotti}},\ }\bibfield  {title} {\bibinfo {title} {Fast evaluation of spherical harmonics with sphericart},\ }\href@noop {} {\bibfield  {journal} {\bibinfo  {journal} {J. Chem. Phys.}\ ,\ \bibinfo {pages} {064802}} (\bibinfo {year} {2023})}\BibitemShut {NoStop}%
\end{thebibliography}%

\end{document}



\section{MACE-OFF Training Details}
Initially, the force weight in the loss was set to 1000 and the energy weight to 40. The learning rate was 0.01 and Adam optimizer with Amsgrad was used. The exponential moving average of the weights was taken in each training step. When the force error converged, the second phase of the training was started with force weight 10 and energy weight 1000 and the learning rate was reduced to 0.00025. Finally, the training was terminated when the energy error also stopped decreasing significantly. All models were trained on a single Nvidia A100 GPU. Training the small model took about 6 days, the medium about 10 days and the large model 14 days. 

After fitting the small MACE model and observing the training errors, we noticed the presence of outliers in the dataset. This is probably caused by errors in the underlying electronic structure calculations, some of which have been documented on the SPICE GitHub repository.\footnote{https://github.com/openmm/spice-dataset} 
To confirm this, we also re-evaluated a selection of the configurations that had the highest error, using the level of DFT used in SPICE, and found that for about a third of the configurations, the recomputed energies and forces agreed well with the MACE prediction and not with the original DFT labels. 
To purify the dataset, we removed from the training set the configurations that had a maximum force error greater than 2~eV/\AA. This meant the removal of just 808 configurations, many of which contained heavy elements, in particular phosphorus and iodine, that might have a more challenging electronic structure. 

In common with the first version of SPICE, close inspection of the SPICE version 2 dataset revealed several nonphysical geometries with correspondingly high DFT forces.  These configurations were removed by applying a maximum force filter of 15 eV/\AA. We also observed that many phosphorous-containing compounds were present among the outliers, as well as configurations with highly distorted geometries, which suggest challenging electronic structure. We found that many configurations had forces that do not sum to zero. To avoid the possibility of these configurations poisoning the dataset, we applied a total force filter of 0.1 eV/\AA. In total this removed 10,372 configurations from the SPICE version 2 subsets.

\clearpage

\section{Test set errors}

The numerical values of the test set errors are displayed in Table~\ref{tab:test_errors}. Intermolecular force errors were obtained as RMSEs between DFT intermolecular forces and MACE predictions. The intermolecular forces were computed by summing over the translational and rotational components for each atom as shown in Ref. \cite{magduau2023machine}. The algorithm can be summarized as follows:

\definecolor{myblue}{RGB}{71,120,207}
\definecolor{mygreen}{RGB}{106,204,100}
\definecolor{myred}{RGB}{213,95,95}

\setlength{\tabcolsep}{1pt}
\begin{table*}[ht!]
    \caption{Summary of test set errors including inter-molecular force errors}
    \label{tab:test_errors}
    \begin{center}
\begin{tabular}{c|c||ccccccccc}
\hhline{~~~--------}
\hhline{~~~--------}
\hhline{~~~--------}
\multicolumn{3}{c}{} & PubChem & \makecell{DES370K\\Monomers} & \makecell{DES370K\\Dimers} & Dipeptides & \makecell{Solvated\\Amino Acids} & Water & QMugs & Tripeptide \\
\hline
\hline
\multirow{9}{*}{\rotatebox[origin=c]{90}{MAE}} & \multirow{3}{*}{\makecell{$\rm E$\\$(\rm meV/at)$}} & 23S & \cellcolor{myblue!90} \textcolor{white} {1.41} & \cellcolor{myblue!90} \textcolor{white} {1.04} & \cellcolor{myblue!90} \textcolor{white} {0.98} & \cellcolor{myblue!90} \textcolor{white} {0.84} & \cellcolor{myblue!90} \textcolor{white} {1.60} & \cellcolor{myblue!90} \textcolor{white} {1.67} & \cellcolor{myblue!90} \textcolor{white} {1.03} & \cellcolor{myblue!90} \textcolor{white} {1.05} \\
 & & 23M & \cellcolor{mygreen!100} \textcolor{black} {0.91} & \cellcolor{mygreen!100} \textcolor{black} {0.63} & \cellcolor{mygreen!100} \textcolor{black} {0.58} & \cellcolor{mygreen!100} \textcolor{black} {0.52} & \cellcolor{mygreen!100} \textcolor{black} {1.21} & \cellcolor{mygreen!100} \textcolor{black} {0.76} & \cellcolor{mygreen!100} \textcolor{black} {0.69} & \cellcolor{mygreen!100} \textcolor{black} {0.57} \\
 & & 23L & \cellcolor{myred!90} \textcolor{black} {0.88} & \cellcolor{myred!90} \textcolor{black} {0.59} & \cellcolor{myred!90} \textcolor{black} {0.54} & \cellcolor{myred!90} \textcolor{black} {0.42} & \cellcolor{myred!90} \textcolor{black} {0.98} & \cellcolor{myred!90} \textcolor{black} {0.83} & \cellcolor{myred!90} \textcolor{black} {0.45} & \cellcolor{myred!90} \textcolor{black} {0.38} \\
\hhline{~----------}
\hhline{~----------}
 & \multirow{3}{*}{\makecell{$\rm F_{total}$\\$(\rm meV/\AA)$}} & 23S & \cellcolor{myblue!90} \textcolor{white} {35.68} & \cellcolor{myblue!90} \textcolor{white} {17.63} & \cellcolor{myblue!90} \textcolor{white} {16.31} & \cellcolor{myblue!90} \textcolor{white} {25.07} & \cellcolor{myblue!90} \textcolor{white} {38.56} & \cellcolor{myblue!90} \textcolor{white} {28.53} & \cellcolor{myblue!90} \textcolor{white} {41.45} & \cellcolor{myblue!90} \textcolor{white} {32.88} \\
 & & 23M & \cellcolor{mygreen!100} \textcolor{black} {20.57} & \cellcolor{mygreen!100} \textcolor{black} {9.36} & \cellcolor{mygreen!100} \textcolor{black} {9.02} & \cellcolor{mygreen!100} \textcolor{black} {14.27} & \cellcolor{mygreen!100} \textcolor{black} {23.26} & \cellcolor{mygreen!100} \textcolor{black} {15.27} & \cellcolor{mygreen!100} \textcolor{black} {23.58} & \cellcolor{mygreen!100} \textcolor{black} {18.74} \\
 & & 23L & \cellcolor{myred!90} \textcolor{black} {14.75} & \cellcolor{myred!90} \textcolor{black} {6.58} & \cellcolor{myred!90} \textcolor{black} {6.62} & \cellcolor{myred!90} \textcolor{black} {10.19} & \cellcolor{myred!90} \textcolor{black} {19.43} & \cellcolor{myred!90} \textcolor{black} {13.57} & \cellcolor{myred!90} \textcolor{black} {16.93} & \cellcolor{myred!90} \textcolor{black} {13.20} \\
\hhline{~----------}
\hhline{~----------}
 & \multirow{3}{*}{\makecell{$\rm F_{inter}$\\$(\rm meV/\AA)$}} & 23S & \cellcolor{myblue!90} \textcolor{white} {0.13} & \cellcolor{myblue!90} \textcolor{white} {0.18} & \cellcolor{myblue!90} \textcolor{white} {3.01} & \cellcolor{myblue!90} \textcolor{white} {0.07} & \cellcolor{myblue!90} \textcolor{white} {16.98} & \cellcolor{myblue!90} \textcolor{white} {19.03} & \cellcolor{myblue!90} \textcolor{white} {0.09} & \cellcolor{myblue!90} \textcolor{white} {0.07} \\
 & & 23M & \cellcolor{mygreen!100} \textcolor{black} {0.13} & \cellcolor{mygreen!100} \textcolor{black} {0.18} & \cellcolor{mygreen!100} \textcolor{black} {1.79} & \cellcolor{mygreen!100} \textcolor{black} {0.07} & \cellcolor{mygreen!100} \textcolor{black} {10.30} & \cellcolor{mygreen!100} \textcolor{black} {10.03} & \cellcolor{mygreen!100} \textcolor{black} {0.09} & \cellcolor{mygreen!100} \textcolor{black} {0.07} \\
 & & 23L & \cellcolor{myred!90} \textcolor{black} {0.13} & \cellcolor{myred!90} \textcolor{black} {0.18} & \cellcolor{myred!90} \textcolor{black} {1.44} & \cellcolor{myred!90} \textcolor{black} {0.07} & \cellcolor{myred!90} \textcolor{black} {8.85} & \cellcolor{myred!90} \textcolor{black} {8.97} & \cellcolor{myred!90} \textcolor{black} {0.09} & \cellcolor{myred!90} \textcolor{black} {0.07} \\
\hline
\hline
\multirow{9}{*}{\rotatebox[origin=c]{90}{Relative MAE}} & \multirow{3}{*}{$\rm E~(\%)$} & 23S & \cellcolor{myblue!90} \textcolor{white} {0.00} & \cellcolor{myblue!90} \textcolor{white} {0.00} & \cellcolor{myblue!90} \textcolor{white} {0.00} & \cellcolor{myblue!90} \textcolor{white} {0.00} & \cellcolor{myblue!90} \textcolor{white} {0.01} & \cellcolor{myblue!90} \textcolor{white} {8.44} & \cellcolor{myblue!90} \textcolor{white} {0.00} & \cellcolor{myblue!90} \textcolor{white} {0.00} \\
 & & 23M & \cellcolor{mygreen!100} \textcolor{black} {0.00} & \cellcolor{mygreen!100} \textcolor{black} {0.00} & \cellcolor{mygreen!100} \textcolor{black} {0.00} & \cellcolor{mygreen!100} \textcolor{black} {0.00} & \cellcolor{mygreen!100} \textcolor{black} {0.00} & \cellcolor{mygreen!100} \textcolor{black} {3.84} & \cellcolor{mygreen!100} \textcolor{black} {0.00} & \cellcolor{mygreen!100} \textcolor{black} {0.00} \\
 & & 23L & \cellcolor{myred!90} \textcolor{black} {0.00} & \cellcolor{myred!90} \textcolor{black} {0.00} & \cellcolor{myred!90} \textcolor{black} {0.00} & \cellcolor{myred!90} \textcolor{black} {0.00} & \cellcolor{myred!90} \textcolor{black} {0.00} & \cellcolor{myred!90} \textcolor{black} {4.18} & \cellcolor{myred!90} \textcolor{black} {0.00} & \cellcolor{myred!90} \textcolor{black} {0.00} \\
\hhline{~----------}
\hhline{~----------}
 & \multirow{3}{*}{$\rm F_{total}~(\%)$} & 23S & \cellcolor{myblue!90} \textcolor{white} {4.65} & \cellcolor{myblue!90} \textcolor{white} {2.72} & \cellcolor{myblue!90} \textcolor{white} {3.46} & \cellcolor{myblue!90} \textcolor{white} {3.55} & \cellcolor{myblue!90} \textcolor{white} {3.16} & \cellcolor{myblue!90} \textcolor{white} {4.50} & \cellcolor{myblue!90} \textcolor{white} {4.12} & \cellcolor{myblue!90} \textcolor{white} {3.66} \\
 & & 23M & \cellcolor{mygreen!100} \textcolor{black} {2.68} & \cellcolor{mygreen!100} \textcolor{black} {1.45} & \cellcolor{mygreen!100} \textcolor{black} {1.91} & \cellcolor{mygreen!100} \textcolor{black} {2.02} & \cellcolor{mygreen!100} \textcolor{black} {1.90} & \cellcolor{mygreen!100} \textcolor{black} {2.41} & \cellcolor{mygreen!100} \textcolor{black} {2.35} & \cellcolor{mygreen!100} \textcolor{black} {2.09} \\
 & & 23L & \cellcolor{myred!90} \textcolor{black} {1.92} & \cellcolor{myred!90} \textcolor{black} {1.02} & \cellcolor{myred!90} \textcolor{black} {1.40} & \cellcolor{myred!90} \textcolor{black} {1.44} & \cellcolor{myred!90} \textcolor{black} {1.59} & \cellcolor{myred!90} \textcolor{black} {2.14} & \cellcolor{myred!90} \textcolor{black} {1.68} & \cellcolor{myred!90} \textcolor{black} {1.47} \\
\hhline{~----------}
\hhline{~----------}
 & \multirow{3}{*}{$\rm F_{inter}~(\%)$} & 23S & \cellcolor{myblue!90} \textcolor{white} {nan} & \cellcolor{myblue!90} \textcolor{white} {nan} & \cellcolor{myblue!90} \textcolor{white} {15.52} & \cellcolor{myblue!90} \textcolor{white} {nan} & \cellcolor{myblue!90} \textcolor{white} {21.66} & \cellcolor{myblue!90} \textcolor{white} {21.19} & \cellcolor{myblue!90} \textcolor{white} {nan} & \cellcolor{myblue!90} \textcolor{white} {nan} \\
 & & 23M & \cellcolor{mygreen!100} \textcolor{black} {nan} & \cellcolor{mygreen!100} \textcolor{black} {nan} & \cellcolor{mygreen!100} \textcolor{black} {9.25} & \cellcolor{mygreen!100} \textcolor{black} {nan} & \cellcolor{mygreen!100} \textcolor{black} {13.14} & \cellcolor{mygreen!100} \textcolor{black} {11.17} & \cellcolor{mygreen!100} \textcolor{black} {nan} & \cellcolor{mygreen!100} \textcolor{black} {nan} \\
 & & 23L & \cellcolor{myred!90} \textcolor{black} {nan} & \cellcolor{myred!90} \textcolor{black} {nan} & \cellcolor{myred!90} \textcolor{black} {7.44} & \cellcolor{myred!90} \textcolor{black} {nan} & \cellcolor{myred!90} \textcolor{black} {11.29} & \cellcolor{myred!90} \textcolor{black} {9.99} & \cellcolor{myred!90} \textcolor{black} {nan} & \cellcolor{myred!90} \textcolor{black} {nan} \\
\hline
\hline
\multirow{9}{*}{\rotatebox[origin=c]{90}{RMSE}} & \multirow{3}{*}{\makecell{$\rm E$\\$(\rm meV/at)$}} & 23S & \cellcolor{myblue!90} \textcolor{white} {2.74} & \cellcolor{myblue!90} \textcolor{white} {1.47} & \cellcolor{myblue!90} \textcolor{white} {1.51} & \cellcolor{myblue!90} \textcolor{white} {1.26} & \cellcolor{myblue!90} \textcolor{white} {1.98} & \cellcolor{myblue!90} \textcolor{white} {2.07} & \cellcolor{myblue!90} \textcolor{white} {1.28} & \cellcolor{myblue!90} \textcolor{white} {1.40} \\
 & & 23M & \cellcolor{mygreen!100} \textcolor{black} {2.02} & \cellcolor{mygreen!100} \textcolor{black} {0.90} & \cellcolor{mygreen!100} \textcolor{black} {0.91} & \cellcolor{mygreen!100} \textcolor{black} {0.85} & \cellcolor{mygreen!100} \textcolor{black} {1.55} & \cellcolor{mygreen!100} \textcolor{black} {0.99} & \cellcolor{mygreen!100} \textcolor{black} {0.89} & \cellcolor{mygreen!100} \textcolor{black} {0.75} \\
 & & 23L & \cellcolor{myred!90} \textcolor{black} {2.48} & \cellcolor{myred!90} \textcolor{black} {0.84} & \cellcolor{myred!90} \textcolor{black} {0.87} & \cellcolor{myred!90} \textcolor{black} {0.70} & \cellcolor{myred!90} \textcolor{black} {1.32} & \cellcolor{myred!90} \textcolor{black} {0.99} & \cellcolor{myred!90} \textcolor{black} {0.58} & \cellcolor{myred!90} \textcolor{black} {0.49} \\
\hhline{~----------}
\hhline{~----------}
 & \multirow{3}{*}{\makecell{$\rm F_{total}$\\$(\rm meV/\AA)$}} & 23S & \cellcolor{myblue!90} \textcolor{white} {61.83} & \cellcolor{myblue!90} \textcolor{white} {26.15} & \cellcolor{myblue!90} \textcolor{white} {28.48} & \cellcolor{myblue!90} \textcolor{white} {36.97} & \cellcolor{myblue!90} \textcolor{white} {53.55} & \cellcolor{myblue!90} \textcolor{white} {39.33} & \cellcolor{myblue!90} \textcolor{white} {62.46} & \cellcolor{myblue!90} \textcolor{white} {71.41} \\
 & & 23M & \cellcolor{mygreen!100} \textcolor{black} {40.51} & \cellcolor{mygreen!100} \textcolor{black} {14.31} & \cellcolor{mygreen!100} \textcolor{black} {16.81} & \cellcolor{mygreen!100} \textcolor{black} {22.25} & \cellcolor{mygreen!100} \textcolor{black} {32.19} & \cellcolor{mygreen!100} \textcolor{black} {21.40} & \cellcolor{mygreen!100} \textcolor{black} {36.73} & \cellcolor{mygreen!100} \textcolor{black} {33.94} \\
 & & 23L & \cellcolor{myred!90} \textcolor{black} {33.34} & \cellcolor{myred!90} \textcolor{black} {10.27} & \cellcolor{myred!90} \textcolor{black} {12.81} & \cellcolor{myred!90} \textcolor{black} {16.19} & \cellcolor{myred!90} \textcolor{black} {26.91} & \cellcolor{myred!90} \textcolor{black} {18.78} & \cellcolor{myred!90} \textcolor{black} {27.17} & \cellcolor{myred!90} \textcolor{black} {22.46} \\
\hhline{~----------}
\hhline{~----------}
 & \multirow{3}{*}{\makecell{$\rm F_{inter}$\\$(\rm meV/\AA)$}} & 23S & \cellcolor{myblue!90} \textcolor{white} {0.78} & \cellcolor{myblue!90} \textcolor{white} {0.44} & \cellcolor{myblue!90} \textcolor{white} {8.70} & \cellcolor{myblue!90} \textcolor{white} {0.13} & \cellcolor{myblue!90} \textcolor{white} {27.11} & \cellcolor{myblue!90} \textcolor{white} {27.59} & \cellcolor{myblue!90} \textcolor{white} {0.17} & \cellcolor{myblue!90} \textcolor{white} {0.12} \\
 & & 23M & \cellcolor{mygreen!100} \textcolor{black} {0.79} & \cellcolor{mygreen!100} \textcolor{black} {0.44} & \cellcolor{mygreen!100} \textcolor{black} {5.54} & \cellcolor{mygreen!100} \textcolor{black} {0.13} & \cellcolor{mygreen!100} \textcolor{black} {16.46} & \cellcolor{mygreen!100} \textcolor{black} {15.05} & \cellcolor{mygreen!100} \textcolor{black} {0.17} & \cellcolor{mygreen!100} \textcolor{black} {0.12} \\
 & & 23L & \cellcolor{myred!90} \textcolor{black} {0.76} & \cellcolor{myred!90} \textcolor{black} {0.44} & \cellcolor{myred!90} \textcolor{black} {4.34} & \cellcolor{myred!90} \textcolor{black} {0.13} & \cellcolor{myred!90} \textcolor{black} {14.24} & \cellcolor{myred!90} \textcolor{black} {13.39} & \cellcolor{myred!90} \textcolor{black} {0.17} & \cellcolor{myred!90} \textcolor{black} {0.12} \\
\hline
\hline
\multirow{9}{*}{\rotatebox[origin=c]{90}{Relative RMSE}} & \multirow{3}{*}{$\rm E~(\%)$} & 23S & \cellcolor{myblue!90} \textcolor{white} {0.00} & \cellcolor{myblue!90} \textcolor{white} {0.00} & \cellcolor{myblue!90} \textcolor{white} {0.00} & \cellcolor{myblue!90} \textcolor{white} {0.00} & \cellcolor{myblue!90} \textcolor{white} {0.01} & \cellcolor{myblue!90} \textcolor{white} {8.30} & \cellcolor{myblue!90} \textcolor{white} {0.00} & \cellcolor{myblue!90} \textcolor{white} {0.00} \\
 & & 23M & \cellcolor{mygreen!100} \textcolor{black} {0.00} & \cellcolor{mygreen!100} \textcolor{black} {0.00} & \cellcolor{mygreen!100} \textcolor{black} {0.00} & \cellcolor{mygreen!100} \textcolor{black} {0.00} & \cellcolor{mygreen!100} \textcolor{black} {0.00} & \cellcolor{mygreen!100} \textcolor{black} {3.96} & \cellcolor{mygreen!100} \textcolor{black} {0.00} & \cellcolor{mygreen!100} \textcolor{black} {0.00} \\
 & & 23L & \cellcolor{myred!90} \textcolor{black} {0.00} & \cellcolor{myred!90} \textcolor{black} {0.00} & \cellcolor{myred!90} \textcolor{black} {0.00} & \cellcolor{myred!90} \textcolor{black} {0.00} & \cellcolor{myred!90} \textcolor{black} {0.00} & \cellcolor{myred!90} \textcolor{black} {3.98} & \cellcolor{myred!90} \textcolor{black} {0.00} & \cellcolor{myred!90} \textcolor{black} {0.00} \\
\hhline{~----------}
\hhline{~----------}
 & \multirow{3}{*}{$\rm F_{total}~(\%)$} & 23S & \cellcolor{myblue!90} \textcolor{white} {5.56} & \cellcolor{myblue!90} \textcolor{white} {2.79} & \cellcolor{myblue!90} \textcolor{white} {4.13} & \cellcolor{myblue!90} \textcolor{white} {3.63} & \cellcolor{myblue!90} \textcolor{white} {3.47} & \cellcolor{myblue!90} \textcolor{white} {4.58} & \cellcolor{myblue!90} \textcolor{white} {4.66} & \cellcolor{myblue!90} \textcolor{white} {5.45} \\
 & & 23M & \cellcolor{mygreen!100} \textcolor{black} {3.64} & \cellcolor{mygreen!100} \textcolor{black} {1.52} & \cellcolor{mygreen!100} \textcolor{black} {2.44} & \cellcolor{mygreen!100} \textcolor{black} {2.19} & \cellcolor{mygreen!100} \textcolor{black} {2.09} & \cellcolor{mygreen!100} \textcolor{black} {2.49} & \cellcolor{mygreen!100} \textcolor{black} {2.74} & \cellcolor{mygreen!100} \textcolor{black} {2.59} \\
 & & 23L & \cellcolor{myred!90} \textcolor{black} {3.00} & \cellcolor{myred!90} \textcolor{black} {1.09} & \cellcolor{myred!90} \textcolor{black} {1.86} & \cellcolor{myred!90} \textcolor{black} {1.59} & \cellcolor{myred!90} \textcolor{black} {1.74} & \cellcolor{myred!90} \textcolor{black} {2.19} & \cellcolor{myred!90} \textcolor{black} {2.03} & \cellcolor{myred!90} \textcolor{black} {1.71} \\
\hhline{~----------}
\hhline{~----------}
 & \multirow{3}{*}{$\rm F_{inter}~(\%)$} & 23S & \cellcolor{myblue!90} \textcolor{white} {nan} & \cellcolor{myblue!90} \textcolor{white} {nan} & \cellcolor{myblue!90} \textcolor{white} {8.95} & \cellcolor{myblue!90} \textcolor{white} {nan} & \cellcolor{myblue!90} \textcolor{white} {20.17} & \cellcolor{myblue!90} \textcolor{white} {20.18} & \cellcolor{myblue!90} \textcolor{white} {nan} & \cellcolor{myblue!90} \textcolor{white} {nan} \\
 & & 23M & \cellcolor{mygreen!100} \textcolor{black} {nan} & \cellcolor{mygreen!100} \textcolor{black} {nan} & \cellcolor{mygreen!100} \textcolor{black} {5.70} & \cellcolor{mygreen!100} \textcolor{black} {nan} & \cellcolor{mygreen!100} \textcolor{black} {12.25} & \cellcolor{mygreen!100} \textcolor{black} {11.00} & \cellcolor{mygreen!100} \textcolor{black} {nan} & \cellcolor{mygreen!100} \textcolor{black} {nan} \\
 & & 23L & \cellcolor{myred!90} \textcolor{black} {nan} & \cellcolor{myred!90} \textcolor{black} {nan} & \cellcolor{myred!90} \textcolor{black} {4.46} & \cellcolor{myred!90} \textcolor{black} {nan} & \cellcolor{myred!90} \textcolor{black} {10.60} & \cellcolor{myred!90} \textcolor{black} {9.79} & \cellcolor{myred!90} \textcolor{black} {nan} & \cellcolor{myred!90} \textcolor{black} {nan} \\
\hhline{-----------}
\hhline{-----------}
\hhline{-----------}
\end{tabular}
\end{center}
\end{table*}

\begin{enumerate}
    \item identify molecules (labeled $j$)
    \item within each molecule $j$ sum over all atomic forces (labeled $k$) to obtain the translational component:
    \begin{equation}
        F^{\rm trans}_j = \sum_{k \in j} f_{k}
    \end{equation}
    \item redistribute the molecular translational force onto individual atoms (labeled $i$) to obtain the atomic translational contributions:
    \begin{equation}
        f^{\rm trans}_i = \frac{m_i}{M_j} F^{\rm trans}_j
    \end{equation}
    where $m_i$ are atomic masses and $M_j$ are molecular masses
    \item similarly, compute the torque on the entire molecule:
    \begin{equation}
        T_j = \sum_{k \in j} f_{k} \times \left(r_{k}-R^{\rm com}_j\right)
    \end{equation}
    \item compute the rotational atomic force contributions that give rise to the given molecular torque:
    \begin{equation}
        f^{\rm rot}_i = m_i \left(r_{i}-R^{\rm com}_j\right) \times (I_j^{\alpha \beta})^{-1} T_j
    \end{equation}
    where $I_j^{\alpha \beta}$ are the molecular moments of inertia
    \item compute the vibrational contribution as the difference:
    \begin{equation}
        f^{\rm vib}_i = f_i - f^{\rm trans}_i - f^{\rm rot}_i
    \end{equation}
    \item the intermolecular force is calculated as the sum of the translational and rotational contributions.
\end{enumerate}

Relative errors were obtained by dividing the absolute errors by the typical DFT force magnitudes:

\begin{equation}
        \mathrm{Rel~MAE} = \frac{\sum_i |f^{DFT}_i-f^{MACE}_i|}{\sum_i |f^{DFT}_i-{\bar f^{DFT}} |}
\end{equation}

\begin{equation}
        \mathrm{Rel~RMSE} = \sqrt{\frac{\sum_i (f^{DFT}_i-f^{MACE}_i)^2}{\sum_i (f^{DFT}_i-{\bar f^{DFT}})^2}}
\end{equation}
where $\bar f^{DFT} = \sum_i f_i^{DFT}$/N is the average DFT force (often zero).

Table \ref{tab:test_errors} shows these errors computed for all three MACE models on each category of atomic configurations found in the test set. Interestingly, even though PubChem, DES370K Monomers, Dipeptides, QMugs and Tripeptides comprise only isolated molecule configurations, we find a small error that is incorrectly attributed to intermolecular interactions. This issue arises because DFT forces in these test data to not obey translational and rotational symmetry, i.e. total molecular force and torque do not sum to zero, even though a single molecule is present in each simulation box. Meanwhile, MACE forces do obey these symmetries, and therefore the difference shows up as an intermolecular error which is identical for all three MACE models.

The code used for this analysis is available at \url{https://github.com/imagdau/aseMolec.git}.

\newpage

\subsection{MACE-OFF24(M) Test Errors}

\begin{table}[]
\begin{tabular}{lcc}
\toprule
                     & MAE Energy (meV/atom)   &   MAE Force (meV/\AA) \\
 \toprule
PubChem              & 1.0          & 22.1        \\
DES370K Monomers     & 0.6          & 9.6         \\
DES370K Dimers       & 0.6          & 9.3         \\
Dipeptides           & 0.5          & 14.4        \\
QMugs                & 0.8          & 23.8        \\
Solvated Amino Acids & 1.3          & 23.2        \\
Amino Acid-Ligand    & 1.5          & 24.2        \\
Solvated PubChem     & 1.2          & 22.8        \\
\bottomrule
\caption{Test set errors for MACE-OFF24(M) }
\end{tabular}
\end{table}

\clearpage


\section{Torsion angle scanning calculations}
The TorsionNet-500~\cite{rai2022torsionnet} and biaryl torsion~\cite{lahey2020biaryl_torsion} benchmark datasets were recomputed at the SPICE~\cite{eastman2023spice} level of QM theory (\textomega B97M-D3(BJ)/def2-TZVPPD~\cite{najibi2018nonlocal, weigend2005balanced, rappoport2010property, grimme2011bj_damp, grimme2010d3}) to assess the accuracy of the MACE models in predicting the potential energy surfaces of rotatable bonds commonly found in drug-like molecules. Starting from the published optimized geometries, a torsion scan was performed using TorsionDrive~\cite{qiu2020torsiondrive} via its interface with QCEngine~\cite{smith2021qcengine}. The dihedral angles were scanned in 15$^{\circ}$ increments and each constrained geometry optimization was carried out with geomeTRIC~\cite{wang2016geometry}, using PSI4~\cite{smith2020psi4} to calculate the DFT energies and forces. For each of the assessed force field models, a separate constrained geometry optimization was performed starting from the DFT reference geometries and holding the target torsion fixed at each grid point value. The final RMSD between the optimized and reference geometry was recorded, along with the energy to calculate the potential energy surfaces.



\clearpage

\section{Lattice enthalpies}

First, the internal energy of the crystalline form was computed under the harmonic approximation:
\begin{align}
\label{eq:cryst_internal_U}
    U(T) &= \left(\frac{\partial \ln{Z}}{\partial \frac{1}{k_B T}}\right)\\
    &= E_{\text{MACE}} + \int_{0}^{\infty} \left[ \frac{\epsilon}{e^{\frac{\epsilon}{k_B T}} - 1} + \frac{\epsilon}{2}\right]\sigma(\epsilon) \text{d}\epsilon \notag
\end{align}
where $Z$ denotes the partition function of the system, and $\sigma(\epsilon)$ represents the degeneracy or phonon density of states as a function of vibrational energy. 
The above expression contains an electronic contribution, computed using MACE, a zero point energy contribution, and a contribution from finite-temperature vibrational energy. To compute the vibrational contribution, we used a $3\times 3\times 3$ supercell for all the molecules considered.  For the gaseous form of the molecules, we employed the ideal gas approximation to compute the enthalpies of the gas as:
\begin{equation}
\label{eq:gas_enthalphy}
    U(T) = e_{\text{MACE}} + E_{\text{ZPE}} + \int_{0}^{T} C_p \text{d}T
\end{equation}
where we used the ideal gas value to evaluate the last term, giving $4RT$ for all system except for the linear carbon-dioxide, where the value is $3.5RT$. 

Combining the above two expressions, we can compute the sublimation enthalpies as:
\begin{align}
\label{eq:subm_H}
    \Delta H_{\text{subm}} &= U_{\text{gas}}(T) - U_{\text{cryst}}(T) \\
        &= E_{\text{latt}} + \Delta E_{\text{vib}} + 4RT \notag
\end{align}

To assess the accuracy of the MACE model we used the X23 set of molecular crystals~\cite{reilly2013understanding}. We first relaxed the cells, followed by phonon and normal mode calculations to obtain the finite temperature vibrational as well as zero point contributions to the enthalpies, following the protocol above. It is important to note that for the ANI-2x force field it was not possible to fully relax the crystal geometries, because the model cannot predict stresses. Hence, we kept the cell fixed at the DFT equilibrium values from Ref.~\cite{reilly2013understanding}. The results are summarized in Table~\ref{tab:X23}. 

\begin{table}[H]
\centering
  \caption{
    \textbf{Sublimation enthalphy in kJ/mol} 
  }
\resizebox{0.45\textwidth}{!}
     {%
\begin{tabular}{m{3.8cm} c c c c c}
\toprule
    System        &  $\Delta H_{\text{sub}}^{0}$ exp. & $\Delta H_{\text{sub}}^{0}$ MACE-OFF23(S) & $\Delta H_{\text{sub}}^{0}$ MACE-OFF23(M) &  $\Delta H_{\text{sub}}^{0}$ MACE-OFF23(L) & $\Delta H_{\text{sub}}^{0}$ ANI-2x  \\
\midrule
    1,4-Cyclohexanedione    &  81.1   &   85.1&  77.7     &  85.1   &  59.7   \\[2pt]
    Acetic acid             &  67.7   &   77.2 &  71.4      &  70.3   &  90.7   \\[2pt]    
    Adamantane              &  61.6   &   88.7  &  65.6     & 69.2   &  67.1   \\[2pt]
    Ammonia                 &  31.2   &    24.4 &   30.9    &26.8  &  23.0   \\[2pt]
    Anthracene              &  101.9  & 118.8    &  95.7     &106.3 &  98.9   \\[2pt]
    Benzene                 &  44.9   &   49.4  &  48.1     &   58.4  &  48.3   \\[2pt]
    Carbon dioxide          &  26.1   &    -35.2 &  -1.44     &4.1  &  -117.9 \\[2pt]
    Cyanamide               &  75.5   &    85.8 &  94.5     &   87.3  &  36.9   \\[2pt]
    Cytosine                &  156.4  &  159.8 & 142.8    &143.2 &  118.6  \\[2pt]
    Ethyl carbamate         &  78.8   &  88.7 &   90.9  &91.4  &  77.0   \\[2pt]
    Formamide               &  71.7   &  82.2 &  75.0   &80.1  &  67.3   \\[2pt]
    Hexamine                &  75.8   &  58.4 &  89.5   &62.3  &  83.5   \\[2pt]
    Imidazole               &  81.4   &  88.7 &  91.6   &86.3  &  86.6   \\[2pt]
    Naphthalene             &  72.6   &  88.9 &  80.1   &83.6  &  72.4   \\[2pt]
    Oxalic acid \textalpha  &  93.7   &  98.5 &  89.5   &89.7  &  44.1   \\[2pt]
    Oxalic acid \textbeta   &  93.6   &  90.0 &  95.2   &93.5  &  45.4   \\[2pt]
    Pyrazine                &  56.3   &  86.4 &  70.0   &69.7  &  53.7   \\[2pt]
    Pyrazole                &  72.4   &  67.5 &  78.6   &73.0  &  75.4   \\[2pt]
    \textit{s}-Triazine     &  55.7   &  58.4 &  64.9   &75.6  &  112.1  \\[2pt]
    \textit{s}-Trioxane     &  56.3   &  75.8 &  59.9   &53.9  &  51.7   \\[2pt]
    Succinic acid           &  123.1   & 132.5 & 119.7    &123.5 &  141.7  \\[2pt]
    Uracil                  &  129.2   & 139.7 & 127.2    &131.8 &  98.9   \\[2pt]
    Urea                    &  93.8   &  121.5 & 114.0    &112.5 &  78.2   \\[2pt]
    \bottomrule
\end{tabular}
}
\label{tab:X23}
\end{table}

\clearpage

\section{Crystal structure geometries}
In Table~\ref{tab:X23_geoms} we compare the MACE-OFF23(L) relaxed crystal structure lattice parameters with the experimental ones reported in Ref.~\cite{dolgonos2019revised_x23}. We confirm that MACE is able to accurately reproduce the lattice constant of molecular crystals purely trained on molecular dimers. Additional trimer training data, or a higher level of reference theory, such as coupled cluster, could probably further improve the agreement with experiments. 
\begin{table}[H]
\centering
  \caption{
    \textbf{Crystal structure geometries}  
  }
\resizebox{0.45\textwidth}{!}
     {%
\begin{tabular}{m{2.5cm} c c c c }
\toprule
    Molecule        &  $T_{\text{exp}}$ &  & Experiment & MACE-OFF23(L)   \\
\midrule
    Acetic acid             &  40   & $a$ & 13.151  &  13.359      \\
                            &       & $b$ & 3.923  &  3.809     \\ 
                            &       & $c$ & 5.762  &  5.542    \\
                            &       & $V$ & 297.27  & 282.01      \\
    Ammonia                 &  2    & $a$ & 5.048  & 4.946      \\
                            &       & $V$ & 128.63  & 120.98      \\
    Benzene                 &  4    & $a$ & 7.351  &  6.59      \\
                            &       & $b$ & 9.364  &  9.303      \\ 
                            &       & $c$ & 6.695  &  6.851    \\
                            &       & $V$ & 460.84 &  420.51     \\
    Naphthalene             &  10   & $a$ & 8.0846  & 7.821      \\
                            &       & $b$ & 5.9375  & 5.836  \\ 
                            &       & $c$ & 8.6335  & 8.430      \\
                            &       & $\beta$ & 124.67 & 125.26   \\
                            &       & $V$ & 340.83  &  314.23     \\
    Pyrazine                & 184   & $a$ & 9.325   & 9.351      \\
                            &       & $b$ & 5.850   & 5.508      \\ 
                            &       & $c$ & 3.733   & 3.545       \\
                            &       & $V$ & 203.64  & 182.57      \\
    Urea                    &  40   & $a$ & 5.565   &  5.330      \\
                            &       & $c$ &  4.684  & 4.670      \\
                            &       & $V$ &  145.06 & 123.68      \\
    \bottomrule
\end{tabular}
}
\label{tab:X23_geoms}
\end{table}

\clearpage

\section{Condensed phase simulations}

Water density simulations were seeded using OpenMM's Modeller functionality, with a padding of 12.5~\AA{} from the central water molecule to the box edge, resulting in box edge of 25~\AA{}.  Simulations were run for 500\,000 steps with a 1~fs timestep, resulting in a 500~ps trajectory. The density was recorded every 100 steps. After an initial 100~ps period of equilibration, the density was calculated by averaging over the remaining 400~ps of trajectory.

Initial PDB files for the remaining organic liquid simulations were taken from Ref.~\cite{horton2019qubekit} and periodic boxes were generated using packmol \cite{martinezPACKMOLPackageBuilding2009a}.  Boxes were prepared containing 64 molecules, with box vectors determined such that the initial density was 80\% of the experimental density.  Structures were minimized with the L-BFGS algorithm prior to MD. Molecular dynamics was performed with OpenMM, using a custom fork of the \texttt{openmm-ml} package to interface MACE models to the MD code.\footnote{https://github.com/jharrymoore/openmm-ml/tree/main}
Simulations were performed in the NPT ensemble. The Langevin equations of motion were integrated with a timestep of 1~fs and a Monte Carlo barostat was used to maintain pressure at 1~atm.  Temperatures were maintained at 298~K except for those compounds whose boiling point was below this value, for which a value of 10~K below the experimental boiling point was used. Dynamics were propagated for 300~ps, and the final density was calculated by averaging over the final 100~ps of the simulation.

Heat of vaporization calculations additionally required simulation of the isolated molecule in the vacuum phase, which were carried out on a nonperiodic molecule.  Final energies for both condensed and vacuum phases were calculated by averaging over the final 100~ps of each trajectory.


Tables~\ref{sitab:denserrors} and~\ref{sitab:dhvaperrors} summarize the overall density and heat of vaporization errors for the MACE-OFF23(S) and MACE-OFF23(M) models.
Although the computational expense of the MACE-OFF23(L) is prohibitive for these condensed phase simulations, we nonetheless benchmarked the large model on a small selection of condensed phase systems.  We saw comparable performance to the medium model, and in particular observed that compounds that had a high prediction error under the medium model had a similar error under the large model. This indicates that this error is driven primarily by the dataset rather than by the size of the model, as previously hypothesized.

To investigate this further, Tables~\ref{sitab:denserrors} and~\ref{sitab:dhvaperrors} additionally show statistics for MACE-OFF24(M), which employs an extended cutoff of 6~\AA{} and has also been trained on additional nonbonded data. We observe a modest decrease in MAE for the heat of vaporization from 2.18 to 1.75~kcal/mol, as well as a small decrease in the accuracy of liquid density predictions. This further reinforces the hypothesis that the poorly predicted densities are a result of under-representation of the functional groups, since the additional SPICE training sets do little to increase sampling of these problematic functional groups.




\begin{table}[h!]
    \centering
    \begin{tabular}{@{}lccc@{}}
        \toprule
         & MAE / g/cm$^3$ & RMSE / g/cm$^3$ & $r$ \\
        \toprule
        MACE-OFF23(S) & 0.23 & 0.38 & 0.20\\
        MACE-OFF23(M) & 0.09 & 0.15 & 0.89 \\
        MACE-OFF24(M) & 0.14 & 0.34 & 0.26 \\
        ANI-2x & 0.21 & 0.36 & 0.26 \\
        \bottomrule
    \end{tabular}
    \caption{\textbf{Summary of density errors}}
        \label{sitab:denserrors}
\end{table}

\begin{table}[h!]
    \centering
    \begin{tabular}{@{}lccc@{}}
        \toprule
        &MAE / kcal/mol & RMSE / kcal/mol & $r$ \\
        \toprule
        MACE-OFF23(S) & 1.48 & 2.19 & 0.86 \\
        MACE-OFF23(M) & 2.18 & 2.53 & 0.87 \\
        MACE-OFF24(M) & 1.75 & 2.81 & 0.78 \\
        ANI-2x & 2.75 & 3.45 & 0.66 \\
        \bottomrule
    \end{tabular}
    \caption{\textbf{Summary of $\Delta H_{vap}$ errors}}
    \label{sitab:dhvaperrors}
\end{table}


\clearpage

\section{Biological Simulations}

\subsection{Ala\textsubscript{3} Free Energy Surface}

Molecular dynamics simulations were performed with OpenMM, as in the previous section. Enhanced sampling simulations were performed with metadynamics. Two collective variables were defined by the two torsion angles on the backbone of the central alanine residue. A 1~ns simulation was performed with a temperature bias factor of 10, a barrier height of 1~kJ/mol and a frequency of 100 steps.

\subsection{Ala\textsubscript{15} peptide folding}

The initial extended structure was generated using PyMOL. The system was propagated with Langevin dybamics using a 1~fs timestep for 500~ps \textit{in vacuo} via the OpenMM interface. 

\subsubsection{Secondary structure prediction with MACE-OFF23(M)}
Compared to MACE-OFF24(M) (see main text), the previous 5~\AA{} model failed to reproduce the experimentally observed oscillation between the \textalpha-helix and the 3\textsubscript{10} helix, instead predicting only the \textalpha-helical structure

\begin{figure}[H]
    \centering
    \includegraphics[width=\linewidth]{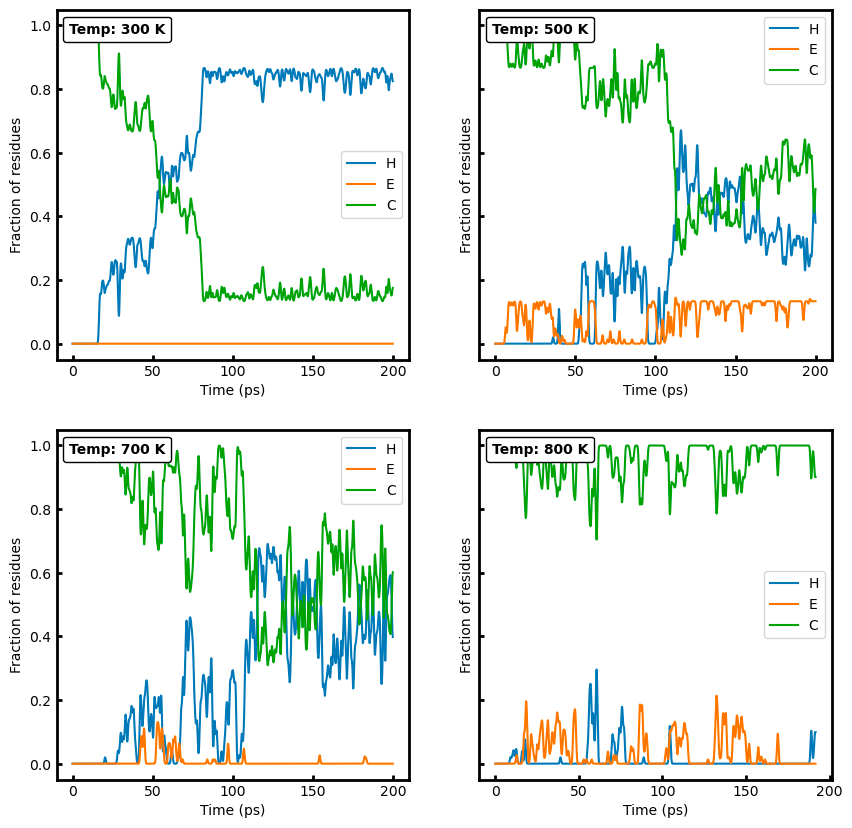}
    \caption{Secondary structure (and temperature dependence) of Ala\textsubscript{15} predicted by MACE-OFF23(M). Single letter codes correspond to standard STRIDE secondary structure labels, where H=\textalpha-helix, E=extended and C=random coil}
    \label{fig:enter-label}
\end{figure}

\subsection{Crambin in explicit solvent}

Crambin was prepared from the PDB structure 1EJG. The initial structure was prepared using pdbfixer~\cite{eastmanOpenMMRapidDevelopment2017} and solvated in an orthorombic box with padding between solute and the box edge set to 1.2~nm.  Dynamics were performed in the NPT emsemble, with pressure maintained at 1~atm with a Monte Carlo barostat, as implemented in OpenMM. A 1~fs timestep was used both for integrating the Langevin equations of motion and for writing the trajectory to disk for post processing. The power spectrum was calculated as the Fourier transform of the velocity autocorrelation function using the Travis program with default settings~\cite{brehmTRAVISFreeAnalyzer2011, brehmTRAVISFreeAnalyzer2020a, thomasComputingVibrationalSpectra2013a, thomasVoronoiDipoleMoments2015}.

\begin{figure}
    \centering
    \includegraphics[width=0.8\linewidth]{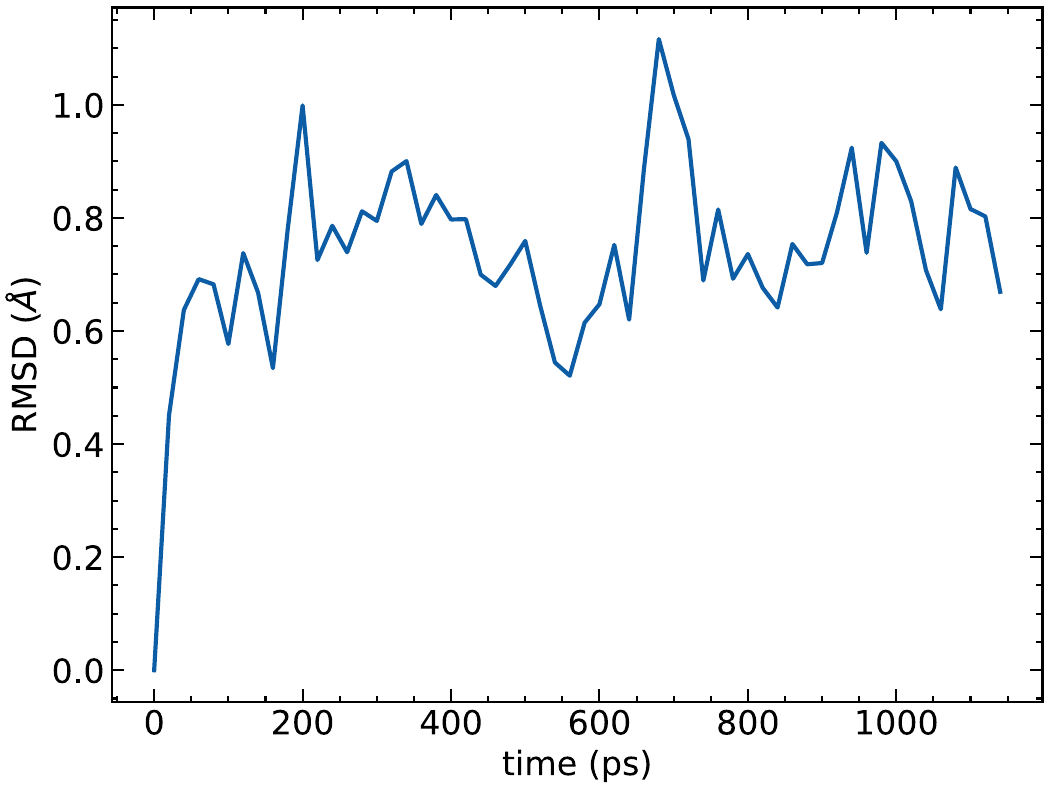}
    \caption{\textbf{Crambin RMSD.} RMSD calculated by averaging two halves of a 1 ns trajectory, relative to the first frame and computed on C$_\alpha$ atoms.}
    \label{sifig:crambin_rmsd}
\end{figure}


\clearpage







\clearpage
\section{Dipole models}
\subsection{MACE-OFF23-$\mu$}

In addition to the series of MACE-OFF force field models, we also developed MACE-OFF23-$\mu$, which predicts the total dipole moment of molecules and can be used, in combination with a regular MACE-OFF model for energies and forces, to estimate vibrational spectra based on the autocorrelation of the dipole moment.
The model architecture is identical to the MACE architecture described in the Methods in the main text, with the only difference being the readout function, which produces an $L=1$ tensor for each atom instead of the atomic site energies.
Given the equivariant order of the readout, the smallest possible model corresponds to the `medium' architecture, which is the one considered in this work.
Other than the readout's equivariant order, all hyperparameters match those of the MACE-OFF23(M) model.
MACE-OFF23-$\mu$ was trained on the subset of SPICE~\cite{eastman2023spice} for which a total dipole moment was available, resulting in the exclusion of the QMugs and water subsets.

\begin{figure}
    \centering
    \includegraphics[width=\columnwidth]{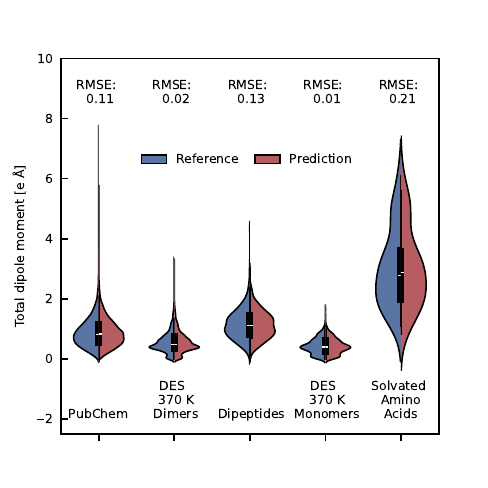}
    \caption{\textbf{Test set prediction and root mean square errors (RMSE).} Violin plots of the reference (blue) and predicted (red) distributions of the dipole moment predicted using MACE-OFF23-$\mu$.  The RMSEs are reported above the violin plots for the individual test sets.  
    }
    \label{fig:dipole_err}
\end{figure}
We assess the accuracy of the MACE-OFF23-$\mu$ model on the held-out SPICE test sets, evaluating its effectiveness for predicting dipole moments and infrared spectra of organic molecules.
Figure~\ref{fig:dipole_err} summarizes this performance by examining the alignment between the violin distributions of the reference and predicted total dipole moment values, as well as the RMSEs.
The small RMSEs and strong alignment of the distributions  indicate high fidelity across the test datasets for which reference dipole moment values were available.
These results suggest promising performance for zero-shot predictions of the dipole moments (and infrared spectra) of unseen organic molecules, as demonstrated in a recent benchmark on out-of-distribution drug-like molecules from the NIST database~\cite{Pracht2024}.
The MACE-OFF23-$\mu$ model is available at: \url{https://github.com/venkatkapil24/MACE-OFF-mu-models}.

\clearpage

\subsection{Computational details}

To compute the vibrational spectrum for the paracetamol form II polymorph, we first performed a 20\,ps path-integral MD simulation using the MACE-OFF23(S) model at 500~K, following the prescriptions of the Path Integral coarse-Grained Simulations (PIGS) approach~\cite{Kapil2023, Musil2022}. The path-integral MD simulations are performed with a 0.50~fs timestep using the BAOAB integration scheme extended to path-integral MD~\cite{Kapil2019} using the path-integral Langevin equation thermostats~\cite{Ceriotti2010} with a time constant of 100~fs.

To generate a potential energy surface that effectively encodes quantum nuclear effects, we followed the PIGS technique. We used the positions and forces on the centroid of the path-integral to fit a MACE model that represents the difference between the potential energy surface and the potential of mean force on the centroid at an elevated temperature of 500~K. The sum of MACE-OFF23(S) and the PIGS model gives a transferable effective potential energy surface that encodes quantum nuclear effects.

To generate a trajectory with quantum nuclear motion, we run MD with the sum of MACE-OFF23(S) and the PIGS model using the \texttt{i-PI}~\cite{ipi} software to propagate the equations of motion and the \texttt{ASE} calculator as the force-provider. The MD simulations were performed with a 0.50 fs timestep using a weak global stochastic velocity rescaling thermostat. To generate classical trajectories we perform MD with the same setup but without the PIGS potential.

We predicted the $L=0$ and $L=2$ spherical harmonic components of polarizability tensors on the trajectories and computed their time correlation functions to obtain the isotropic and anisotropic Raman spectra~\cite{Kapil2023}. We estimate the powder Raman spectrum as a linear combination of the isotropic and the anisotropic Raman spectra~\cite{raimbault2019parac_raman_AbInitio}.

The model architecture was identical to the MACE architecture used for the energy models, with the only difference being the readout function producing a vector and a tensor (dipole and polarisability) for each atom, instead of the atomic site energies. A relatively small MACE model is already capable of achieving high accuracy predictions, and the selected models used the hyperparameters displayed in Table~\ref{tab:mace_dielectric}.

\begin{table}[ht]
    \centering
    \caption{Hyperparameters of the MACE models for dipoles and polarizabilities}
    \label{tab:mace_dielectric}
    \begin{tabular}{lcc}
        \toprule
        & Paracetamol & Water  \\
        \toprule
        Cutoff radius (\AA) & 5.0 & 6.0 \\ \hline
        Chemical channels & \multicolumn{1}{c}{\multirow{2}{*}{16}} & \multicolumn{1}{c}{\multirow{2}{*}{32}} \\
        $k$ (Eq.~1) & & \\ 
        max L (Eq.~5) & 2 & 2 \\
        \bottomrule
    \end{tabular}
\end{table}

\subsection{Vibrational spectroscopy of paracetamol \label{ss:spectra_parac}}

Raman spectroscopy is one of the most widely used techniques for characterizing molecular crystals. Unlike IR spectroscopy, which only detects vibrational modes that distort dipoles, Raman spectroscopy is more sensitive to collective modes governed by weak, non-bonded interactions in a broad range of molecular materials. The low-frequency region of the Raman spectrum (e.g., the THz regime) gives a vibrational fingerprint of the intermolecular interactions. Thus, it is widely used to differentiate between polymorphs of molecular crystals. Meanwhile, the high-frequency Raman spectrum probes intramolecular modes and their coupling to low-frequency modes. \\

Here, we test the ability of the MACE-OFF23(S) model to predict the Raman spectrum of the ``Form II" polymorph of paracetamol. To compare MACE-OFF23(S) directly with experiments, we rigorously incorporate quantum nuclear and non-Condon effects using our ML-aided framework~\cite{Kapil2023}. We incorporate quantum nuclear effects by fitting an effective potential energy surface~\cite{Musil2022} (using MACE) to calculate quantum nuclear corrections to the MACE-OFF23(S) model within the path-integral coarse-grained simulations (PIGS) method. To incorporate non-Condon effects, we fit a separate equivariant MACE model to the first-principles polarizability of paracetamol polymorphs (data taken from Refs.~\citenum{Raimbault2019Using, raimbault2019paracetamol_SAGPR}). The remaining steps of our quantum nuclear simulations mimic an entirely classical calculation involving a $NVE$ molecular dynamics simulation, prediction of the isotropic and anisotropic components of the polarizability tensor, and calculation of their time correlation functions. For further details we refer the reader to Ref.~\citenum{Kapil2023}.\\ 

\begin{figure*}
    \centering
    \includegraphics[width=0.75\textwidth]{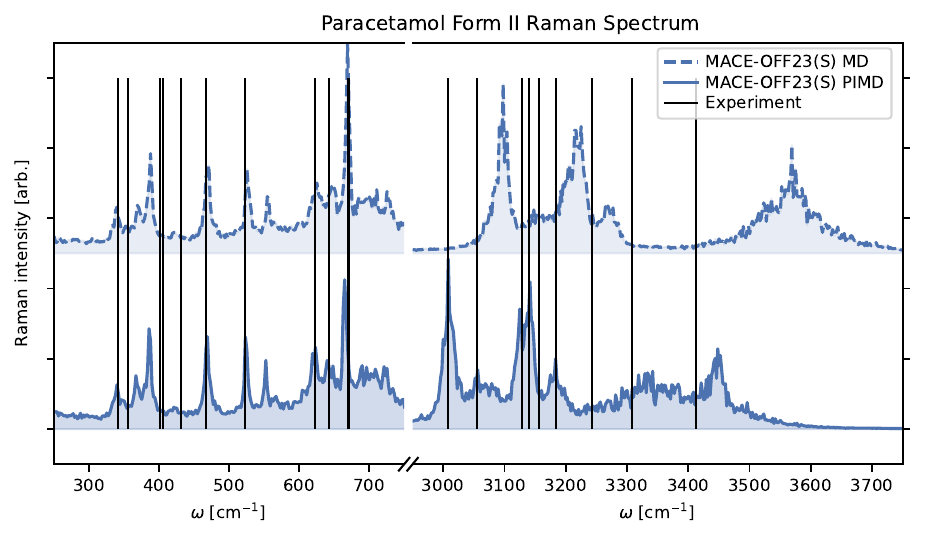}
    \caption{\textbf{Powder Raman spectrum of paracetamol form II~\cite{Raimbault2019Anharmonic}.} Spectrum computed at ambient conditions using the MACE-OFF23(S) model for the potential energy surface, a MACE model of the polarizability, and a MACE model that incorporates quantum nuclear effects on the potential energy surface using the PIGS approach~\cite{Musil2022}. The black lines represent experimentally determined band positions~\cite{Kolesov2011}, scaled by 3\% to aid visual comparison with the predicted spectra. We focus on high and low frequency features of the spectrum where experimental peaks are well separated.}
    \label{fig:mace_parac_Raman}
\end{figure*}

As shown in Figure~\ref{fig:mace_parac_Raman}, we predict both the high- and low-frequency regions of the Raman spectrum of paracetamol form II with an overall good agreement with the experimental band positions~\cite{Kolesov2011}. Since the experiment captures the Raman spectra along different crystal directions while we estimate the ``powder'' Raman spectrum~\cite{raimbault2019paracetamol_SAGPR}, we do not compare band intensities. We find that the classical predictions (based on MACE classical MD) are consistently shifted with respect to the experiment. At the same time, quantum nuclear predictions (encoded by the PIGS method) play an essential role in improving the agreement between theory and experiments. Moreover, we note that a broad band at around 3300~$\text{cm}^{-1}$ is only captured at the level incorporating quantum nuclear effects. The low-frequency modes also agree semi-quantitatively with the experiments and do not require a quantum nuclear description. The 2--3 \% overall shift in the vibrational frequencies notwithstanding, the MACE-OFF23(S) combined with quantum nuclear effects and non-Condon effects shows promising capabilities for the characterization of molecular crystals. 

\subsection{Vibrational spectroscopy of liquid water\label{ss:spectra_dipole_water}}
\begin{figure}
    \centering
    \includegraphics[width=1.0\linewidth]{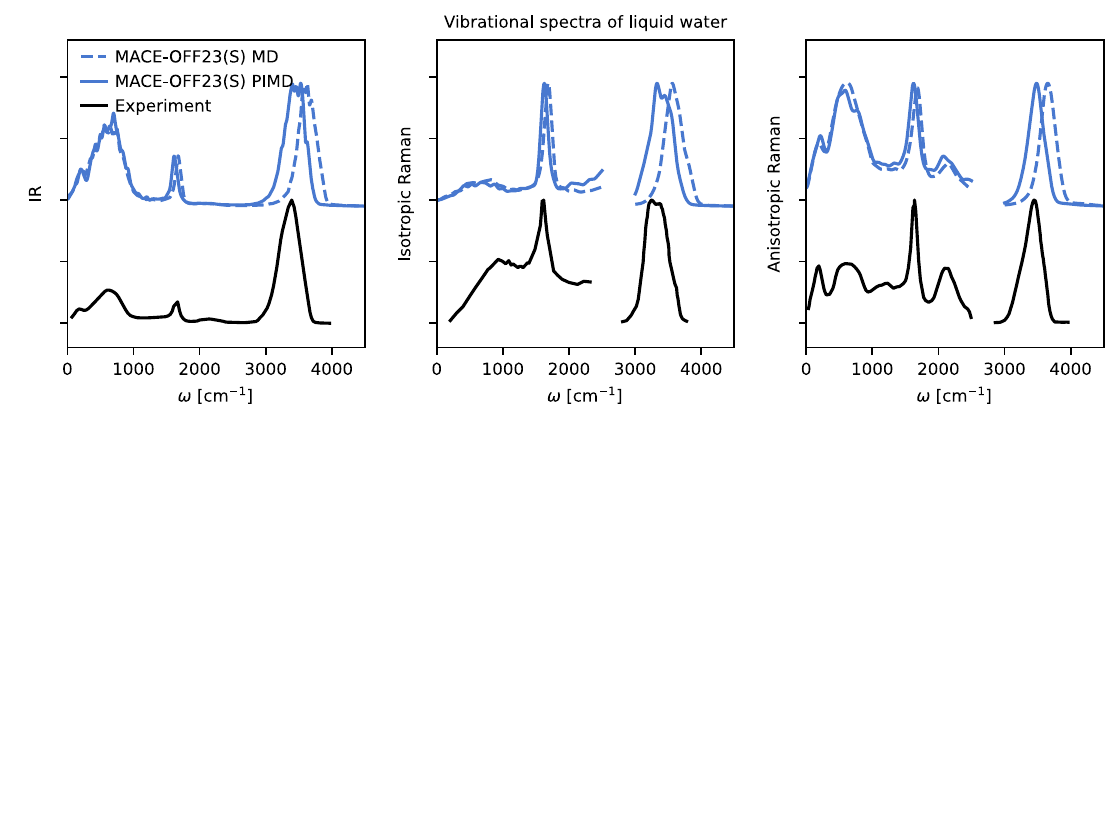}
    \caption{IR, isotropic Raman and anisotropic Raman spectra of water at ambient temperature and density using the MACE-OFF23(S) model for the potential energy surface and a single MACE model of water's dipole and polarizability~\cite{Kapil2023}. The black curve represents the experiment~\cite{morawietz2018}. The dashed blue curve is obtained from classical MD, while the solid curve incorporates quantum effects using the PIGS method~\cite{Musil2022}.}
    \label{fig:water_vib}
\end{figure}

Next, we characterize the dynamical properties of water by computing the vibrational density of states, including the IR and Raman spectra selection rules, using MACE models of dipoles and polarizabilities~\cite{Kapil2023}. Given the significant impact of quantum nuclear motion on the dynamics of water~\cite{Musil2022}, we incorporate quantum nuclear effects using the PIGS approach~\cite{Musil2022} (discussed in Section~\ref{ss:spectra_parac}), which has a classical cost and has recently been shown to describe quantum nuclear effects in water accurately~\cite{Ceriotti2016}. As shown in Figure~\ref{fig:water_vib}, after incorporating quantum nuclear effects, the MACE-OFF23(S) predictions agree with the experiments across the entire frequency range. We note the presence of an overall 2-3 \% blue shift with respect to the experiments, consistent with the spectra for paracetamol form II (see Section~\ref{ss:spectra_parac}). MACE-OFF23(S) even qualitatively captures subtle features of the spectra, including the 0 -- 1000~$\text{cm}^{-1}$ hydrogen bonding fingerprints in the anisotropic Raman spectrum, and the bimodal nature of the isotropic Raman stretching band arising from the presence of hydrogen-bonded defects in room temperature water.  

\clearpage

\section{Computational performance in LAMMPS}

\begin{figure*}[!ht]
    \centering
    \includegraphics[width=0.9\textwidth]{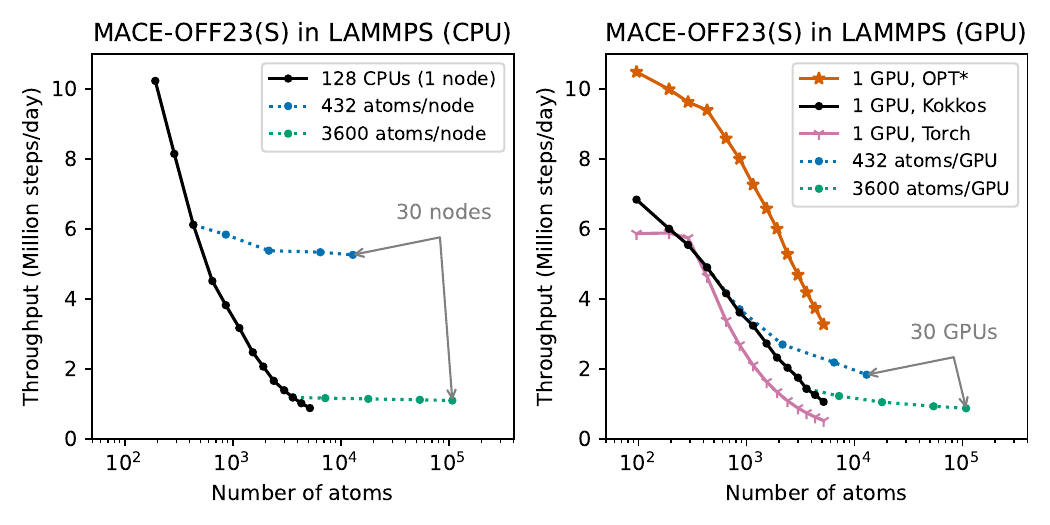}
    \caption{\textbf{Throughput in LAMMPS.} Performance of molecular dynamics in LAMMPS for water in the NVT ensemble at 1~g/cm$^3$ and 300~K with a 1 fs timestep. All simulations used variants of the MACE-OFF23(S) model.
    \newline\newline
    The left panel highlights CPU performance achievable on the UK-based Archer2 CPU supercomputer, using the new pure-C++ MACE implementation. The solid black curve demonstrates the throughput of a single 128-core node as a function of the number of atoms. The two dotted curves demonstrate weak scaling with 432 atoms/node or 3600 atoms/node. All calculations used double precision floating point numbers.
    \newline\newline
    The right panel demonstrates the performance of the GPU evaluators. The three solid curves provide single-GPU results (determined with NVIDIA A100 80GB GPUs) for the native Torch-based MACE-OFF23(S), a custom Kokkos-based MACE implementation, and the \texttt{cuda\_mace} implementation marked as OPT*. The asterisk indicates that these latter simulations used single precision floating point numbers, with error correction for the most numerically sensitive terms, in contrast to the remaining calculations which were performed with double precision. Finally, the two dashed curves demonstrate the performance when multiple GPUs are employed, using the pure-Kokkos implementation, to access larger systems with domain decomposition. More work is needed to improve this weak scaling.
    }
    \label{fig:lammps-throughput}
\end{figure*}

\begin{figure*}[!ht]
    \centering
    \includegraphics[width=0.9\textwidth]{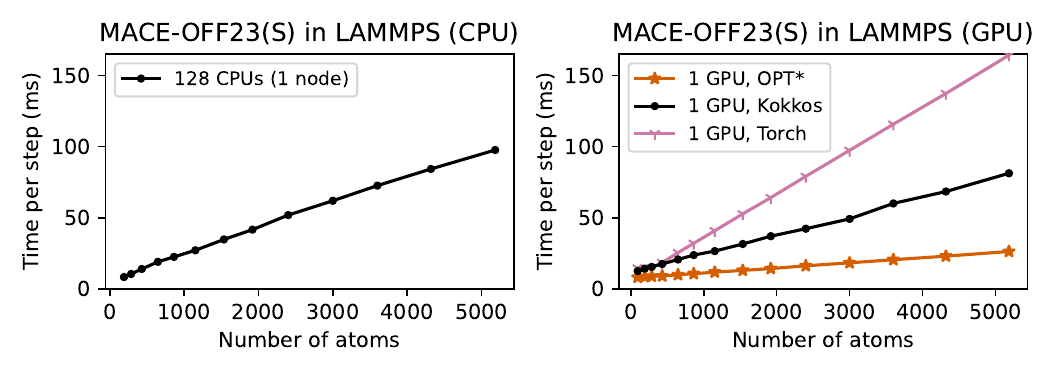}
    \caption{\textbf{Evaluation speed in LAMMPS.} Data from Figure~\ref{fig:lammps-throughput}, replotted to show the average duration of each timestep.
    }
    \label{fig:lammps-step}
\end{figure*}

\clearpage

\bibliography{references.bib}             
